\documentclass[aps,prx,twocolumn,superscriptaddress,notitlepage]{revtex4-1}

\usepackage{graphicx}
\usepackage{epsfig}
\usepackage{epstopdf}
\usepackage{float}
\usepackage{subfigure}
\usepackage{bbold}
\usepackage{tikz}
\usetikzlibrary{calc}

\usepackage{dcolumn}
\usepackage{latexsym}
\usepackage{amssymb}
\usepackage{amsmath}
\usepackage{amsfonts}
\usepackage{wasysym}
\usepackage{bm}

\usepackage[colorlinks,bookmarks=false,citecolor=blue,linkcolor=red,urlcolor=blue]{hyperref}
\usepackage{verbatim}

\usepackage{booktabs}

\usepackage{epsfig} 
\usepackage{epic} 
\usepackage{eepic} 
\usepackage{units} 
\usepackage{url} 
\usepackage{longtable} 
\usepackage{mathrsfs} 
\usepackage{multirow} 
\usepackage{bigstrut} 
\usepackage{amssymb} 
\usepackage{graphicx} 
\usepackage{setspace} 
\usepackage{xspace} 
\usepackage{amsmath} 
\usepackage{siunitx} 
\usepackage{booktabs} 

\usepackage[a4paper,top=2cm,bottom=3cm,left=3cm,right=3cm,marginparwidth=1.75cm]{geometry}
\usepackage{subfigure}
\usepackage{braket}
\usepackage{extarrows} 
\usepackage{amsfonts}
\usepackage{indentfirst}

\usepackage{algorithm}
\usepackage{algpseudocode}

\usepackage{indentfirst}
\usepackage{listings}

\usepackage{chemformula}

\newcommand{\ud}{\mathrm{d}}

\newcommand{\lefta}{\left\langle}
\newcommand{\righta}{\right\rangle}

\newcommand{\E}{\mathcal{E}}

\newcommand{\RNum}[1]{\uppercase\expandafter{\romannumeral #1\relax}}
\newcommand{\tx}[1]{\textmd{#1}}
\newcommand{\onefrac}[1]{\frac{1}{#1}}
\newcommand{\refeq}[1]{\textmd{Eq.\,}(\ref{#1})}
\newcommand{\reffg}[1]{\textmd{Fig.\,}\ref{#1}}

\newcommand{\nx}{\mathbf{n}_1}
\newcommand{\ny}{\mathbf{n}_2}
\newcommand{\Ny}{N_2}
\newcommand{\Nx}{N_1}
\newcommand{\kx}{k_1}

\newcommand{\kv}{\mathbf{k}}
\newcommand{\nphi}{n_\phi}
\newcommand{\Nephi}{N^e_\phi}
\newcommand{\Nbphi}{N^b_\phi}
\newcommand{\fsz}{\small}

\newcommand{\clrbl}[1]{\color{blue}{1}}

\def\be{\begin{equation}}
\def\ee{\end{equation}}
\def\bea{\begin{eqnarray}}
\def\eea{\end{eqnarray}}

\def\vec{\mathbf}
\def\mc{\mathcal}

\usepackage{array}
\newcommand{\PreserveBackslash}[1]{\let\temp=\\#1\let\\=\temp}
\newcolumntype{C}[1]{>{\PreserveBackslash\centering}p{#1}}
\newcolumntype{R}[1]{>{\PreserveBackslash\raggedleft}p{#1}}
\newcolumntype{L}[1]{>{\PreserveBackslash\raggedright}p{#1}}

\usepackage{color}

\definecolor{darkblue}{rgb}{0,0.02,0.45}
\definecolor{darkred}{rgb}{0.45,0.02,0}

\makeatletter
\newcommand{\thickhline}{%
\noalign {\ifnum 0=`}\fi \hrule height 0.7pt
\futurelet \reserved@a \@xhline
}
\newcolumntype{"}{@{\hskip\tabcolsep\vrule width 0.7pt\hskip\tabcolsep}}
\makeatother

\usepackage{times}

\begin{document}

\title{   Further insights  into the thermodynamics of the Kitaev honeycomb model}

\author{Kexin Feng}
\affiliation{School of Physics and Astronomy, University of Minnesota, Minneapolis, MN 55455, USA}
\author{Natalia B. Perkins}
\affiliation{School of Physics and Astronomy, University of Minnesota, Minneapolis, MN 55455, USA}
\author{F. J.  Burnell}
\affiliation{School of Physics and Astronomy, University of Minnesota, Minneapolis, MN 55455, USA}
\date{\today}

\begin{abstract}
 Here we revisit the thermodynamics  of the Kitaev quantum spin liquid realized on  the honeycomb lattice. We address two main questions:
First, we   investigate whether  there are observable thermodynamic signatures  of the topological Majorana boundary modes of the Kitaev honeycomb model.
We argue that  for the time-reversal invariant case the residual low-temperature entropy is the primary thermodynamic signature of these  Majorana edge modes, and verify using large-scale Monte Carlo simulations that this residual entropy is present in the full Kitaev model.  When   time-reversal symmetry is  broken,  the  Majorana edge modes are potentially observable in  more direct thermodynamic measurements such as the specific heat, though only at temperatures well below the bulk gap.  %
Second, we study the energetics, and the corresponding thermodynamic signatures, of the flux excitations  in the  Kitaev model. 
Specifically, we study the flux interactions on both  cylinder and torus  geometries numerically,  and quantify their impact
on  the thermodynamics  of the Kitaev spin liquid  by using a polynomial fit for the average flux energy as a function of flux density and extrapolating it to the thermodynamic limit.  By comparing this model to Monte Carlo simulations, we find that flux interactions have a significant quantitative impact on the shape and the position of  the low-temperature peak in the specific heat. 
 \end{abstract}

\maketitle

\section{Introduction}\label{Sec:intro}

The Kitaev model on  the honeycomb lattice~\cite{Kitaev2006}  and related models have  recently attracted much interest in both theoretical and experimental communities~\cite{Knolle2017,Trebst2017,Takagi2019,Motome2019}.  Strikingly, these 2D and 3D Kitaev models have  exact quantum spin liquid (QSL) ground states~\cite{Kitaev2006},  and  can be potentially realized  in
Mott insulating magnets  on tri-coordinated two- (2D) and three-dimensional  (3D) lattices with the strong spin-orbit coupling~\cite{Jackeli2009,Jackeli2010,Rau2016,Trebst2017,Winter2017,Takagi2019}.  

One hallmark of a quantum spin liquid is fractionalization.  Spin excitations in the Kitaev model are fractionalized into two types of quasiparticles:  itinerant spinon-like excitations, which are described by the Majorana fermions which are gapless or gapped depending on the coupling parameters, and localized gapped $\mathbb{Z}_2$ fluxes, also referred to as visons~\cite{Kitaev2006}.
Much effort has been  devoted to searching for traces of such fractionalization in  the spin dynamics of
 $\alpha$-RuCl$_3$~\cite{plumb2014alpha,Majumder2015,Johnson2015,Sears2015,kasahara2018majorana} and H$_3$LiIr$_2$O$_6$~\cite{Kitagawa2018}, which are believed to be proximate to the Kitaev QSL~\cite{Jackeli2009}; evidence suggestive of Majorana fermions has been found in these compounts at temperatures up to 100K.
 Though most of these materials  have magnetically ordered ground states, and hence are not spin liquids,  the idea is that  even if residual long range magnetic order sets in below a certain temperature, the fractionalized quasiparticles of the nearby  QSL phase may still lead to characteristic signatures  in the dynamical response spectrum reminiscent of the nearby QSL ~\cite{Rousochatzakis2018,Knolle2017,Motome2019}.
 One promising route  to look for such signatures  is  by using various dynamical probes, such as inelastic neutron
scattering~ \cite{Banerjee2016, Knolle2014a, Knolle2015,udagawa2019spectroscopy}, Raman scattering with visible light
 \cite{Knolle2014, Sandilands2015,Sandilands2016,Nasu2016, Perreault2015,Perreault2016a,Perreault2016b,Yiping2020,Loosdrecht2019,Dirk2020}, resonant inelastic X-ray scattering
 \cite{Halasz2016,Halasz2017,Halasz2019} and through the phonon dynamics~\cite{Ye2020,Brenig2019}.

 Another important probe of fractionalization in the Kitaev QSL is  thermodynamics \cite{nasu2014vaporization,Nasu2015,Do2017,Kato2017,Motome2019,Eschmann2019}. Not surprisingly, the emergent fractionalized quasiparticles  of the Kitaev QSL reveal themselves in the thermodynamic behavior in a peculiar manner.  In the 2D Kitaev QSL,  two characteristic crossovers are seen in the specific heat, indicating a two-stage release of magnetic entropy.  The first is  associated with itinerant fermionic excitations; the second with the localized $\mathbb{Z}_2$ fluxes~\cite{nasu2014vaporization,Nasu2015}.  In 3D Kitaev QSLs, flux freezeout is associated with a phase transition in which topological order is lost.   Under certain conditions in three-dimensional tri-coordinated lattices, coexistence between the low-temperature chiral QSL phase and crystalline ordering of the $\mathbb{Z}_2$ fluxes has also been observed numerically  ~\cite{Kato2017}.

The goal of the present work is to revisit the thermodynamics of the Kitaev model on the honeycomb lattice with  two main foci.
 First, we scrutinize the impact  of the boundary geometry  on the specific heat.  In the presence of a boundary, one distinctive signature of the Kitaev spin liquid phase is the existence of topologically protected Majorana boundary modes~\cite{kells2010zero,matsuura13,o2016classification,Perreault2016a,Perreault2016b,Schaffer2016}.  Because these protected boundary modes are associated with  fractionalized Majorana fermion excitations, they are more difficult to detect than the boundary modes characteristic of topological insulators and superconductors, and thermodynamic measurements have been proposed~\cite{o2016classification} as one approach.   Our second focus is to study the impact of flux  interactions on the thermodynamics of the Kitaev model, and present a quantitative model of the low-temperature specific heat that we argue captures the main features of the crossover in the thermodynamic limit.

For the first question, our main results are as follows.  
We show that  in principle, the Majorana boundary modes can be   observed in the thermodynamics.  Specifically, we find that when time-reversal symmetry is broken, or with unbroken time-reversal symmetry on very narrow nanoribbons, the Majorana boundary modes lead to a low-temperature power law in the specific heat that differs from the $C(T) \sim T^2$ power-law behavior of the bulk.    When time-reversal symmetry is unbroken, however, the temperatures at which these power laws occur in currently available materials are well below the range accessed by current experiments.  In this case, the Majorana boundary modes can be detected only indirectly, from their contribution to the residual low-temperature entropy.  
We  expect that this residual entropy can be detected using methods similar to those employed in spin ice, where the macroscopic ground state degeneracy contributes to a finite entropy density at ultra-low temperatures ~\cite{wang2006artificial, ramirez1999zero, bramwell2001spin},  which has been detected  
using highly accurate measurements of the specific heat \cite{ramirez1999zero, lau2006zero, higashinaka2003anisotropic}.   For Kitaev materials, such experiments offer an alternative to resonant Raman scattering, which can also be used to detect the fractionalized Kitaev boundary modes under certain conditions  \cite{Perreault2016b, Perreault2015}.

For the second focus, we study in detail several aspects of the energetics of flux excitations in the Kitaev model that have not been scrutinized in the literature.  First, we show that on a lattice with open boundaries, the energy cost of a single flux is significantly reduced near the boundary relative to its bulk value; this has a noticeable impact on the specific heat, as it leads to a lower onset temperature for flux excitations on the cylinder relative to the torus.  
Second, we study the energetics of fluxes at finite density.  
We use the results to construct a model of the flux-only contribution to the specific heat in the thermodynamic limit, and show that this has good quantitative agreement with the low-temperature peak in the specific heat obtained from the Monte Carlo (MC) simulations in finite systems.  
This affirms the conclusion of previous numerical work~\cite{Nasu2015} that in the 2D Kitaev model there is no finite-temperature phase transition, but rather a cross-over 
from a low-temperature region with vanishingly small flux density to an intermediate temperature region where fluxes (but not fermionic excitations) have proliferated. 
 
 Our model is based on an analysis of the
  distribution  of the energies of different flux configurations  at fixed flux density.   We show that for sufficiently large torus lattices, this distribution is sharply peaked and is approximately independent of lattice size.  Therefore, we can model flux thermodynamics by numerically fitting the average energy as a function of flux density.   A similar fit is obtained on the cylinder, by separately accounting for flux densities in the bulk and on the boundary.  These models shed light on the flux's contribution to  the low-temperature specific heat, and help identify the role of the boundary fluxes therein. 
Since the best-fit energy depends only on the flux density,  we refer to it  as to the pseudo-potential energy (PPE), by analogy with the local density approximation (LDA) in density functional theory \cite{Hohenberg1964,Jones1989}. 
 In Section \ref{Sec:interactions}, we  show that some features of  the multi-flux interactions can be understood by looking into the microscopics of the  two-flux interactions.

 The paper is organized as follows.  In Sec.\ \ref{sec: Model} we briefly review the exact solution of the Kitaev model \cite{Kitaev2006}. In Sec.\ \ref{Edge modes}, we analyze the energy spectrum of the fermionic boundary modes in the flux-free sector and identify the signatures of these modes in the low-temperature scaling behavior of the specific heat. In Sec.~\ref{finiteTedgemodes}, we show that the  residual entropy from the fermionic boundary modes can be observed in MC simulations of the Kitaev model. 
 In Sec.\ \ref{chap: flux energetics}, we study the flux energetics  in the  time-reversal symmetric case. Using  best-fit polynomials to describe the PPE, we propose phenomenological flux models for both torus and cylinder lattices, and use these 
to  describe the flux thermodynamics in the thermodynamic limit.   
Next, Sec.\ \ref{sec: mag_flux_energetics} focuses on the flux energetics when time-reversal symmetry is  broken. We analyze how the specific heat changes with varying magnitude of the time-reversal breaking term.  Finally, to better understand the resulting flux PPE models, in Section \ref{Sec:interactions} we examine the two-flux interactions, and show that
they  capture the essence of the flux energetics in the multi-flux systems  Finally, in Sec.~\ref{sec: summary}, we summarize the main results of this paper.


\section{The model}\label{sec: Model}
\begin{figure}
    \centering
       \includegraphics[width =0.9
    \linewidth]{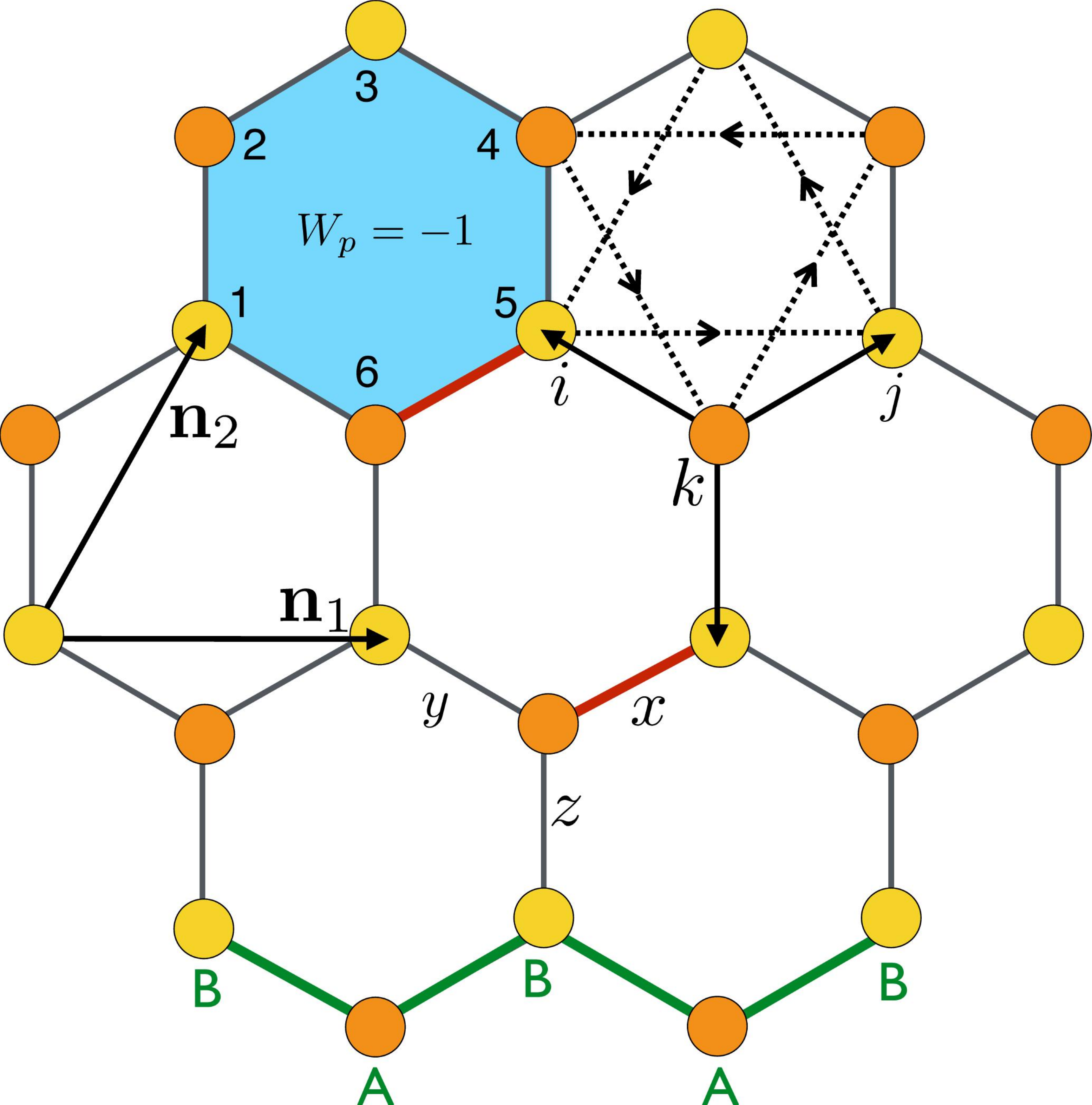}
    \caption{Kitaev model on the honeycomb lattice. The unit vectors $\nx=(1,0)$, $\ny=(\frac{1}{2},\frac{\sqrt{3}}{2})$.
      The blue hexagon shows a $\pi$ flux, i.e. the eigenvalue of  $\hat{W}_p = \sigma_1^x \sigma_2^y \sigma_3^z \sigma_4^x \sigma_5^y \sigma_6^z$ is equal to -1. The convention for the sign of the NN couplings on the bonds  labelled by $x$, $y$, and $ z$ is as follows: an arrow pointing from site $k$ to $i$ means  $u_{ki}$ on the corresponding bond $(k,i)$ in \refeq{eq: Hamiltonian} is positive. The sites $ (i,k,j)$  are an example of the NNN triplet used in Eq. (\ref{eq: Hamiltonian}).   The solid green line shows the zigzag edge of the finite size system.}
    \label{fig:lattice}
\end{figure}

The extended Kitaev model on a honeycomb lattice is given by the Hamiltonian~\cite{Kitaev2006}:
\begin{equation}
    H  = -\sum_{\langle i, j\rangle_\alpha}J^{\alpha} \sigma_i^{\alpha} \sigma_j^{\alpha} - \kappa \sum_{\langle i,j,k \rangle} \sigma^\alpha_i \sigma^\beta_j \sigma^\gamma_k,
 \label{eq-H}
\end{equation}
where  $J^{\alpha}$  in the first term denotes the nearest neighbor (NN) Kitaev interaction  on  the corresponding bond of type $\alpha=x,y,z$ (see Fig.\ \ref{fig:lattice}), and  $\sigma^\alpha_{\bf r}$ are the Pauli matrices. The second term is a three-spin interaction on the three adjacent sites (see Fig.\ \ref{fig:lattice}) of strength $\kappa$, which breaks time-reversal symmetry (TRS).
It mimicks the effect of a magnetic field but preserves exact solubility of the model~\cite{Kitaev2006}.   

At the heart of exact solvability is the macroscopic number of local symmetries in the plaquette operators $[H, \hat{W_p}] = 0$, where $\hat{W_p}$ are defined as $\hat{W}_p = \sigma_1^x \sigma_2^y \sigma_3^z \sigma_4^x \sigma_5^y \sigma_6^z$ (see Fig.\ \ref{fig:lattice}). 
These operators commute with the Hamiltonian and take eigenvalues of $\pm 1$. 
Thus the Hilbert space can be divided into eigenspaces of $\hat{W_p}$, and the ground state is the one with all $W_p$ equal to one, which is also referred to as the flux-free sector.

Using Kitaev's representation of spins in terms of  Majorana fermions~\cite{Kitaev2006}, $\sigma_j^\alpha = i b_j^\alpha c_j$, we  rewrite the spin Hamiltonian (\ref{eq-H})  as
\begin{align}
\label{eq: Hamiltonian}
H\! & \!= \!\!&\!\!-\!\! \sum_{\langle i j \rangle_{\alpha}} \!J^{\alpha_{ij}} i \hat{u}_{ij} c_i c_j \! -\! \kappa\!\sum_{\langle \langle i j \rangle \rangle} \!\! i \hat{u}_{ik} \hat{u}_{jk} c_i c_j .
\end{align}
where  the bond operators are   defined as $\hat{u}_{ij}=i b_i^{\alpha} b_j^{\alpha}$,  and $\langle i j \rangle$ and $\langle \langle i j \rangle \rangle$ denote NN and  next NN  (NNN) bonds, respectively.   When $\kappa\neq 0$ and time reversal  symmetry is broken, \refeq{eq: Hamiltonian} is  closely  related  to  the  Haldane  model  for  the anomalous quantum Hall effect~\cite{Haldane1988}. 

The Hilbert space of the fermionic model is larger than that of the spin model; the latter is recovered when we impose the constraint $b^x_j b^y_j b^z_j c_j |\Psi \rangle=|\psi \rangle$ at each site $j$ of our lattice.  
This constraint commutes with the Hamiltonian, as well as with the  bond operators $\hat{u}_{ij}$.  Thus the  eigenvalues  $u_{ij} = \pm 1$ of the operators $\hat{u}_{ij}$ are constants of motion of the model (\ref{eq: Hamiltonian}), and can be understood as $\mathbb{Z}_2$ gauge fields.  This picture captures the fact that not all choices of $\{ u_{ij} \}$ correspond to distinct physical states of the spin model, and only those that are gauge inequivalent should be treated as distinct.  
In the fermionic representation, $\hat{W_p} = \prod_{(i,j)\in \text{edge}(p)} \hat{u}_{ij}$,   
and $\hat{W_p}$ can be understood as a gauge invariant Wilson loop operator around a single plaquette $p$, with the eigenvalue $W_p = -1$ corresponding to a (gapped) $\pi$-flux excitation on the plaquette in question.  This is why in the following we  will also use notation $ \{\phi_p \}$  to denote 
a particular  flux sector, described (in a given gauge) by choosing  a particular configuration of the eigenvalues $\{{u}_{ij}\}$.

\begin{figure}[!b]
    \centering
       \includegraphics[width=.5\columnwidth]{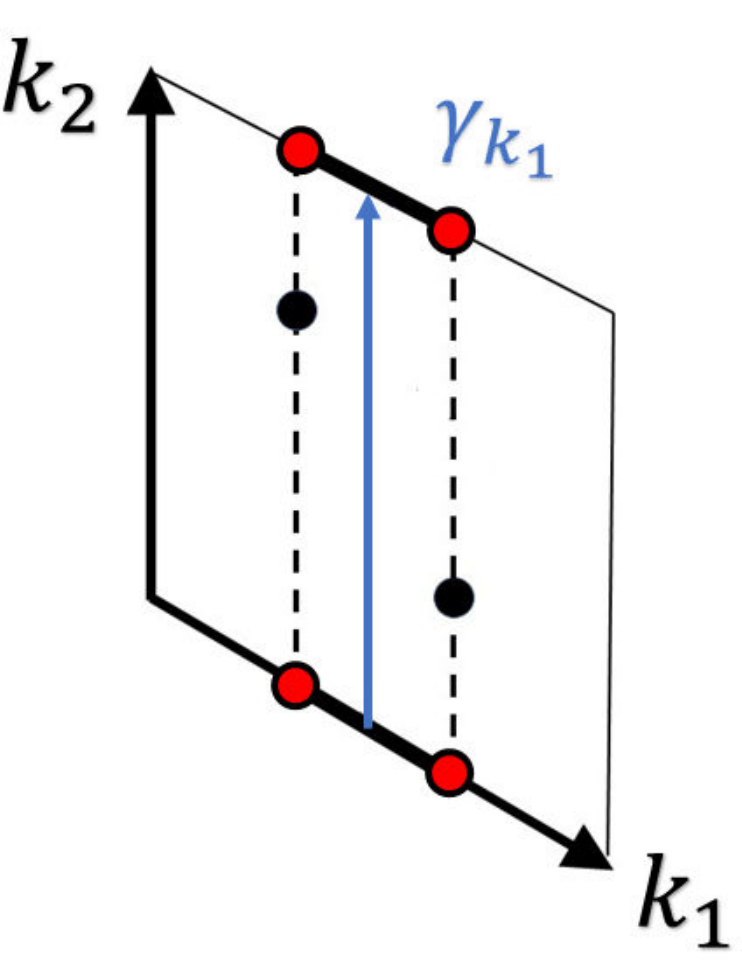}
    \caption{The 2D Brillouin zone with two Dirac points. $\gamma_{k_1}$ (blue vertical line) is the integration loop path along which the winding number is calculated. The bold segment on the $\mathbf{k_1}$-axis denotes the region where the gapless edge mode exists (in the limit of large system size) when the lattice is made open in the $\nx$ direction. It is referred to as topologically nontrivial region.}
    \label{fig: 2d BZ}
\end{figure}

\subsection{Topological band structure of the Kitaev Spin Liquid}
\label{sec: edge mode review}

One interesting feature of the Kitaev spin liquid is that, with $u_{ij}$ chosen such that the hopping matrix element from the $A$ sublattice to the $B$ sublattice is $+i$ , for $\kappa =0$ Eq. (\ref{eq: Hamiltonian}) describes a  band structure in  the symmetry class BDI whose band crossings are topologically protected~\cite{o2016classification}, leading to a protected, zero energy band of states at the system's boundary in the thermodynamic limit~\cite{matsuura13}.  These topologically protected boundary (or edge) modes are characteristic of the Kitaev QSL {\it phase}, rather than the specific Kitaev Hamiltonian.  When  time-reversal symmetry is broken for $\kappa \neq 0$,  the bulk band structure becomes gapped with a non-vanishing Chern number, leading to topologically protected chiral edge modes. We briefly review the nature and origins of these Majorana edge modes, as this understanding is crucial for the remaining discussion.


{
To understand the topology of the band structure, it is helpful to write the Majorana fermion Hamiltonian Eq.~(\ref{eq: Hamiltonian}) in the zero-flux sector   as $H = \sum_{\kv} \vec{c}_{\kv}^T H_{\kv}  \vec{c}_{\kv}$, where $\vec{c}_{\kv}^T = (c_{A, \kv}, c_{B, \kv})$, and }
 \begin{equation} \label{HCH2}
 H_\kv = \vec{d}_\kv \cdot \boldsymbol{\sigma} \ , 
 \end{equation}
  where  $\boldsymbol{\sigma}$ is a Pauli matrix over the 2 sublattices of the honeycomb lattice  \cite{Schaffer2016,Perreault2016b}.   
Explicitly, for $\kappa=0$, we have 
\begin{align} \label{HCH}
d^x_\kv &=-J^x \sin k_2 - J^y\sin (k_2- k_1)  ,\nonumber \\
d^y_\kv &=-J^z -J^x \cos k_2 - J^y \cos (k_2- k_1)  , \nonumber \\
d^z_\kv &=0,
\end{align}
where $k_i \equiv {\bf k} \cdot {\bf n}_i$.
The Fermi surface then consists of the pair of Dirac points obtained by taking $d^x_\kv =0$ and $d^y_\kv =0$.  
Time-reversal symmetry and particle-hole symmetry constrain the $\vec{d}_{\bf k} $ vector to lie in the plane of  the 2D  honeycomb lattice, and the Hamiltonian  (\ref{HCH2}) can be used to define the topological winding number
 \cite{Schaffer2016, zhao2013topological,o2016classification}:   
\begin{align}\label{nu}
    \nu[\gamma_{k_1}] &= \frac{1}{4\pi i} \int  \ud k_2 \,  \text{Tr}\,\left[ H^{-1}_{\mathbf k} \sigma^z \partial_{k_2} H_{\mathbf k} \right] \nonumber 
    \\
    &= \frac{1}{2\pi} \int  \ud{k_2} \,\text{Im}\, \left[ \frac{\partial_{k_2} \Gamma_{\mathbf k}}{\Gamma_{\mathbf k}} \right] ,
\end{align}
where $\gamma_{k_1}$ is a path that traverses the Brillouin zone  in the $k_2$-direction at fixed $k_1$.  For $J^x = J^y = J^z=J$, the winding number $\nu[\gamma_{k_1}]=1$ if $\kx \in [\frac{2\pi}{3}, \frac{4\pi}{3}]$, and is 0 elsewhere.  

On a cylinder with open boundaries in the ${\bf n}_2$ direction and periodic boundaries in the ${\bf n}_1$ direction, the effective Hamiltonian at a fixed value of the conserved momentum $\kx$  describes a 1D Majorana chain with open boundary conditions.  The topologically non-trivial winding implies that each  1D Majorana chain at fixed $\kx \in [\frac{2\pi}{3}, \frac{4\pi}{3}]$ hosts zero-energy topologically protected  Majorana boundary modes \cite{ryu2010topological,bernevig2013topological,Schaffer2016,Perreault2016b}.


Breaking  time-reversal symmetry by taking $\kappa \neq 0$ introduces a diagonal term $d^z_\kv\sigma_z$ into the bulk Hamiltonian   (\ref{HCH2})  with 
\begin{equation} \label{Eq:dz}
d^z_\kv =2 \kappa ( \sin (k_2) + \sin (k_1-k_2 )- \sin (k_1 )  ) \ .
\end{equation}
  The bulk band structure becomes gapped and acquires  non-vanishing Chern number~\cite{Kitaev2006}.  This destroys the boundary flat-bands described above and leads instead to chiral edge modes and a corresponding thermal Hall conductance. 

\section{Thermodynamic Signatures of Majorana edge modes}\label{Edge modes}

 The topological boundary modes of the Kitaev spin liquid are a distinctive signature of the spin liquid phase at temperatures well below the flux gap, where flux excitations are exponentially suppressed~\footnote{Though strictly speaking the classification of nodal topological phases requires translation invariance, it is easy to check numerically that 0-energy states localized at the boundary remain a feature of the spectrum with weak disorder -- and in fact persist up to strong disorder.}.
 In this section, we discuss under what conditions these gapless  edge  modes can be experimentally detected with low-temperature thermodynamic probes by studying the fermionic contribution to the specific heat -- i.e. by studying the specific heat of the model (\ref{eq: Hamiltonian}) in the absence of fluxes.  We expect this to be a good description of the actual specific heat of the Kitaev QSL at temperatures well below the flux gap.  Throughout this section, we work on a cylindrical lattice
 with open boundaries in the ${\bf n}_2$ direction and periodic boundaries in the ${\bf n}_1$ direction (see Fig.\ \ref{fig:lattice}).  We let $\Ny$ denote the number of unit cells in the $\ny$ direction.  Since the system is translationally invariant in  the ${\bf n}_1$ direction, we set the distance $a$ between two A sites to $a = 1$, and diagonalize the Hamiltonian  (\ref{HCH2}) for each $\kx$ to obtain the dispersion.

\subsection{Specific heat of fermions}

First, we briefly review how the low-temperature fermionic specific heat is calculated.  Suppose that the fermionic dispersion in an energy window  $\epsilon_{\text{min}} \leq \epsilon \leq \epsilon_{\text{max}}$ is well approximated by a power law of the form $\epsilon-\epsilon_{\text{min}}  \sim k^\alpha$ ($\alpha \geq 1)$.   The density of states  in this energy window is  given by
\begin{align}
    D (\epsilon) &={\mc A} \int \delta\left(\epsilon-\epsilon_{k}\right) \ud k \nonumber \\  
   &={\mc A} (\epsilon- \epsilon_{\text{min}})^{-(1-\onefrac{\alpha})},\,\alpha \geq 1.
   \end{align}
   where $ {\mc A}$ is a normalization constant, which in general will depend on the choice of the interval $(\epsilon_{\text{min}},\epsilon_{\text{max}})$.  
The corresponding contribution to the fermionic specific heat can be evaluated via:
\begin{align}
\label{eq: C_fermion1}
C &={\mc A} \onefrac{T^2}\int_{\epsilon_{\text{min}}}^{\epsilon_{\text{max}}}  D(\epsilon)\frac{\epsilon^2 e^{\beta\epsilon}}{(e^{\beta\epsilon}+1)^2} \ud\epsilon \\
&= {\mc A} T^{\onefrac{\alpha}} \int_{ x_0}^{ \beta \epsilon_{\text{max}}} (x -x_0)^{-(1-\onefrac{\alpha})}  \frac{x^2 e^{x} }{ (e^{x} + 1)^2 } \ud x
\label{eq: C_fermion}
\end{align}
where  $\beta=1/T$, and $x_0=\beta \epsilon_{\text{min}}$.  
Thus we find
\begin{align} 
     C &=  {\mc A} I\, T^{\onefrac{\alpha}}  \label{eq: power law 0}   
\end{align}
where 
\begin{align} \label{eq:Idef}
I =  \int_{\epsilon_{\text{min}}}^{\epsilon_{\text{max}}}  (x - \beta \epsilon_{\text{min}})^{-(1-\onefrac{\alpha})}  \frac{x^2 e^{x} }{ (e^{x} + 1)^2 } \ud x
\end{align}
To obtain the low temperature limit, we take $\epsilon_{\text{min}} \ll T$, in which case we can safely treat $\epsilon_{\text{min}} $ as $0$.  In this case, for $T \ll  \beta \epsilon_{\text{max}}$, we have 
\begin{align} \label{Eq:Int}
I &\approx  \int_{0}^{\infty} x^{1+\onefrac{\alpha}}  \frac{ e^{x} }{ (e^{x} + 1)^2 } \ud x 
\nonumber \\
&=
\sum_{l=0}^{\infty}(-1)^l G(\onefrac{\alpha}+2, l+1).
\end{align}
Here we define $G(\nu, l) = \int_0^{\infty}x^{\nu-1} (\cosh{x})^{-l}\ud x$, which is converging for $\nu>0, l\in \mathbb{N}$.
When the temperature increases,  $\beta \Delta^{\text{fs}}$ becomes finite and
 the integral in Eq. (\ref{eq: C_fermion}) needs to be computed more accurately.  
 
 Our calculation gives the contribution of states within the energy window $(\epsilon_{\text{min}}, \epsilon_{\text{max}})$ to the specific heat, for any temperature.  However, at temperatures comparable to $\epsilon_{\text{max}}$, other, higher energy states also begin to contribute to the specific heat via $C = \frac{{\mc A}}{T^2} \int_{\epsilon_{\text{max}}}^{\infty} D(\epsilon) \frac{ \epsilon^2 e^{\beta \epsilon} }{(e^{\beta \epsilon} + 1)^2} d \epsilon$.  
 
 In what follows, we will discuss the appropriate energy window, and corresponding values of $\alpha$ and $n_{\epsilon_{\text{max}}}$, that describe the topological boundary modes of both the nodal topological Kitaev QSL with $\kappa =0$, and the gapped topological Kitaev QSL obtained for $\kappa >0$.  

\begin{figure*}
    \includegraphics[width=0.85\textwidth]{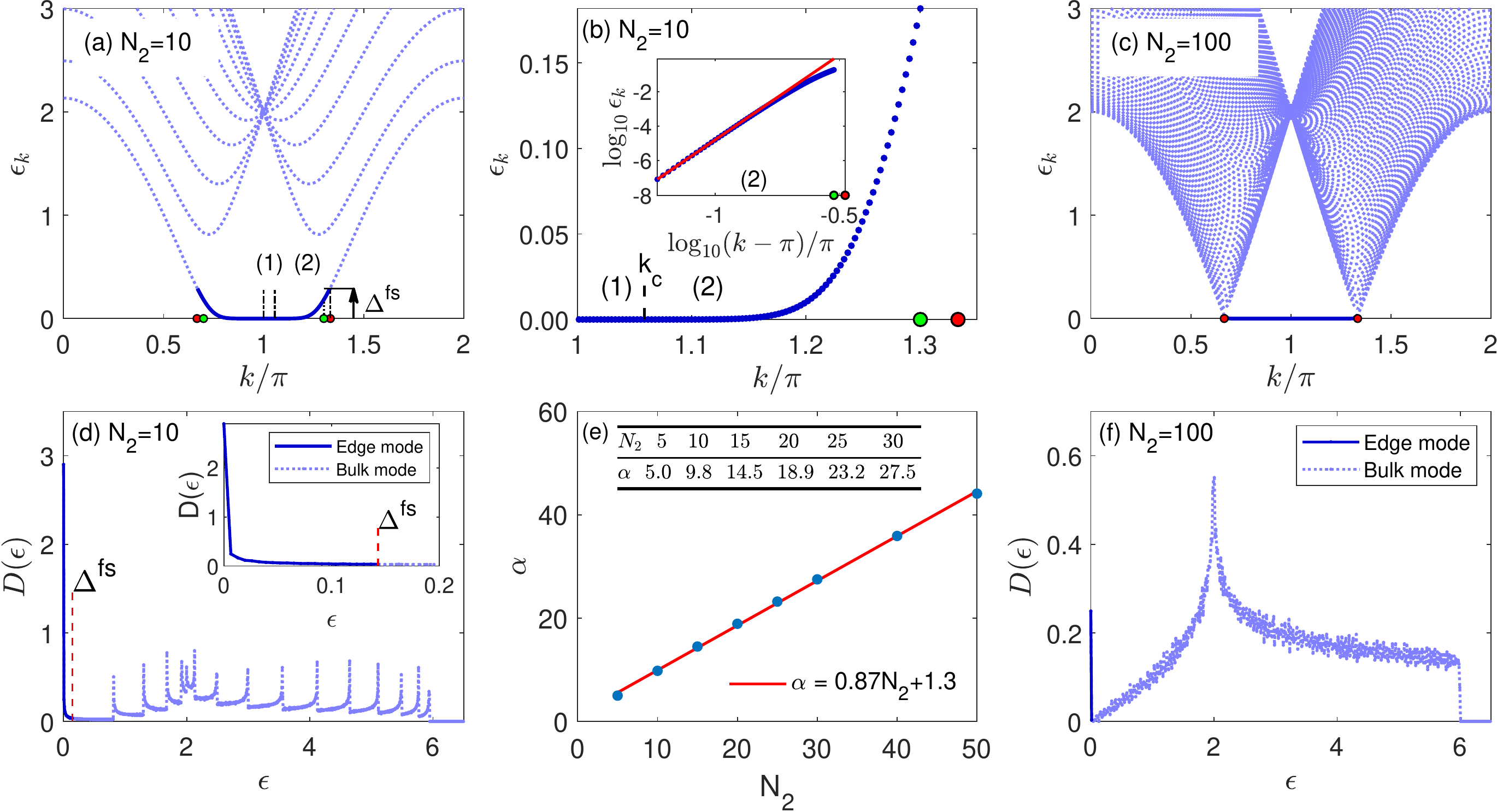}
    \caption{ \label{fig:cylband}
     Fermionic  spectrum and density of states (DOS) of the Majorana Fermion Hamiltonian (\ref{eq: Hamiltonian}) with $\kappa=0$ obtained on the cylinder lattices  with  $N_2=10$ ((a) and (d))  and   with $N_2=100$ ((c) and (f)).  The nearly-flat lowest energy band in (a) and (c) appears as a near-zero energy peak in the DOS.      Here  red dots indicate the position of the bulk Dirac points projected to $k_1$ (i.e. $\kx = \frac{2 \pi}{3},\frac{4 \pi}{3}$), and $\Delta^{\text{fs}}$ corresponds to the energy of the lowest band at these projected Dirac points. 
 Green dots indicate the analytical solution~\cite{wakabayashi2010electronic}  for  the boundary of the  $k_1$ region inside which the states in the lowest band are exponentially localized to the edges.  
In (a) and (b), the regions (1) and (2)  denote  $[\pi, k_{c}]$ and $[k_{c}, 4 \pi /3]$, respectively. $\Delta^{\text{fs}}$ and $k_c$ for different values of $\Ny$ are  given in Table \ref{tab: kc}. 
    (b) A zoomed-in view of  the lowest band obtained on the cylinder lattice  with $N_2=10$.  The inset shows the zoomed-in  edge mode spectrum in the log-log scale (blue dots), with the red line  showing an approximate power-law fit to  the dispersion in the leftmost one fifth of region (2).  We use this to approximate the dispersion throughout region (2) by a power law, noting however that this deviates substantially from the exact dispersion before the boundary of the region containing exponentially localized edge modes (green dot).  (e) The $\Ny$-dependence of the scaling powers of the edge mode dispersion, obtained as shown in (b).  Blue dots show the exact values (also quoted in the Table in the inset).  The red line $\alpha = 0.87 \Ny + 1.3$ is the linear fit of these data points. } 
\end{figure*}

 \subsection{ Time-reversal symmetric Kitaev QSL ($\kappa=0$)}

We first consider $\kappa=0$, where the edge state consists of a boundary Majorana flat band in the thermodynamic limit.   Clearly the entropy associated with this flat band never freezes out; in the thermodynamic limit its thermodynamic signature is a residual zero-temperature entropy, which we discuss in detail in Sec. \ref{Sec:Sinfinity}.   However, when the lattice size is finite, the edge modes localized at opposite boundaries have finite overlap, and the boundary bands have finite-size splitting everywhere except at $\kx = \pi$.  Here we determine when this finite-size splitting is large enough to  be observed in thermodynamic measurements.  We quote all energies  in units of $J$, and set the Boltzmann constant $k_B=1$.

As described above, in order to calculate the specific heat, we first need to understand the low-energy spectrum of our model.  To this end,
 Fig.\ \ref{fig:cylband}  illustrates the effects of finite-size on the fermionic band structure, which is particularly clear from the comparison of the fermion energy spectrum for cylinder lattices with $\Ny = 10$ and $\Ny = 100$ shown in Figs. \ref{fig:cylband} (a) and (c), respectively. 
 The finite $\Ny$ manifests itself in two main ways.  First, the Dirac cones are lifted, and the purely bulk bands (i.e. the second and higher sub-bands) acquire a finite-size gap, which decreases  as $1/\Ny$.   As can be seen in Fig. \ref{fig:cylband} (c), for $\Ny = 100$ the resulting bulk gap is extremely small compared to the bandwidth. Second, states in the lowest sub-band localized at opposite edges of the system have finite overlap, leading to a finite-size splitting that falls off exponentially in $
\Ny$.
In Fig. \ref{fig:cylband} (b), in order to present the details of this finite-size effect, we show the zoomed-in dispersion of the the lowest band for $\Ny = 10$.

We numerically  introduce the scales $\epsilon_{\text{max}}$ and $\epsilon_{\text{min}}$  for this band structure as follows.  
We take   $\epsilon_{\text{max}}$ to be the energy of the lowest sub-band at the projected Dirac points $\kx = \frac{2 \pi}{3},\frac{4 \pi}{3}$ (shown by  red dots in Fig.\ \ref{fig:cylband} (a)-(c)), which is non-zero due to finite-size effects. Thus, we denote  $\epsilon_{\text{max}} = \Delta^{\text{fs}}$. 
Note that states in the lowest sub-band become delocalized in the bulk at a slightly lower energy than this, at the momentum indicated by the green dot in Fig. \ref{fig:cylband} (see  Ref.\ \onlinecite{wakabayashi2010electronic} for exact expressions); however this difference vanishes as $\Ny$ increases, and is already small for $\Ny=10$.  
We 
 take $\epsilon_{\text{min}} = 10^{-7} J$ to be a low-temperature cutoff below which thermodynamic measurements cannot be performed.  For the value $J \approx 100K$, this corresponds to a temperature scale of $10^{-5}$K.   
 The corresponding momentum, $k_{c}$, depends on the value of $\Ny$ (see Table \ref{tab: kc}), approaching $4 \pi/3$ as $\Ny$ increases.  

To understand the contribution of modes in the energy window $ (\epsilon_{\text{min}},\epsilon_{\text{max}})$ to the specific heat, 
 we approximate 
the dispersion in the lowest sub-band in the region $k_{c} \leq \kx \leq 4 \pi/3$ (region (2) in Fig. \ref{fig:cylband} (a) and (b)) with a power law: $\epsilon_{\text{edge}}^{\text{fs}} \sim k^\alpha,\, \alpha \geq 1$.     Fig.\ \ref{fig:cylband} (b) (inset) shows this fit on log-log scale for $\Ny=10$; the best-fit power $\alpha$  for different values of  $\Ny$  is presented in Fig. \ref{fig:cylband} (e).   We emphasize that because the dispersion is not an exact power law (see Ref.  \cite{wakabayashi2010electronic}), the precise value we obtain for $\alpha$ depends on the range of momenta that we include in our scaling.   Here, we obtain $\alpha$ by fiting only the left-most fifth of region (2), meaning that our power law is a good approximation for the dispersion at the lowest energies, but becomes less accurate as we approach $\Delta^{\text{fs}}$.  Irrespective of this choice, however, we find that $\alpha$ is large compared to $1$, and  grows linearly with $\Ny$, reflecting the fact that for $\pi \leq \kx < 4 \pi/3$, the lowest sub-band becomes flatter as $\Ny$ increases.  For the range of momenta that we fit to, we find $\alpha \approx 0.87\,N_2 + 1.3$, with exact values shown in the inset of Fig. \ref{fig:cylband} (e).

  \begin{table}
    \begin{center}
    \begin{tabular}{l|cccccccccccc}
    \hline
    $\Ny$ & \fsz 5 & \fsz 10  & \fsz 25  & \fsz 50  & \fsz 100  & \fsz 500 & \fsz 1000 \\ 
    \hline
    $k_{c}/ \pi$ & \fsz 1.011 & \fsz 1.06  & \fsz 1.166  & \fsz 1.235  & \fsz 1.277 & \fsz 1.319 & \fsz 1.326\\
    \hline
    $r_c$ & \fsz 0.967 & \fsz 0.82 & \fsz 0.502 & \fsz 0.296  & \fsz 0.169  & 0.042 & 0.021\\
    \hline
    $\Delta^{\tx{fs}}$ &  \fsz 0.569 & \fsz 0.299 & \fsz 0.123 & \fsz 0.062 & \fsz 0.031 & \fsz 0.006 & \fsz 0.003  \\
    \hline
    \end{tabular}
    \caption{The quantities $k_c, r_c$ and $\Delta^{\tx{fs}}$ for a variety of cylinder widths.  Here $\Ny$ is the number of unit cells along $\ny$ direction on a cylinder lattice.  The momentum $k_{c}$, which indicates the boundary between regions (1) and (2), is given in units of $\pi$. $r_c$ is the ratio of the length of the region (2)  to the length of the topologically nontrivial region $[2\pi/3, 4\pi/3]$. $\Delta^{\tx{fs}}$ is the fermionic energy at the projected Dirac point. }
    \label{tab: kc}
  \end{center}
\end{table}

\begin{figure}
    \includegraphics[width=1\columnwidth]{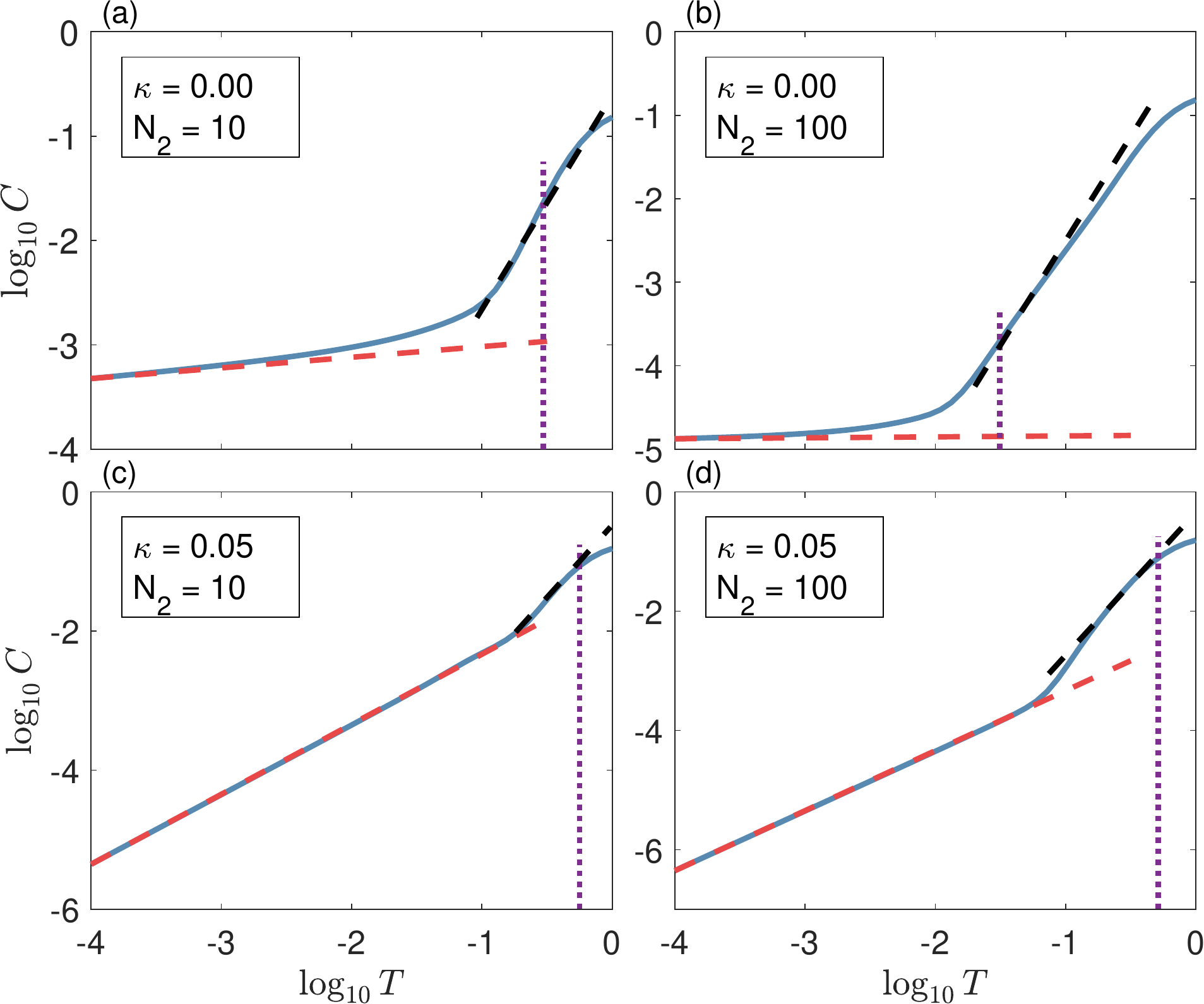}
    \caption{The total fermionic specific heat, $C_{\tx{tot}}=C_{\tx{edge}}+C_{\tx{bulk}}$, in the flux-free sector for different $\kappa$ and $\Ny$. 
    The  inverse of the slope of the red dashed line gives the power of the low-temperature scaling behavior of the specific heat: (a) $\alpha = 9.8$, (b) $\alpha = 88.3$, (c) $\alpha = 1$, and (d) $\alpha = 1$.  The black dashed line shows the expected bulk fermionic contribution to the specific heat with $\alpha = 2$. The purple dotted line marks the temperature  corresponding to $\Delta^{\tx{fs}}$, which is virtually indistinguishable from $\epsilon_{\text{max}}$ on a logarithmic scale. } 
    \label{fig: fermionic_C}
\end{figure}

We may now use Eq. (\ref{eq: power law 0}) to evaluate the  contribution of the boundary modes to the specific heat.  The total number of states in  region $(2)$ is approximately
  $\frac{1}{3} N_1 r_c $, where $N_1$ is the number of unit cells in the $\nx$ direction.  Here $r_c =  (4 -3 k_{c}/ \pi )$ is the length of region (2) divided by $\pi/3$ (which is half the length of the topologically nontrivial region), and $\frac{1}{3}N_1$ gives the number of states in the topologically non-trivial region.  
  The remaining boundary states, contained in  the  region (1) in Fig. \ref{fig:cylband} (a) and (b), do not contribute to the specific heat in the temperature range of interest (where $T$ is large compared to $\epsilon_{\text{min}}$); instead  they contribute to the zero-temperature entropy.      Therefore, for $\epsilon_{\text{min}} \ll T \ll  \epsilon_{\text{max}}$ the specific heat obtained in
   Eq.(\ref{eq: power law 0} ) becomes
\begin{align}
     C_{\tx{edge}} &=  \onefrac{3}  r_c N_1 I T^{\onefrac{\alpha}}  \label{eq:CvKappa0}   
\end{align}
where  $I$ is given 
 by Eq. (\ref{Eq:Int}).  Note that as $\Ny\rightarrow \infty$, $r_c$ (and hence the length of region (2)) approaches $0$, as shown in Table \ref{tab: kc}, while $T^{1/\alpha}$ approaches unity.  Thus for fixed $N_1$, the edge's contribution to the specific heat vanishes as $\Ny$ increases.

At temperature scales on the order of $\Delta^{\text{fs}}$, the low-temperature fermionic specific heat also has a contribution from the bulk states, which in practise dominates over the contribution from the edge.  
This is shown in Fig.\ \ref{fig: fermionic_C}(a) and (b), which plot the total fermionic specific heat as a function of temperature for $\kappa =0$ and $\Ny = 10$ and $100$, respectively.  The solid blue line is obtained from the exact fermionic density of states in the absence of fluxes, shown for $\Ny=10$ and $100$ in Figs.\ \ref{fig:cylband}(d) and (f), respectively.  At temperature scales larger than $ \Delta^{\text{fs}}$, the specific heat is approximately proportional to $T^2$ (black dashed line), which is the usual bulk power law from the Dirac cone.  
We see a cross-over from this bulk power law to the boundary power law in Eq. (\ref{eq:CvKappa0}), at a temperature scales of approximately  $0.1 \epsilon_{\text{max}}$, below which the specific heat changes much more slowly with temperature.   The red dashed line indicates our prediction of the low temperature specific heat based on the power $\alpha$ shown in Fig. \ref{fig:cylband}(e).

\begin{figure}[t!]
\centering
\includegraphics[width=1\columnwidth]{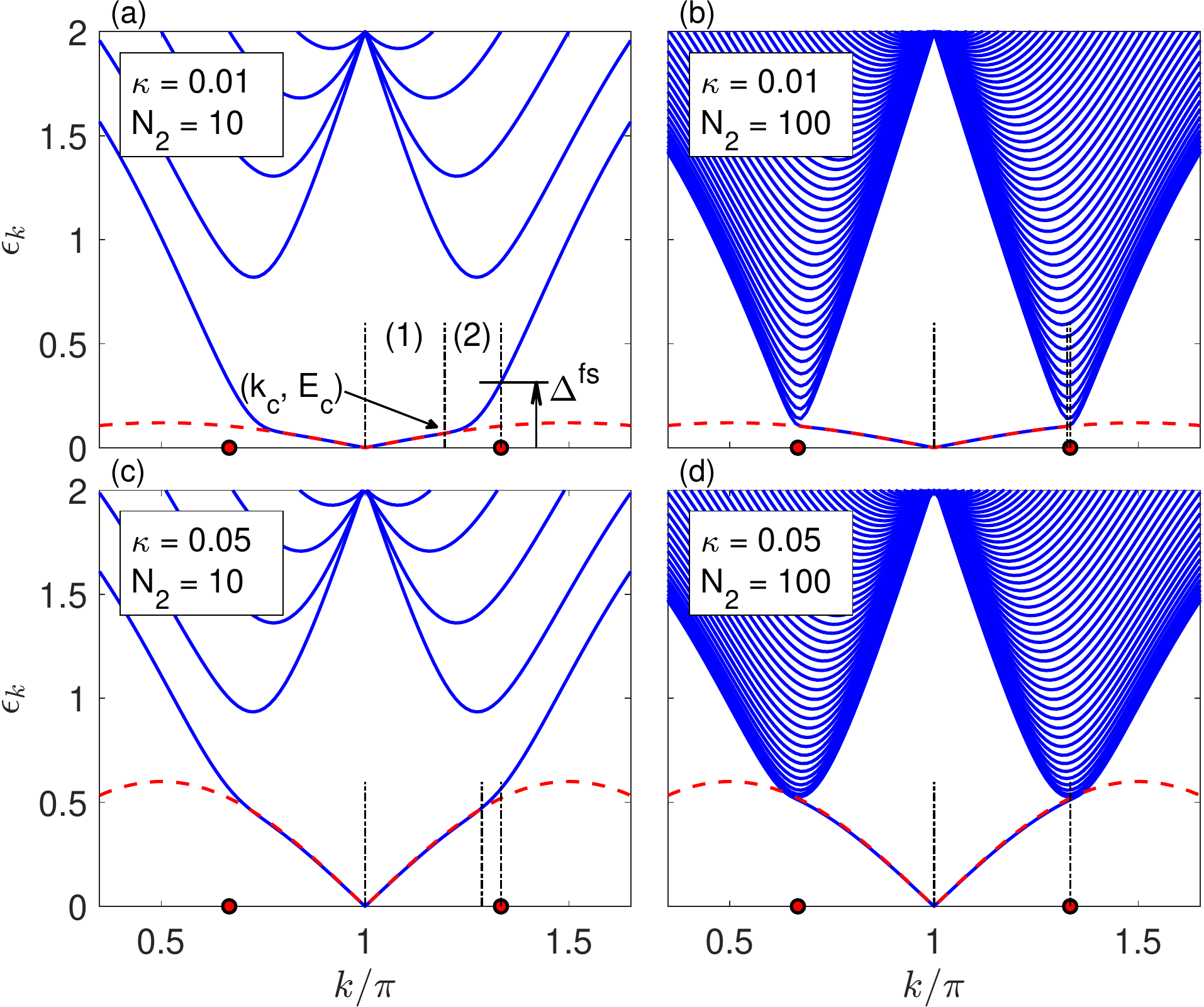}
 \caption{\label{fig: mag_fermi_band} The fermionic energy spectrum for different $\kappa$ and $N_2$. The spectrum obtained by numerically diagonalizing the Majorana Fermion Hamiltonian (\ref{eq: Hamiltonian}) with $\kappa\neq0$ is shown with blue lines. The magnetic dispersion of the edge mode obtained from the perturbative analytical expression \refeq{eq: mass} is shown by the
  red dashed line.
  The right half of the topologically nontrivial region $\kx \in [2\pi/3, 4 \pi/3]$ is  divided into a region (1) spanning the interval $[\pi, k_{c}]$ and a region (2), containing $[k_{c}, 4\pi/3]$. The division is such that in region (1) the deviation between two dispersions is less than 1\%, 
   and in region (2)   the deviation between two dispersions is more than 1\%, so the finite-size effects are essential.   $\Delta^{\tx{fs}}$ is the energy of the lowest sub-band at the projected Dirac point (red dots). $\Delta^{\text{fs}}$ and $k_{c}$ for different values of $N_2$ are  given in Table \ref{tab: kc_N2_mag}. 
}
\end{figure}

\subsection{System with broken time-reversal symmetry ($\kappa\neq 0$)} \label{sec:kappanonzero}

  In the  extended Kitaev model (\ref{eq-H}),   time reversal symmetry is broken by the $\kappa$-term, which
 introduces the diagonal term (\ref{Eq:dz}) into the bulk Hamiltonian   (\ref{HCH2}). 
This diagonal term opens a bulk energy  gap equal to $\Delta_{\text{bulk}}=6\sqrt{3}\, \kappa$ in the thermodynamic limit~\cite{Kitaev2006}.
 It also introduces energy dispersion to the edge modes.
For small $\kappa$, the dispersion of the corresponding edge modes in the thermodynamic limit can be obtained by perturbation theory in $\kappa/J$ \cite{Kitaev2006, Perreault2016b}: 
\begin{align}
    \epsilon^{\text{edge}}_{k}  = 12\kappa |\sin k|, \ k \in \left[ 2\pi/3, 4\pi/3\right]. \label{eq: mass}
\end{align}
We will refer to this dispersion as the magnetic dispersion.  
Note that near the gapless point $k=\pi$, $ \epsilon^{\text{edge}}_k$ is linear in $k$, with a velocity proportional to $\kappa$. At the projected Dirac point, $\epsilon^{\text{edge}}_{4\pi/3} = 6\sqrt{3}\kappa$; this agrees with the bulk energy gap $\Delta_{\tx{bulk}}$, and in the thermodynamic limit, the boundary mode merges with the bulk bands at this point.  

However,  Eq. (\ref{eq: mass}) does not account for finite-size corrections that modify both the bulk gap and the dispersion of the boundary modes.    In order to illustrate how the magnitude of these corrections depends  on the magnitude of the time-reversal symmetry breaking term and the size of the system, in Fig. \ref{fig: mag_fermi_band}  we plot the fermionic energy spectrum for different $\kappa$ and $N_2$, where 
the  numerical band structures are shown by solid blue lines and  the magnetic dispersions  computed  from \refeq{eq: mass} are shown  by red dashed lines.  
Two observations are in order here.  First, we see that for all values $\kappa$ and $\Ny$, the magnetic dispersion is essentially indistinguishable from the true dispersion over much of the range $2 \pi/3 < \kx < 4 \pi/3$, with deviations most noticeable in the immediate vicinity of the projected Dirac points. 
 Second, as expected, the finite-size effects decrease with increasing $\Ny$ (Fig.\ \ref{fig: mag_fermi_band}(b)); however they also decrease with increasing $\kappa$ (Fig.\ \ref{fig: mag_fermi_band}(c-d)).  

As for the $\kappa=0$ case, it is convenient to divide the interval $\kx \in [\pi, 4 \pi/3]$ into two regions, as shown in Fig.\ \ref{fig: mag_fermi_band}(a).   Region (1) consists of the interval $[\pi, k_{c})$, on which the boundary dispersion \refeq{eq: mass} deviates from the numerical dispersion of the lowest sub-band by less than 1\%; here finite-size effects are negligible and the gap of the edge mode is dominated by $\kappa$.  In region (2), consisting of the interval $[k_{c}, 4\pi/3]$, finite-size effects cannot be neglected, as they lead to a correction of $1\%$ or more relative to \refeq{eq: mass}.    In this region, the dispersion of the edge mode can still be approximated by the power law,  $\epsilon^{\text{edge} (2)}_{k} \sim k^{\alpha_2}$; numerically, we find $\alpha_2=(2.14, 2.02, 1.94,1.7)$ for  the system sizes $\Ny=(10,25,30,50)$. 
Table \ref{tab: kc_N2_mag} shows the extent of these finite-size effects for a variety of system sizes with $\kappa = 0.01$.   Specifically, it lists the momentum $k_{c}$ at which the deviation from \refeq{eq: mass} exceeds $1 \%$; the corresponding energy $\epsilon_{\text{min}}$; the ratio $r_1 = 3(k_{c}/ \pi  - 1)$ of the length of region (1) to the length of the interval $[\pi, 4 \pi/3]$; and the minimum energy $\Delta^{\text{fs}}$ at the projected bulk Dirac point.
We see that by $\Ny = 100$, finite-size corrections in all of these quantities are on the order of a few percent, and by $\Ny=500$ they are less than a fraction of a percent.  
At larger values of $\kappa$,  the finite-size corrections will be further reduced relative to those shown in Table \ref{tab: kc_N2_mag}, as is clear from Fig. \ref{fig: mag_fermi_band}(c).

When the finite-size corrections can be neglected, the energy dispersion is well approximated by
${\epsilon}_{k}^{\text{edge}(1)} \approx 12\kappa |k- \pi|$, and 
 the  DOS is  given by $D_1 \sim \kappa^{-1}$ leading to the scaling behavior of the specific heat  given by
\begin{align}\label{eq: region1}
    C^{(1)}_{\tx{edge}} &\sim \frac{1}{3} N_1 \kappa^{-1} T,\quad T < \epsilon_{\text{min}},
\end{align}
in agreement with Ref.\ \onlinecite{Perreault2016b}. Therefore, a low-temperature specific heat $C/T \sim \kappa^{-1}$ at scales below the bulk gap is a signature of the edge mode in the presence of  time reversal symmetry breaking. Unlike the time-reversal symmetric case, this signature is robust in the thermodynamic limit.

We note that for very small $\Ny$, finite-size effects can lead to observable deviations from the scaling (\ref{eq: region1}) for temperatures below the bulk gap, but above the energy scale $\epsilon_{\text{min}}$.  
In region (2), where $\epsilon_\text{min} \leq \epsilon^{\text{edge}(2)}_{k} \leq \epsilon_{\text{max}} = 6 \sqrt{3} \kappa$, 
the corresponding DOS scales as $D_2\sim (\epsilon - \epsilon_{\text{min}})^{1/\alpha_2 -1}$,  and contributes a term to the specific heat that scales as:
\begin{align}
    C^{(2)}_{\tx{edge}}  
    =   \onefrac{3} N_1(1-r_1) I T^{1/\alpha_2} 
\end{align}
where the constant $I$ is given by Eq. (\ref{Eq:Int}), with $\epsilon_{\text{max}} = 6 \sqrt{3} \kappa$. 

 \reffg{fig: fermionic_C}(c-d) show the total fermionic specific heat for $\kappa =0.01$. At temperatures well below the bulk gap, the specific heat fits well to the analytical result in \refeq{eq: region1} for both $\Ny=10$ (panel c) and $\Ny=100$ (panel d). 
 At temperatures on the order of $10 \%$ of the bulk gap $\Delta_{\text{bulk}} = 6 \sqrt{3} \kappa$, bulk states begin to dominate the fermionic specific heat, which fits reasonably well to the $T^2$ dependence
 expected for a $2D$ Dirac cone.

\begin{figure}[t!]
\centering
\includegraphics[width=1\columnwidth]{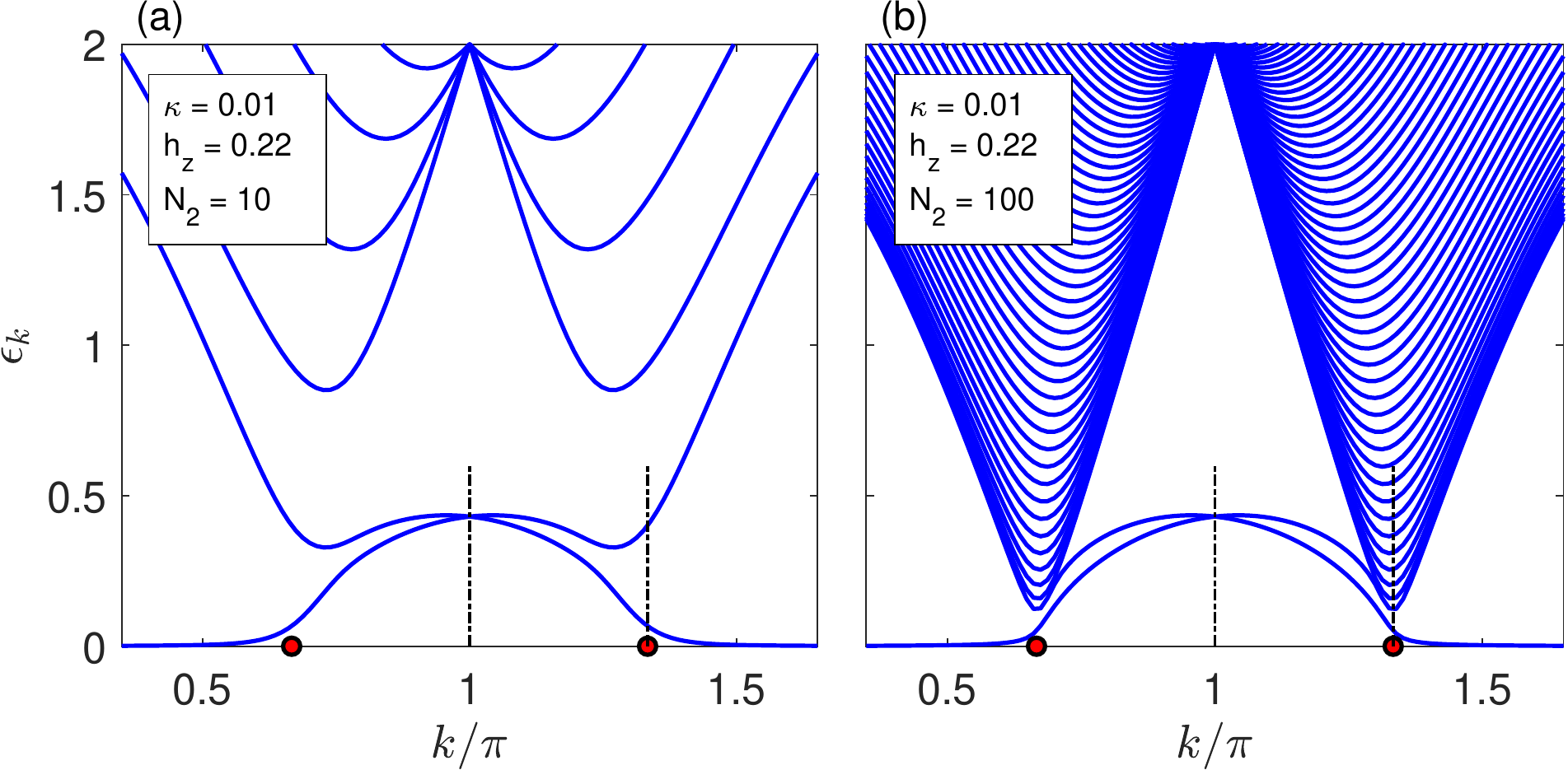}
 \caption{\label{fig: mag_fermi_band_edge} { The fermionic energy spectrum  for various values of $\kappa=0.01$ and   (a) $N_2=10$ and (b) $N_2=100$    computed in the presence of the   Zeeman coupling $h$ between the dangling edge fermions and the itinerant fermions.  Here we set the external filed components  $h_x = h_y = h_z = \kappa^{1/3}$.
 The red dots show the projected Dirac points.}
}
\end{figure}

{
In the pure Kitaev model, a further correction to the edge spectrum at finite magnetic field comes from the Zeeman coupling to spins on the boundary \cite{Kitaev2006}.  In the absence of magnetic field, these edge spins are associated with a non-topological zero-energy boundary flat band of ``dangling" Majorana fermions, which happen to be completely decoupled from the bulk of the system (see Sec. \ref{Sec:Sinfinity}).  
 Since the resulting flat band is non-topological, we expect that in real materials, terms that perturb the bulk Hamiltonian away from the Kitaev limit will gap out these states; hence we have ignored them in our analysis thus far.  
However, it is worth noting that if we consider the Zeeman term as the {\it only} coupling of our system to these dangling boundary fermions, the edge spectrum is modified significantly, since the zero-energy trivial flat band hybridizes strongly with the chiral boundary modes that we have focused on here.  This is shown in \reffg{fig: mag_fermi_band_edge}, where we include a coupling $h_z \sum_i c_i b^z_i$ between   the dispersing Majorana $c_i$ and the danging Majorana $b^z_i$ on each site $i$ on the edge,  via a Zeeman field $h_z = \kappa^{1/3}$.   We see that the velocity of the chiral edge mode is drastically reduced through hybridization with the boundary flat band, and that the gapless point is moved from the center to the edge of our Brillouin zone.  Though this does not alter the qualitative nature of the low-energy power law arising due to the chiral boundary modes, it does significantly alter the energy scale at which it sets in.  
}

\begin{table}
  \begin{center}
    \begin{tabular}{l|ccccccccccccccc} 
    $\Ny$ & \fsz 5 & \fsz 10 & \fsz 25 & \fsz 50  & \fsz 100 & \fsz 500 \\
    \hline
    $k_{c}/\pi$ & \fsz 1.101 & \fsz 1.201 & \fsz 1.285 & \fsz 1.315 & \fsz 1.328 &1.333\\ 
    \hline
    $\epsilon_{\text{min}}$ & \fsz 0.038 & \fsz 0.072 & \fsz 0.095 & \fsz 0.101 & \fsz 0.104 & \fsz 0.104\\
    \hline
    $r_1$ & \fsz   0.302 & \fsz 0.604 & \fsz 0.856 & \fsz 0.944 & \fsz 0.984 & \fsz 1.000\\ 
    \hline
    $\Delta^{\text{fs}}$ & \fsz  0.576 & \fsz 0.314 & \fsz 0.159 & \fsz 0.12 & \fsz 0.108 & \fsz 0.104  \\
    \end{tabular}
    \caption{ $k_{c}$, $\epsilon_{\text{min}}$, $r_1$  and $\Delta^{\text{fs}}$   computed for different  $\Ny$ and taking $\kappa = 0.01$. $k_{c}$ corresponds to the point where the energy calculated by numerical diagonalization and the energy from the perturbative analysis  are different by 1\%, as shown in Fig.\ \ref{fig: mag_fermi_band}. $r_1$ is the ratio of the number of  states in region (1) to the number of states in the whole topologically nontrivial region. $\Delta^{\text{fs}}$ is the fermionic energy at the projected Dirac point. }
    \label{tab: kc_N2_mag}
  \end{center}
\end{table}

\subsection{Specific heat as a probe of topological boundary modes}

We now turn to the question of whether  the Majorana edge modes can be detected experimentally using realistic specific heat measurements.  

As observed above, for $\kappa =0$ this requires measurements to be made at temperatures below the finite-size gap $\Delta^{\text{fs}}$.  
The best chance of detection is therefore in nanoribbons with the smallest possible $\Ny$, which increases both the temperature scale at which the specific heat becomes dominated by $C_{\text{edge}}$, and the magnitude of the edge contribution at the higher end of this range.  To the best of our knowledge, the smallest $\alpha$-$\ch{RuCl_3}$ monolayer nanosheet has a size of around $10\textup{$\mu$m} \times 10 \textup{$\mu$m} \times 2 \textup{nm}$. Given that the spacing of the ions in $\alpha$-RuCl$_{3}$ is around $10 \textup{\AA}$ \cite{bastien2019spin, roslova2019detuning}, the number of the unit cells of the nanosheet is estimated to be $\Nx\times\Ny \sim 10^{4}\times 10^{4}$.  The corresponding high temperature cutoff is set by $\Delta^{\text{fs}}\sim 2\times10^{-4}$ which is around $2\times 10^{-2}K$, given that $J \approx 100K$~\cite{Banerjee2016,sandilands2015scattering}. To the best of our knowledge, the lowest accessible temperature for the specific heat measurements is on a scale of 0.1K, \cite{Do2017,wolter2017field} which is 3 orders of magnitude larger than $\Delta^{\text{fs}}$. 
We conclude that  direct observation of the Majorana boundary modes in the specific in  $\alpha$-$\ch{RuCl_3}$ without time-reversal symmetry breaking is likely out of reach for current experiments.  

For $\kappa >0$, on the other hand,  the  linear-in $T$ specific heat predicted in Eq. (\ref{eq: region1}) up to a temperature scale  corresponding to approximately $0.1 \Delta_{\text{bulk}}$, followed by a cross-over to the quadratic  temperature dependence of the bulk, could potentially be observed even  for  realistic sample sizes.   For example, with $\Ny\sim 10^4$ and $\kappa=0.01$, we have $\Delta_{\text{bulk}}= 6 \sqrt{3} \kappa =0.104 J$, which gives bulk gap of $10 K$, and a temperature scale of  $1 K$ for $T$-linear specific heat to be observed $\alpha$-RuCl$_3$.  We caution, however, that in real experiments time-reversal symmetry is typically broken by applying a magnetic field; in this situation our analysis applies only for $\kappa$ much smaller than the bulk flux gap.

{ Note that in realistic samples, the edges will not be of the perfect zig-zag type considered here, but rather will include a degree of boundary disorder.  While disorder only at the boundary cannot eliminate the gapless chiral boundary modes found at $\kappa \neq 0$, the fate of the boundary states of the nodal topological phase for $\kappa=0$ is much less well understood.  However, a few remarks on its likely impact are in order here.  Disorder can include both random steps (or more generally, segments of armchair edge) interspersed with segments of zig-zag edges, as well as random vacancies along the zig-zag edge.   In the former case, the armchair edge segments do not contribute to the boundary states \cite{wakabayashi2009electronic, wakabayashi2010electronic}.  If the zig-zag segments are long on the scale of the lattice constant, we expect that the system can be heuristically understood as a series of short, disconnected zig-zag ribbons, such that much of the phenomenology described here will be qualitatively unchanged at energy scales that are larger than the typical segment length.  If the zig-zag segments are short, then we expect the effect of disorder to be qualitatively similar to adding a high density of vacancies on the edge.   The latter can be viewed as a particular type of (strong) hopping disorder at the boundary of the zig-zag nanoribbon.  
The random-hopping problem has been much studied in 1D systems with linear dispersion, and with the same symmetries as the Kitaev model (see, e.g. \cite{Dyson,MotrunichDamleHuse,Gruzberg2005}).  It is known that in this case, the disorder does not gap out the low-energy modes, but instead leads to a low-energy divergence in the density of states, which can be either power-law or Dyson-like $(\rho(\epsilon) \sim \epsilon^{-1} | \ln^3 \epsilon|$.  Thus we expect that the system will retain a divergent low-energy density of states and a corresponding contribution to the zero-temperature entropy; however,  the low-temperature specific heat, which is sensitive to the precise nature of this divergence, will evidently be sensitive to such disorder.  
To fully understand this dependence  requires further systematic analysis, similar to that performed in Refs. \cite{ZschockeVojta,kao2020vacancy}, where the finite-size effects were properly accounted for. }


\begin{figure*}
    \includegraphics[width=1\textwidth]{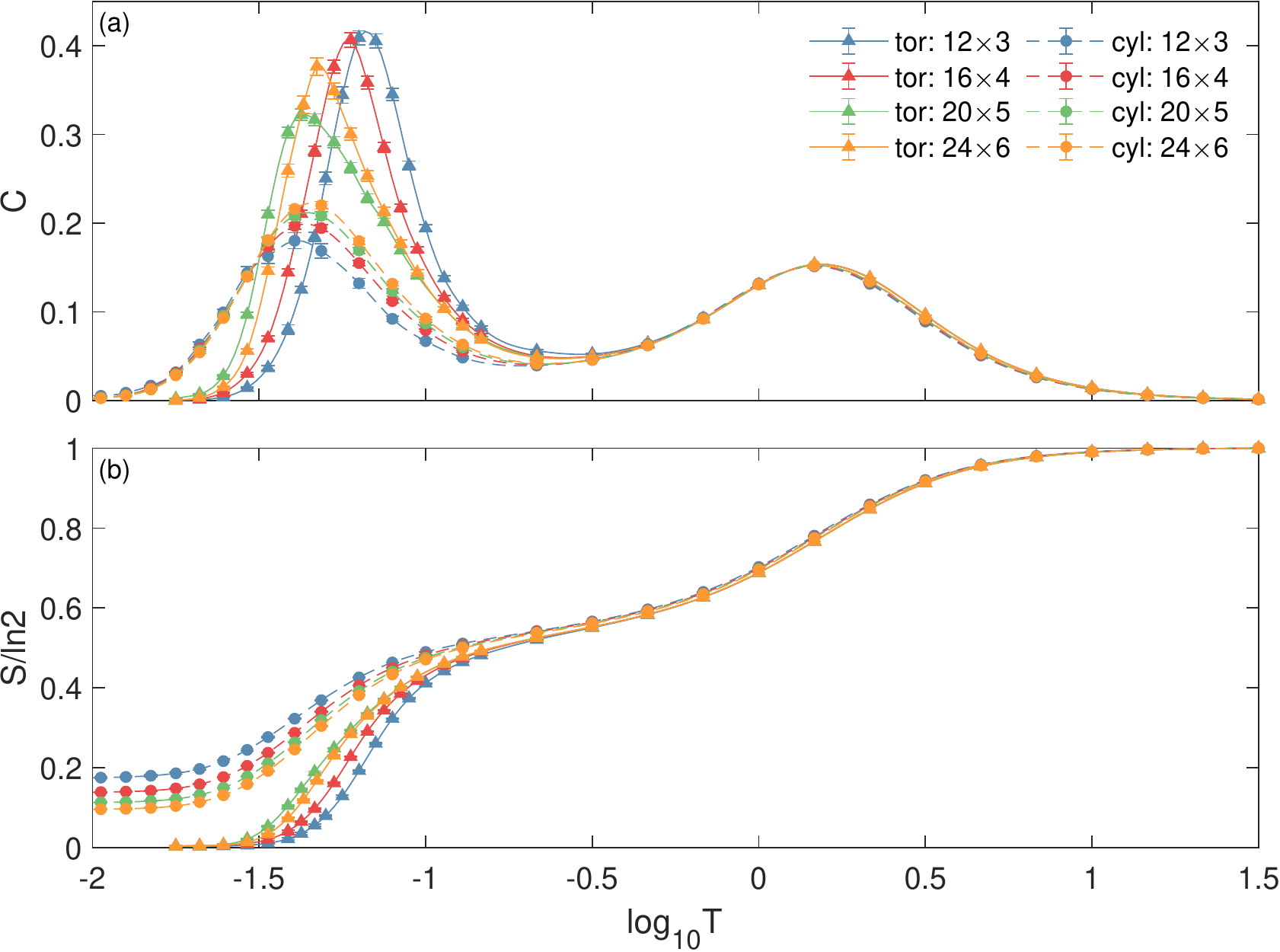}
    \caption{Specific heat (a) and entropy (b) on torus and cylinder lattices with different $\Nx \times \Ny$ sizes. For cylinder lattices, $\nx$ is periodic direction. We choose aspect ratio to be 4 to have longer open edges and more edge modes.\label{fig:CS_cyl_tor}}
    \end{figure*}

\section{Residual entropy and gapless edge modes in the time-reversal symmetric Kitaev model}\label{finiteTedgemodes}

In the previous section we discussed the fermionic contribution to the specific heat, which we expect to dominate at temperatures well below the temperature scale associated with the flux gap.   In the time-reversal invariant case, we argued that the boundary modes do not give a measurable contribution to the specific heat for realistic sample sizes; instead their primary thermodynamic signature is a residual low-temperature entropy $S_0$.  Here we derive the value of $S_0$ in the absence of fluxes, and compare this to  $S_0$ for the full Kitaev spin liquid.   To this end,  we will compute the specific heat by
 using the MC method for the Kitaev spin liquid developed by  Nasu, Udagawa  and Motome ~\cite{nasu2014vaporization}. The entropy is  then given by
\begin{align}
    S (T) &= S_{\infty} - \int_{T}^{T_{\rm max}} C \  \ud \ln T, \label{eq: integrate_C}
\end{align}
where $S_{\infty} $ is the entropy in the limit $T \to \infty$, i.e. the maximum entropy of our system, and we choose  $T_{\tx{max}}=10^{1.5} $, as above this the specific heat is vanishingly small.   Recall  that throughout we work in units where $J=1$.  In the limit $T\rightarrow 0$,  the residual low-temperature entropy is estimated numerically using $S_0= S(T_{\text{min}})$, where $T_{\text{min}} = 10^{-2}$ is the lowest temperature simulated.
  In order to extract an approximate finite-size scaling of $S_0$,   we compare numerical results on tori (with no boundary modes) and cylinders (with boundary modes) of different sizes.  
   
 \subsection{$S_{\infty}$  in the system with open boundaries}\label{Sec:Sinfinity}
 
At this point a note on $S_{\infty}$  is in order.
For a system of $N_s$ spin-$1/2$ degrees of freedom,  $S_{\infty}= \ln \Omega=N_{s}\ln 2$, where $\Omega$ is the number of states.
However, on a lattice with open boundary conditions, this does not correspond exactly to the infinite-temperature entropy of our flux-fermion model.   This is because the Kitaev spin model  (\ref{eq-H}) contains extra static degrees of freedom at the boundary~\cite{Kitaev2006},  that are not included in the  flux-fermion model (\ref{eq: Hamiltonian}).

 To see this,  consider a zig-zag boundary with $A$ sites on the boundary (such a zigzag edge is shown by the solid green line in Fig.\ \ref{fig:lattice}). For a given site on the boundary, the Hamiltonian includes only two terms: 
\begin{equation}
H_{\text{edge}} = \sum_i \sigma^x_{i,A} \sigma^x_{i,B} + \sigma^y_{i,A} \sigma^y_{i-1,B}
\end{equation}
Thus in addition to the conserved plaquette operators $W_p$,  there is a sub-extensive set of operators
\begin{equation}
W^{(b)}_i = \sigma^y_{i,A} \sigma^z_{i,B} \sigma^x_{i+1,A},
\end{equation}
which commute with the Hamiltonian (\ref{eq-H}) for every site $i$ on the boundary.  In the Majorana fermion representation, by applying the constraint 
 operator  $\hat{D}_j=b_j^xb_j^yb_j^zc_j$,  $W_i^{(b)}$ operator can be expressed as $W^{(b)}_i = - i \hat{u}_{iA,iB} \hat{u}_{(i+1)A,iB}b^z_{i,A} b^z_{i+1,A} $.  
It is always possible to work in a gauge where all $\hat{u}_{ij}$ on the boundary are $+1$, where $W^{(b)}_i = - i b^z_{i,A} b^z_{i+1,A} $.  Since the ``dangling" fermions $b^z_{i,A}$ (for unit cells $i$ on the edge) do not enter the Hamiltonian at all, the two possible values of $  W^{(b)}_i$ correspond to states with the same energy.   This leads to a $2$-fold degeneracy in the spectrum for every pair of dangling fermions (i.e. every pair of unit cells) on the boundary -- i.e. a total of $2^{N_1}$ zero-energy states from both boundaries combined.   

These zero-energy states, which we refer to as ``dangling fermion states", are not included in the  model (\ref{eq: Hamiltonian}), but must be accounted for in calculating $S_\infty$.  Specifically, each pair of unit cells on the zig-zag boundary contributes an additional $\ln 2$ to the entropy of the spin model, which we must subtract to obtain $S_{\infty}$.  Thus, on a cylinder with $ N_1$ unit cells on each of the two zig-zag boundaries we find that 
\begin{equation}
  S_{\infty} = ( N_{s} - N_1) \ln 2 .
  \end{equation}

 \subsection{Results of Monte Carlo simulations}
\label{app: algo}
 The  thermodynamics of the Kitaev honeycomb model  
  has been studied previously by  Nasu, Udagawa and Motome~\cite{nasu2014vaporization, Nasu2015}. 
   In particular, using MC simulations they have  determined the  basic structure of the specific heat of the Kitaev honeycomb model  on finite-size systems with $L \times L$ unit cells, using both periodic and open boundary conditions. It was found that  the specific heat $C(T)$ has a two-peak structure, with
  the low-temperature peak at $T=T_L$ associated with the freeze-out of flux excitations, and  a high-temperature peak  from the fermionic excitations at  $T=T_H$.     Since this  is important for  our discussion, we review the derivations of the relevant thermodynamic quantities for the Kitaev model in Appendix \ref{App:Kitaevthermodynamics}.
  
   Here we extend the analysis of  Refs. \cite{nasu2014vaporization,Nasu2015} and present a detailed  study of  the specific heat and entropy of the Kitaev model based on extensive  MC simulations on various finite-size lattices  with periodic  (tori) and semi-periodic (cylinders) boundary conditions.  Our main focus is on exploring the  contribution of the Majorana fermion boundary modes  to the specific heat and the entropy.  As such,  we focus on systems with $\Nx = 4 \Ny$, leading to a long boundary in the cylinder's case, thereby maximizing effects of the edge states.  A short description  of  the  implementation of the MC algorithm~\cite{nasu2014vaporization}   is outlined in Appendix \ref{MC}. 
 
The MC results for the specific heat 
 are presented in Fig. \ref{fig:CS_cyl_tor} (a).  
 As in \cite{nasu2014vaporization, Nasu2015}, we observe
a two-peak structure for the specific heat on both torus and cylindrical lattices, with 
 a high-temperature peak that is basically  insensitive to both the system size and boundary conditions.  This is expected since the bulk band structure for these high-energy modes is qualitatively similar in all cases.  In contrast, the low-temperature peak ($T_L$) is markedly different  between the torus and the cylindrical lattices.  Notably, the onset temperature for this lower peak on a cylindrical lattice  is lower than on a  comparably sized torus lattice.  This  indicates a smaller average flux gap due to the presence of open boundaries.  Finite-size effects are also significant for the lower peak, primarily due to interactions between fluxes that are enhanced by small system size.  We will discuss both of these effects in detail  in Sec.~\ref{chap: flux energetics}.
 Fig.~\ref{fig:CS_cyl_tor} (b) presents the corresponding entropy, computed  by using Eq. (\ref{eq: integrate_C}).  The two peaks in $C(T)$ lead to a two-step  entropy release, with entropy release near $T_L$ predominantly due to the proliferation of flux degrees of freedom, while that near $T_H$ stems from the proliferation of high-energy itinerant fermionic degrees of freedom.  
 
Fig.\ref{fig:CS_cyl_tor} (b) also clearly shows the difference between torus and cylindrical lattices. 
On torus lattices, the entropy decreases to nearly zero at the lowest temperatures.  This is expected: in the thermodynamic limit all the physical degrees of freedom except the states very near the gapless Dirac points are frozen out at temperatures small compared to the flux gap.  On finite-sized tori the freeze-out is more pronounced, since the Dirac points acquire a gap on the order of $1/\Ny$ unless $\Nx$ and $\Ny$ are both divisible by 3 (see Appendix \ref{App:extrapolation}), leading to a freeze-out of all degrees of freedom below this scale.  
In contrast, the residual entropy on  cylindrical lattices  is non-zero due to the band of low-energy fermionic edge modes.  
Specifically, with a low-temperature cutoff at a scale $T_{\text{min}}$,  the residual entropy $S_0$ is approximately given by the number of states with energy less than $T_{\text{min}}$.   For the model (\ref{eq: Hamiltonian}), and $T_{\text{min}} \sim \Delta^{\text{fs}}$, this is approximately:
\begin{align}
S_0= \ln \Omega_0,
       \label{eq: entropy_formula}
\end{align}
where $\Omega_0$ is the number of  states associated with the zero-energy edge mode,  $\Omega_0 \approx 2^{\onefrac{3} \Nx}$.  Note that for the small values of $\Ny$ simulated here, $\Delta^{\text{fs}} > T_{\text{min}} = 10^{-2}$, and we find $S_{0} \lessapprox  \onefrac{3}\Nx \ln 2$. 
  To obtain the residual entropy associated with the original Kitaev spin model (\ref{eq-H}), we must add the contribution from dangling gauge fermions discussed in Sec. \ref{Sec:Sinfinity} giving
\begin{align}
S_0^{\text{spin}}= \ln \Omega_0 +  \Nx\ln 2.
       \label{eq: entropy_formulaspin}
\end{align}

\begin{table}[!h]
    \begin{center}
        \begin{tabular}{c|c|c|c}
        $\Nx\times \Ny$   &$s_0^{\text{spin}}(\textrm{I})$ & $s_0^{\text{spin}}(\textrm{II})$& $s_0^{\text{spin}} (\textrm{III})$ \\
        \hline
        12$\times$3   & 0.222  &   0.176      &   0.174 $\pm$ 0.002 \\
        16$\times$4   & 0.167  &   0.136      &   0.138 $\pm$ 0.001 \\
        20$\times$5   & 0.133  &   0.112      &   0.113 $\pm$ 0.001 \\
        24$\times$6   & 0.111  &   0.095      &   0.096 $\pm$ 0.002 \\
        \end{tabular} 
    \end{center}
\caption{\label{table: R.Entropy} Residual entropy per site  divided by $ \ln 2$, i.e. $s_0^{\text{spin}} = S_0^{\text{spin}}/(N_s \ln 2)$, computed for cylindrical lattices  with open boundaries in  the $\ny$ direction.   $S_0^{\text{spin}} (\textrm{I})= \frac{4}{3}N_1 \ln2$  is obtained by neglecting  finite size effects and assuming all states in the lowest sub-band on the interval $\kx \in [2 \pi/3, 4 \pi/3]$ contribute to the residual entropy. $S_0^{\text{spin}}(\textrm{II})$ is  obtained from Eq. (\ref{eq: entropy_formulaspin}) by numerically counting the number $\Omega_0$ of  fermionic states below  the energy scale of $\epsilon_{\text{min}}$. $S_0^{\text{spin}} (\textrm{III})$ is obtained by integrating the specific heat from high temperature $T_{\tx{max}}$ down to $T_{\tx{min}}$,  according to   \refeq{eq: integrate_C}.}
\end{table}

A quantitative comparison between these analytical estimates and the MC results is given in Table \ref{table: R.Entropy}, which compares various ways of evaluating $s_0^{\text{spin}}\equiv S_0^{\text{spin}}/(N_s \ln 2)$.  In the Table, 
 $S_0^{\text{spin}} (\textrm{I})= \frac{4}{3}N_1 \ln2 $ is the residual entropy of the spin model in the thermodynamic limit, i.e., assuming that all edge states contribute to $S_0$;   $S_0^{\text{spin}}(\textrm{II})$ is  the residual entropy obtained from  \refeq{eq: entropy_formula} by taking $\Omega_0$ equal to the number of fermionic states with energy less than $T_{\text{min}}$;  $S_0^{\text{spin}}(\textrm{III})$ is  the residual entropy obtained from the MC results using \refeq{eq: integrate_C}.   In  both $S_0^{\text{spin}}(\textrm{I})$ and $S_0^{\text{spin}} (\textrm{II})$ we have added the contribution $\Nx\ln 2$ due to dangling fermions, to match the residual entropy of the spin model.
As expected we find an excellent agreement between $S_0^{\text{spin}} (\textrm{II})$ and $S_0^{\text{spin}} (\textrm{III})$, even when finite size effects are significant. This illustrates how, in principle, a measurement of the residual entropy $S_0^{\text{spin}}$ for different sample geometries can be used to establish the presence of the topological boundary modes characteristic of the Kitaev spin liquid. {

We emphasize that because the boundary flat band is topologically protected provided that time-reversal symmetry is unbroken, the residual low-temperature entropy is a feature of the entire gapless Kitaev spin liquid phase, and thus is a robust experimental signature of the gapless Kitaev spin liquid.    Ref. \cite{li2020universal} verified this explicitly, showing 
that the residual entropy survives  in the presence of  non-Kitaev interactions that are finite but not too strong.  }

\section{Flux energetics in the time-reversal symmetric Kitaev model ($\kappa=0$)}
\label{chap: flux energetics}

The MC results discussed in the previous section reveal significant differences in the low-temperature specific heat peak associated with flux freezeout on the cylinder relative to the torus.  
In this section, we give a physical interpretation for these differences, by studying the energetics of  flux excitations in both torus and cylinder lattices.  We use this to model how flux energetics impact thermodynamic quantities in both geometries, for both finite and infinite systems.   Here we  will focus on the time-reversal symmetric Kitaev model, $\kappa=0$, and consider the effects of  finite $\kappa\neq 0$ in Section \ref{sec: mag_flux_energetics}.

\subsection{One-flux gap energy in torus and cylinder}
\label{subsec: one_flux gap}

 \begin{figure}[!b]
\subfigure{
    \centerline{
         \includegraphics[width = .5\textwidth]{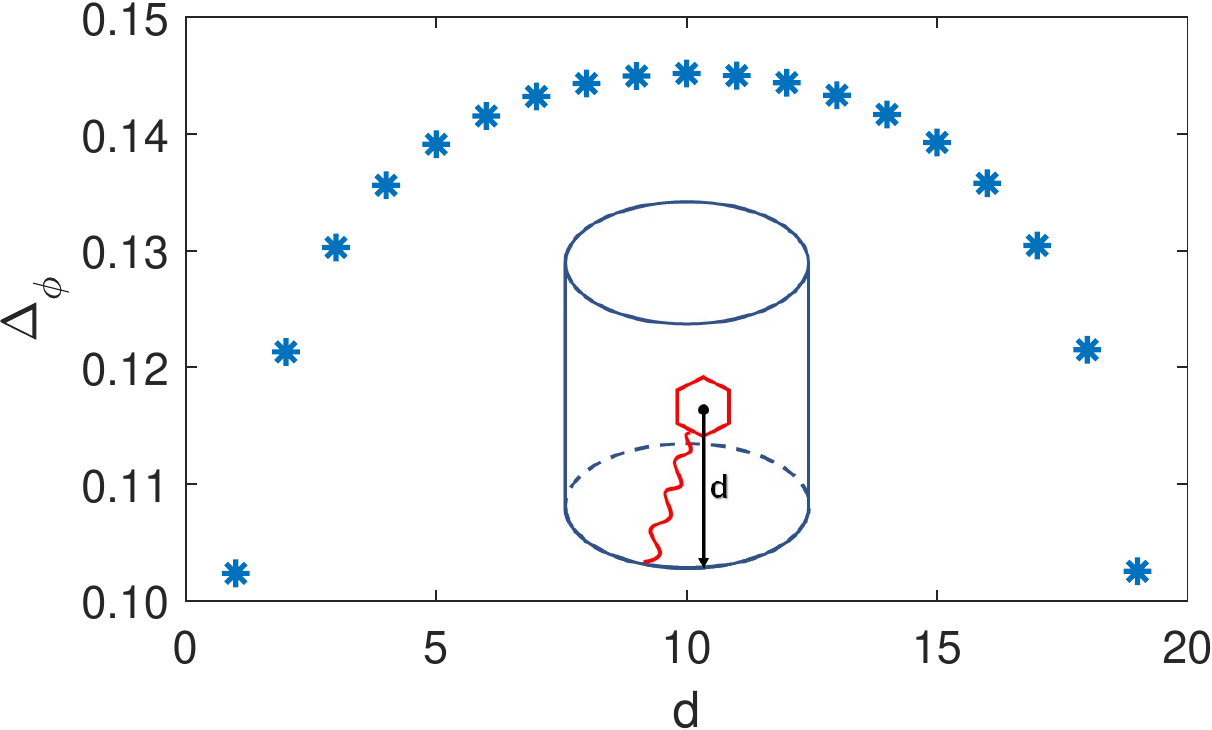}
    }
}
\caption{\label{fig: one_flux_cyl} Position-dependence of the one-flux gap on a $\Nx\times\Ny = 40 \times 20$ cylinder. Here $d$ indicates the distance between the flux and  the edge. }    
\end{figure}

We start by computing the  one-flux energy gap  $\Delta_\phi$  in  toroidal and cylindrical geometries.   On the torus, the  one-flux energy gap   was analyzed by Kitaev [\onlinecite{Kitaev2006}], who associated it with a half of the two-flux energy $\Delta_\phi=\frac{1}{2}E^{(0)}_{2\phi}$, where  $E^{(0)}_{2\phi}$ is defined to be  the energy of the flux configuration $\{\phi_p\}=2\phi$  with the maximum separation between the two fluxes (see \refeq{eq: flux_energy}).
  He also showed that      $E^{(0)}_{2\phi}$
 displays oscillatory   period-3 modulations as the lattice size increases,  due to the fact that on a $L \times L$ lattice, for $L$ divisible by $3$, the Dirac cone is located on the reciprocal lattice point, while for other values of $L$ it is not (see Appendix \ref{App:extrapolation}).   By extrapolating   $E^{(0)}_{2\phi}$ to the thermodynamic limit,  Kitaev obtained an estimate for the flux gap $\Delta^{\infty}_\phi = 0.154$.

\begin{figure}[!h]
    \includegraphics[width = 0.99\columnwidth]{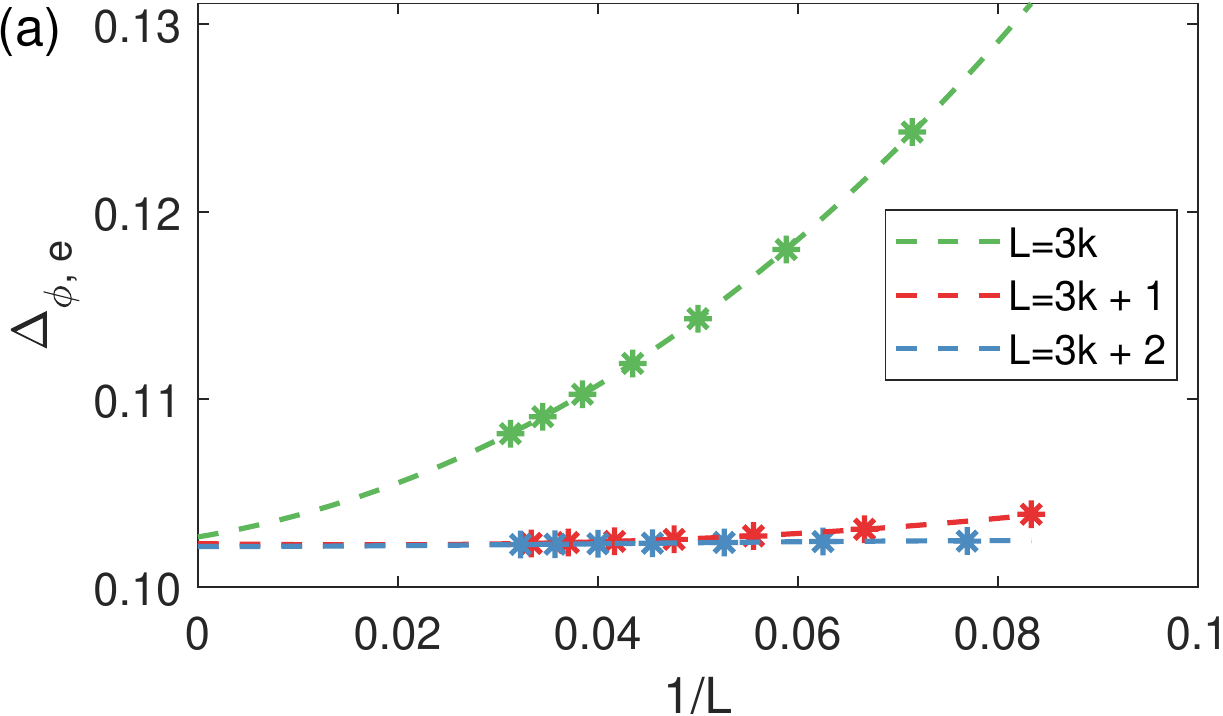}
    \includegraphics[width = 0.99\columnwidth]{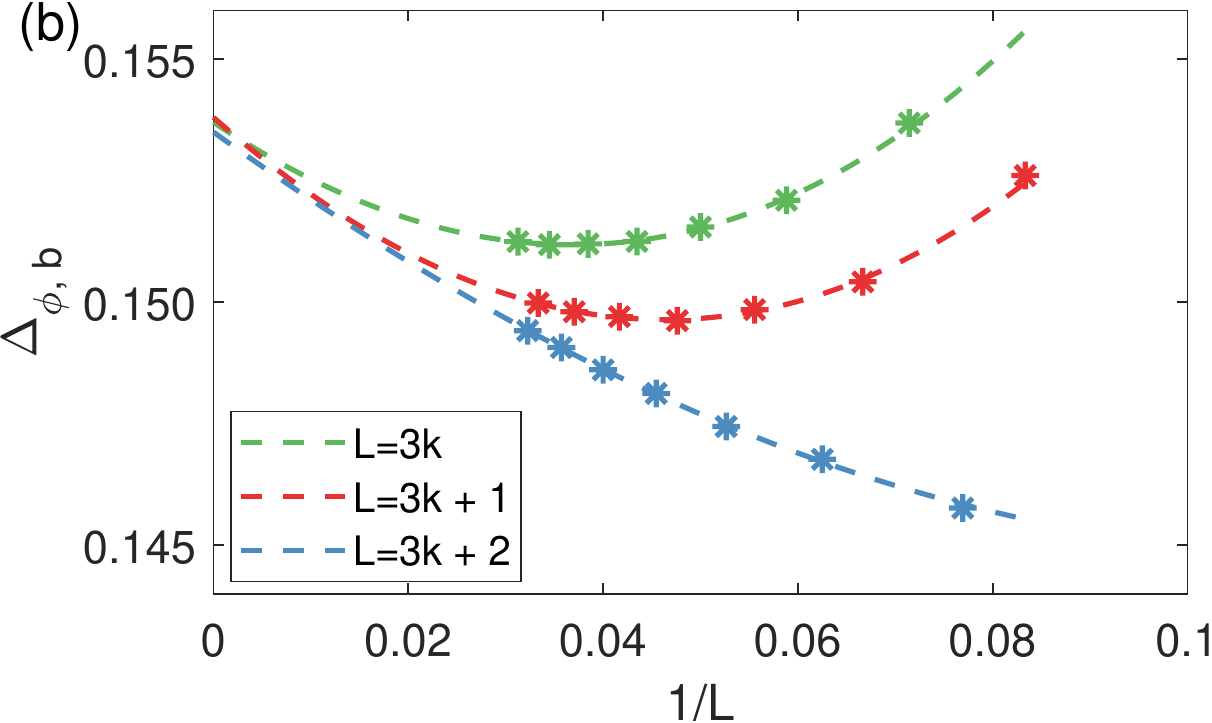}
\caption{One-flux energy  as function of the inverse lattice size $1/L$
 computed  when the  flux is located  (a) on a boundary plaquette, and  (b) on  a plaquette maximally distant from the edge, for different cylinder systems with $\Nx = \Ny = L$.  Due to the periodic dependence on system size modulo 3,   we use separate polynomial fits to the series $L=3k$ (green),  $L=3k+1$ (red), and $L=3k+2$ (blue), and extrapolate these in the limit $1/L \to 0$ to estimate the gap for large system sizes.  We find that the energy gaps for the edge  and bulk fluxes 
extrapolate to (a) $\Delta^\infty_{\phi, \tx{e}} = 0.102$ and  (b) $\Delta^\infty_{\phi, \tx{b}} = 0.154$, respectively. }
\label{fig: one_flux_gaps_cyl}
\end{figure}

We follow the same method and compute the one-flux gap in cylinder lattices.  In this case, however, the open boundary allows us to  create just one flux and study how  the one-flux energy  depends  on the distance of the flux to the open boundary.  Fig.\ \ref{fig: one_flux_cyl} shows the one-flux energy gap at different distances from the boundary computed  on a $\Nx\times\Ny = 40 \times 20$ cylinder.  We can clearly see that  the flux gap decreases as the flux approaches to the open boundary, and  is significantly lower (by over 30\%) on the boundary than deep in the bulk. This decrease in energy can be seen as an effective attractive interaction between the fluxes and the boundary, due to  the  Majorana boundary modes~\cite{lahtinen2011interacting}. 
  To see the finite size scaling of the one-flux gap,  in Fig.\ \ref{fig: one_flux_gaps_cyl}
we plot the flux gap when it is created on the plaquette on the edge (a) and in the middle of the cylinder (b) calculated for different system sizes $L$.   By extrapolation to thermodynamic limit
 separately for $L=3k$,  $L=3k+1$ and $L=3k+2$ (see details in Appendix \ref{App:extrapolation}),
  we get  the edge gap equal to $\Delta^{\infty}_{\phi,\tx{e}} = 0.102$ and  the bulk gap equal to $\Delta^{\infty}_{\phi,\tx{b}} = 0.154$. The later, as expected, is the same as  on the torus lattices.

 To summarize, our  numerical analysis suggests  that, on average, the one-flux gap in cylinder lattices is smaller than that on comparable torus lattices, and decreases as the flux approaches the boundary. This finding is in agreement with our MC results of the specific heat (see \reffg{fig:CS_cyl_tor}) which shows that 
 the onset temperature for flux excitations on cylinders are lower than those on tori.  
 
\begin{figure*}
    \includegraphics[width = 1\textwidth]{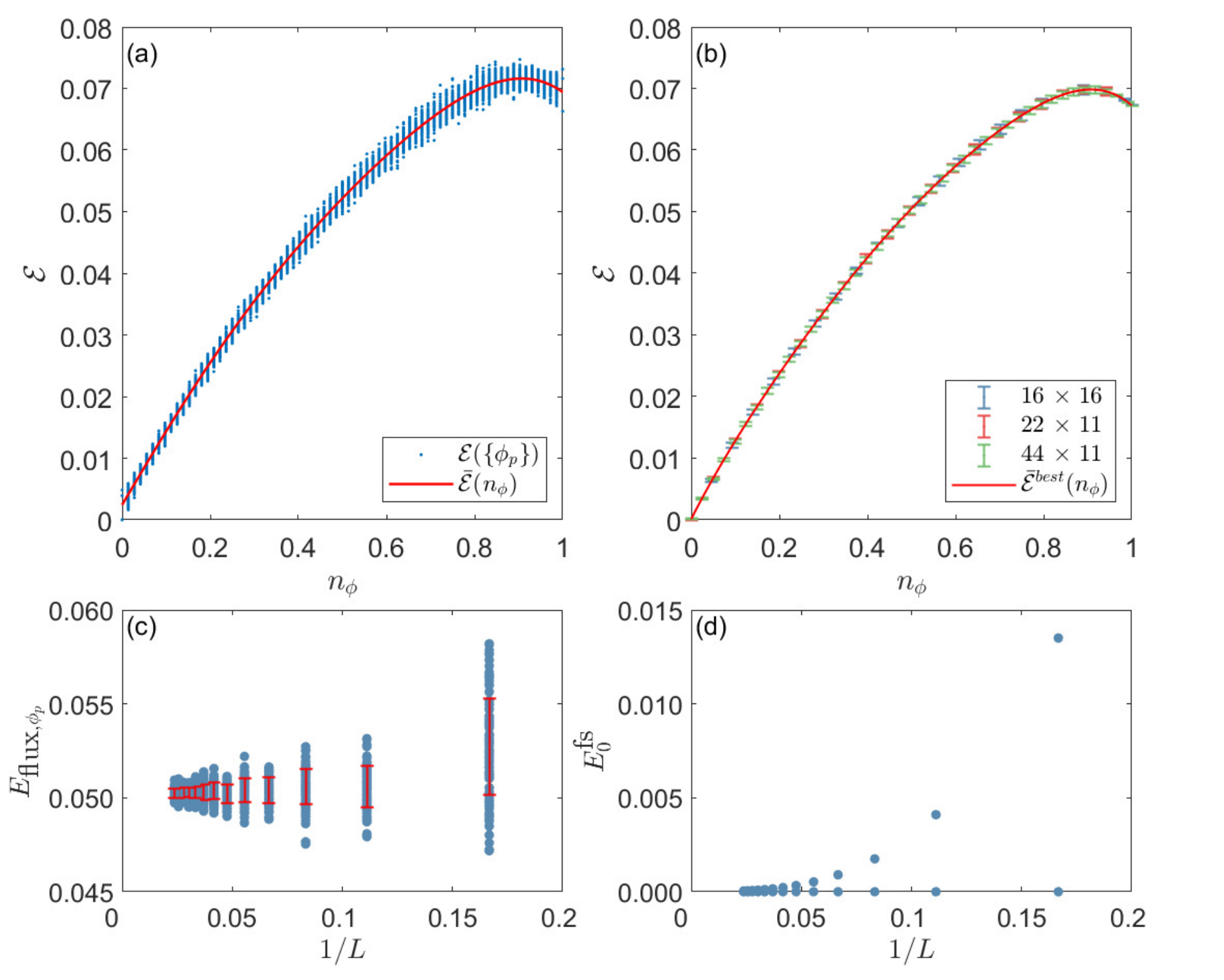}
    \caption{
    (a) Distribution of the flux energy density $\E(\nphi)$ for different flux densities $\nphi$ on the torus lattices 
   with $\Nx\times\Ny=$ 24$\times$6, generated by taking 100 randomly generated flux configurations at each flux density.   
    The best-fit polynomial { $\bar{\E}(\nphi) \simeq -0.143 \nphi^5 + 0.274 \nphi^4 - 0.181 \nphi^3 - 0.004 \nphi^2 + 0.121 \nphi + 0.002$ } is shown by the red curve. The standard deviation of the residual errors is 0.14.  
    (b) The flux energy densities on the lattices of  $\Nx\times\Ny= 16\times16$ (blue),  $22 \times11$ (green) and $44\times 11$ (yellow), from a random sampling of 60 flux configurations.  Vertical bars at each flux number show one standard deviation of energy. The best-fit polynomial that simultaneously fits the three sets of data is { $\bar{ \E }^{\tx{best}}(\nphi) \simeq  -0.321\nphi^6 +0.850\nphi^5 - 0.898\nphi^4+ 0.475 \nphi^3 -0.180\nphi^2 + 0.141\nphi $}.  (c) Distribution of flux energies for various lattices sizes $L$ for a fixed flux density $\nphi=0.5$, from a sampling of 60 flux configurations. The red error bars show one standard deviation of the flux energies.   
    (d) The finite-size energy splitting of the ground state energy $E^{\tx{fs}}_0$ as function of the inverse lattice size $1/L$, where $\Nx=\Ny=L$. 
    \label{fig: fluxenergydensity} }
\end{figure*}

\subsection{Phenomenological flux energy models}
\label{subsec: flux_band_PPE}

To understand the energetics of systems with many fluxes, we  analyze the distribution of the flux energy  $ E^{(0)}_{\phi_p}$  (defined as   the lowest-energy state with a given flux configuration  $\phi_p$ -- see \refeq{eq: flux_energy}) for different flux configurations $\{\phi_p\}$ at fixed flux density on torus and cylinder lattices.  
{We
perform a polynomial fit to the average energy obtained
over many random flux configurations at each
flux density.}
 We call this fit the  flux pseudopotential energy  (PPE).
{  By comparing the resulting curves on lattices  of different sizes and geometries, we argue that in the limit of large system size, the PPE on the torus describes a universal dependence of flux energy density on flux density.  On the cylinder the dependence on geometry is not universal, but we find evidence of a universal dependence of total flux energy to total flux density on cylinders of fixed width in the thermodynamic limit.}

\subsubsection{ Flux pseudopotential energy on a torus}\label{subsubsec: torus_PPE}

For a given flux configuration, the flux energy  density is given by $E^{(0)}_{\phi_p} /N_p $, where $N_p$ is the number of plaquettes and $N_\phi$ is the number of fluxes.  
 For fixed flux density $\nphi=N_\phi / N_p$, we define $\E(\nphi)$ to be the distribution of $ E^{(0)}_{\phi_p} /N_p $ for different flux configurations with the same flux number $N_\phi$.    We begin by studying how $\E(\nphi)$ depends on  $\nphi$ for torus lattices.  
An example is shown in Fig.~\ref{fig: fluxenergydensity} (a), which plots  $\E(\nphi)$ as a function of $\nphi$  on the $\Nx\times\Ny= 24\times 6$  torus lattice.  To generate this distribution we sample 100 random flux configurations at each flux density $n_\phi$; each blue dot represents the energy of one such configuration.  The solid red line represents a   fifth-order polynomial best fit to the average $\bar{ \E}(\nphi) $  at each flux density $\nphi$, given by $\bar{\E}(\nphi) \simeq -0.143 \nphi^5 + 0.274 \nphi^4 - 0.181 \nphi^3 - 0.004 \nphi^2 + 0.121 \nphi + 0.002$. 
We call this polynomial fit the flux pseudopotential energy (PPE). We choose to fit to fifth-order polynomials as this is the lowest order that consistently fits the data, based on the distribution of residual errors.

 The PPE curves for smaller tori, where the finite-size splitting of topological sectors is large, are not universal, and can differ significantly depending on the system size and aspect ratio.   To understand why, recall that  in the absence of fluxes the Kitaev model has four topologically distinct sectors on the torus~\cite{Kitaev2006}.
  These sectors are energetically degenerate in the thermodynamic limit; however on finite-size tori the splitting between them can be appreciable, as is evident from the distribution of energies for the four configurations with $n_{\phi} = 0$ in Fig.~\ref{fig: fluxenergydensity} (a).  This splitting is shown for $L \times L$ tori in Fig.~\ref{fig: fluxenergydensity} (d), as a function of $1/L$.  
  For comparison, Fig.~\ref{fig: fluxenergydensity}  (c) shows how the width of the distribution $\E(\nphi = 0.5)$ depends on the system size $L$ for an $L \times L$ torus.  
  We see that for $L \lessapprox 20$, the differences between the energies of these four sectors rapidly becomes comparable to the width of the distribution of flux energies at $n_{\phi}=0.5$, suggesting that the difference between topological sectors accounts for a significant part of the finite-size broadening of this distribution at small $L$.  
  
  For $L \gtrapprox 20$, in contrast, differences between the energies of these four sectors are small compared to the width of the distribution of flux energies.  
Though the distribution narrows with increasing $L$ in this range, it does so only very slowly.  Indeed we expect this distribution to reach a plateau of finite width as $L \rightarrow \infty$, since the interactions between fluxes lead to variations in the energies of different flux configurations at fixed $n_{\phi}$.  
Our numerics suggest that for $L\gtrapprox 20$ the distribution of energies  converges to  { a universal function, independent of both system size and aspect ratio, which captures the main features of the flux energetics as a function of flux density in the thermodynamic limit. 
In order to find this universal  function, we  consider tori  with  $\Nx\times\Ny=$ 16$\times$16,  22$\times$11 and 44$\times$11 (for which  the aspect ratios $\Nx / \Ny$  are 1,  2,  and 4, respectively) and for each of the systems obtain $\E(\nphi)$  (shown in  Fig.~\ref{fig: fluxenergydensity} (b)) generated from 60 random  flux configurations.} The height of the bars at each value of $n_\phi$ represents the standard deviation of the corresponding energy distribution.   We see that the overlap between the three curves is large, indicating that the dependence on aspect ratio (as well as system size, not shown here) is negligible.  All three curves are well described by the best-fit polynomial (shown by the red line)
{
\begin{align}\nonumber
    &\bar{ \E }^{\tx{best}}(\nphi) \simeq  -0.321\nphi^6 +0.850\nphi^5 - 0.898\nphi^4\\
    &+ 0.475 \nphi^3 -0.180\nphi^2 + 0.141\nphi  \,.\label{fbest}
\end{align}
}
 We find that the sixth-order polynomial gives better fit for the distribution of energies  shown in Fig.\ref{fig: fluxenergydensity} (b) than the
 fifth-order polynomial.  
 We thus use the functional form  (\ref{fbest}) as an approximate universal PPE fit, which describes the average flux energy density in the thermodynamic limit.

The PPE polynomial \refeq{fbest} gives us { a useful tool to obtain insights}   into the nature of the flux interactions on the torus.  First, the concave shape of the flux PPE as a function of the flux density indicates that the  flux interactions are, on average, attractive; our analysis suggests this remains true at all flux densities. 
Second, it is natural to try to identify the coefficient of the linear term in the best-fit curve with an average flux gap $\bar{\Delta}_{\phi}$, since on a large torus $2 \bar{\Delta}_{\phi}$ should give a good approximation of the average energy of a configuration of two fluxes.  The value obtained from the universal PPE in Eq. (\ref{fbest})   is $0.14$, which is smaller than $\Delta^\infty_\phi= 0.154$.  We attribute this to the attractive interactions between fluxes: $\Delta^\infty_\phi$ is measured by putting the two fluxes at the maximum separation, while $\bar{\Delta}_{\phi}$ is obtained from the 2-flux energy averaged over all separations.  As interactions between  the fluxes are attractive on average, the latter is smaller. 

\begin{figure}[!htbp]
    \includegraphics[width = 1.0\columnwidth]{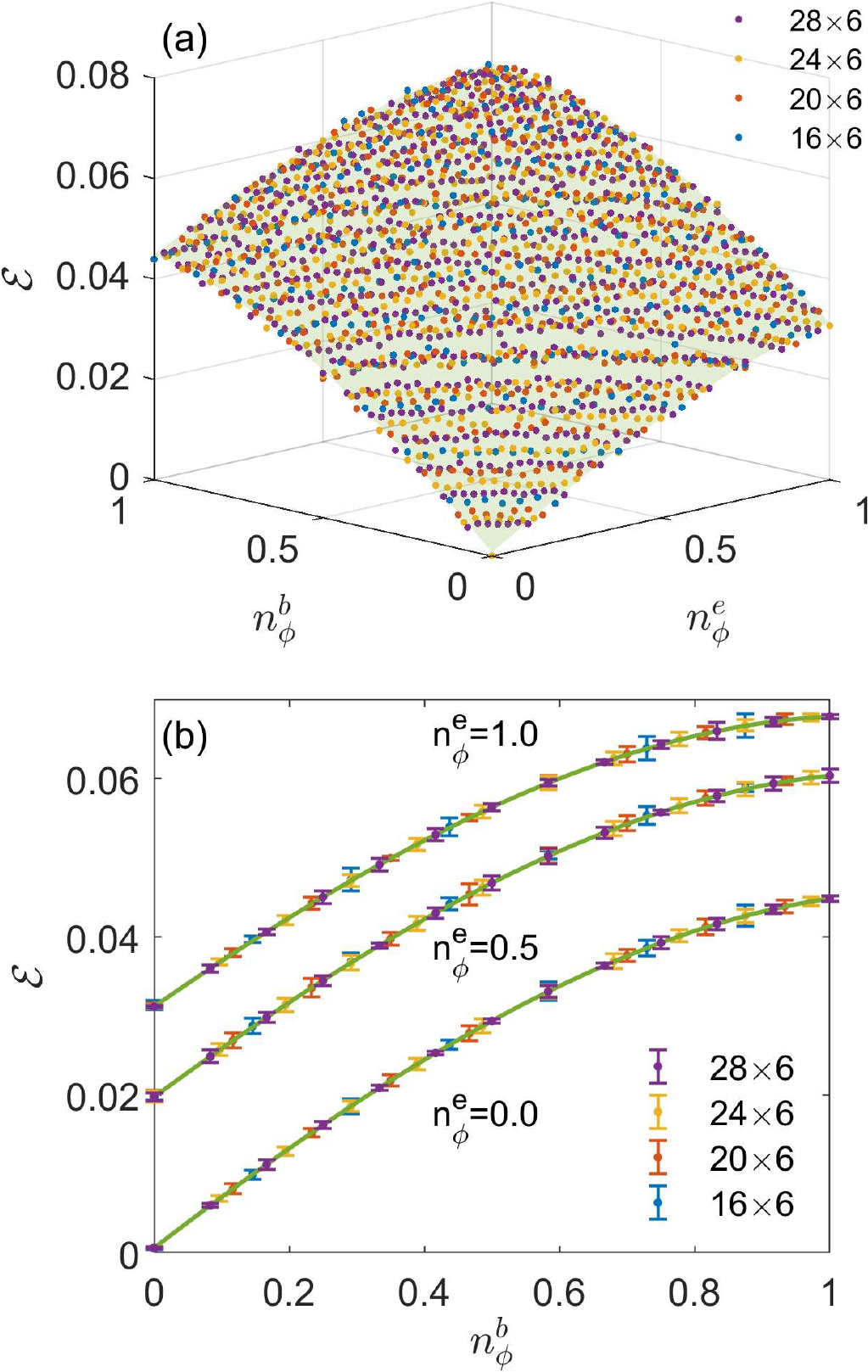}
    \caption{ \label{fig:24_6_cyl}
(a) Distribution of the flux energy density $\E$  as a function of edge ($n^e_\phi = N^e_\phi/N^e_p$)  and bulk  ($n^b_\phi = N^b_\phi/N^b_p$) flux densities computed on cylindrical  lattices with $\Nx\times\Ny=28\times6$ (purple), $24\times6$ (yellow),  $20\times 6$ (orange),  and  $16\times6$ (blue). 
 The best-fit polynomial surface is  shown. (b) The line cuts of the best-fit surface obtained for  fixed edge flux  densities $n^e_\phi = 0, 0.5, 1.0$.  Vertical bars indicate one standard deviation in energies.}
\end{figure}

\subsubsection{ Flux pseudopotential energy on a cylinder}\label{subsubsec: cyl_PPE}
 
  In order to study how the flux energies are distributed on the cylinder lattices,  we again sample
 different flux configurations for fixed flux  densities.  However, since the energy of a single flux significantly depends  on its proximity to the edge of the lattice (see Fig.\ \ref{fig: one_flux_cyl}),  we now do this for a pair of two independent flux densities: one for the edge, $n^e_\phi = N^e_\phi/N^e_p$,  and one for the bulk, $n^b_\phi = N^b_\phi/N^b_p$.  Here  $N^e_\phi$ is the edge flux number, counting the fluxes on the two outermost layers of plaquettes, and  $N^b_\phi=N_\phi-N^e_{\phi}$ is the bulk flux number.  $N^b_p$ and $N^e_p$ count the  number of plaquettes on the edges and in the bulk, respectively.   The energy density is defined as  $\mathcal{E}(n_{\phi}^{e},n_{\phi}^{b}) = E^{(0)}_{\phi_p} (N^e_{\phi},N^b_{\phi})/N_p$, where  $E^{(0)}_{\phi_p}(N^e_{\phi},N^b_{\phi})$ is the flux energy of the configuration  with $N^e_\phi$ edge fluxes and $N^b_\phi$ bulk fluxes.
 
 On the cylinder, we do not expect the flux PPE to be universal even for large system sizes, since it depends on the fraction of plaquettes that lie on the boundary.  However, for fixed $\Ny$ (and hence a fixed boundary-to-area ratio), we do find good agreement to a universal PPE, as shown in Fig.\ref{fig:24_6_cyl}.
  The distributions of flux energy densities $\E(n^e_\phi, n^b_\phi ) $  computed on cylindrical  lattices with $\Nx\times\Ny=$ 28$\times$6, 24$\times$6,  20$\times$6,  and  16$\times$6 are shown in Fig.\ref{fig:24_6_cyl} (a), with 30 different flux configurations sampled randomly for each pair ($n^e_{\phi}, n^b_{\phi})$. 
As for  the torus lattices, due to the attraction (on average) between fluxes, the average energy density $\bar{\E}(n^e_\phi ,n^b_\phi )$  lies  on a concave-down surface.
  
  To obtain the flux PPE surface appropriate to cylinders with $\Ny=6$, we perform a two-variable polynomial fit  to the average of all of the  energy distributions shown in Fig.\ref{fig:24_6_cyl} (a). 
    This gives the best-fit polynomial surface shown in Fig.\ref{fig:24_6_cyl} (a), described by the equation:
\begin{align} \label{Eq:CylPPE}
    & \bar{\E}^{\text{best}}(n_{\phi}^{e},n_{\phi}^{b}) = 0.001 + 0.042 n_{\phi}^{e} + 0.066n_{\phi}^{b} \\\nonumber &-0.005 (n_{\phi}^{e})^2 
     -0.006 n_{\phi}^{e} n_{\phi}^{b} -0.011 (n_{\phi}^{b})^2 
    \\\nonumber & -0.007 (n_{\phi}^{e})^3
     -0.001 (n_{\phi}^{e})^2n_{\phi}^{b} -0.001 n_{\phi}^{e}(n_{\phi}^{b})^2\\\nonumber & -0.010 (n_{\phi}^{b})^3.
\end{align}
Using this polynomial, we can also determine the  edge and bulk flux gaps as
\begin{align}
 &\Delta_{\phi,e} = \frac{\partial \bar{\E}^{\text{best}}(n_{\phi}^{e},n_{\phi}^{b})}
 {\partial n_\phi^e}\big|_{n_{\phi}^{e}, n_{\phi}^{b}=0} \cdot \frac{N_p}{N^e_p} = 0.105,\label{1}\\&
 \Delta_{\phi,b} = \frac{\partial \bar{\E}^{\text{best}}(n_{\phi}^{e},n_{\phi}^{b})}{\partial n_\phi^b}\big|_{n_{\phi}^{e}, n_{\phi}^{b}=0} \cdot \frac{N_p}{N^b_p} = 0.110.\label{2}
\end{align} 
  While the estimate of the edge flux gap is close to the extrapolated result shown Fig.\ref{fig: one_flux_gaps_cyl} (a),
  the bulk gap energy is lower than the energy obtained from the extrapolation shown in Fig.\ref{fig: one_flux_gaps_cyl} (b).  As for the torus lattice, one contribution to this difference is that the two-flux interactions are attractive on average.   In addition, however, for the small $\Ny$ shown here many of our ``bulk" fluxes are in fact quite close to the edge, leading to a significant further reduction.

 The line cuts of the best-fit surface obtained for  fixed edge flux  densities $n^e_\phi = 0, 0.5, 1.0$ are shown in Fig.\ref{fig:24_6_cyl} (b), where the height of the bars at each value of $n^b_\phi$ represents the standard deviation of the corresponding energy distribution for a given value of  $n^e_\phi$.   All system sizes shown are well fit within error bars by the PPE given in Eq. (\ref{Eq:CylPPE}).   

 \subsection{Flux contribution to specific heat ($\kappa=0$) }\label{subsubsec: flux_model}

 We now leverage our phenomenological flux  models to predict the contribution of fluxes to the specific heat in the thermodynamic limit of  the time-reversal invariant Kitaev model.  In particular, since  the low-temperature peak in the specific heat is mainly due to the flux excitations, its position and   shape  are closely related to the flux energetics, which  we can now describe  in the thermodynamic limit using our numerically obtained  best-fit polynomials. We will also compare the predictions of the specific heat  based on the PPE polynomials with the results of the  MC simulations in analogous boundary conditions  and analyze how well they capture the position and shape  of the low-temperature peak in the specific heat.

\begin{figure*}
    \includegraphics[width = 1\textwidth]{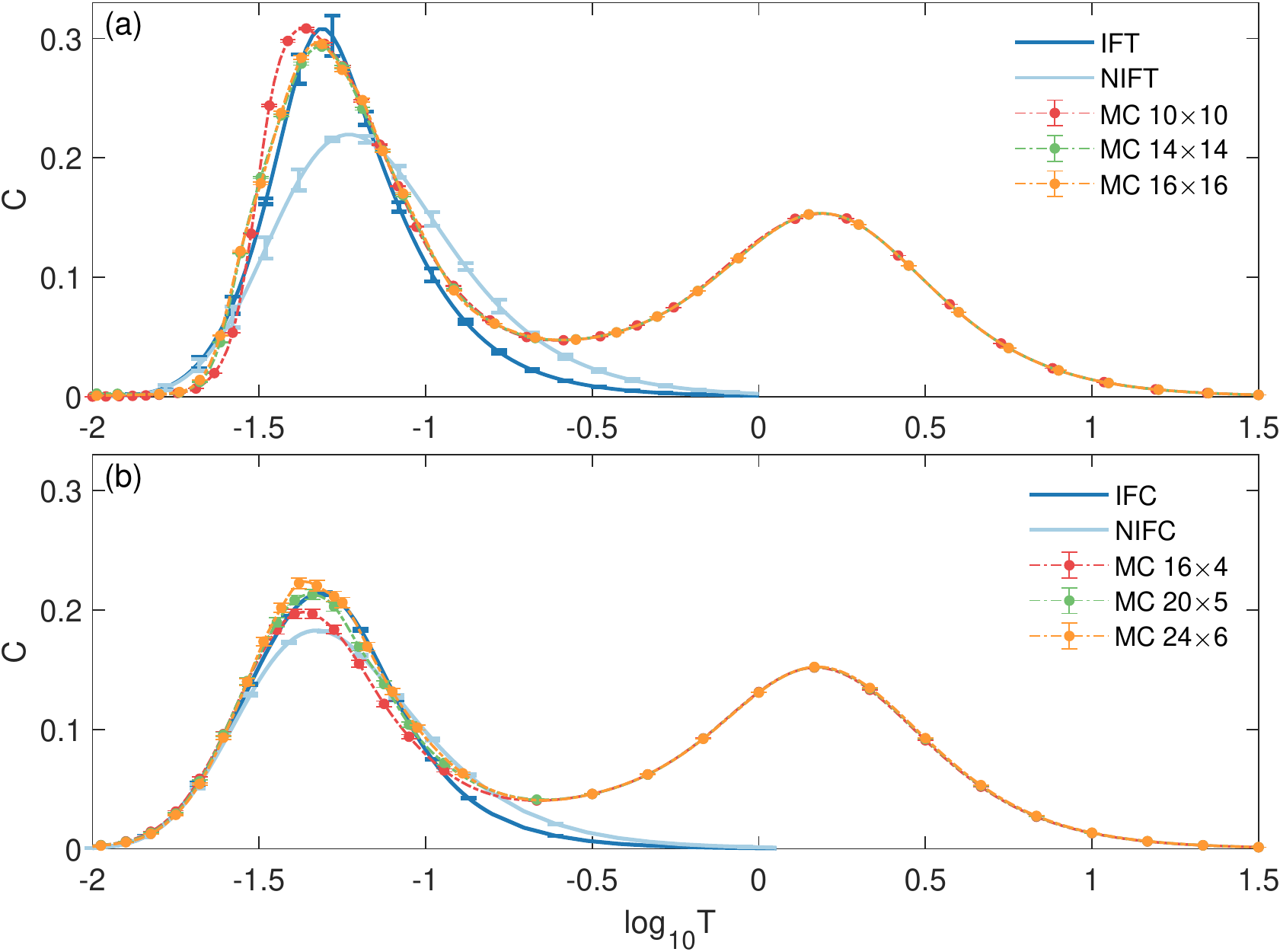}
    \caption{\label{fig: thermolimit_C} Comparison between IFT, NIFT, and MC predictions of the specific heat.  (a) The specific heat per site with IFT and NIFT models, computed according to \refeq{eq: poly_C}, compared to Monte Carlo simulations on various $L\times L$ tori.
    (b) The specific heat per site with IFC and NIFC models, computed according to \refeq{eq: exact_sum}, compared to Monte Carlo simulations. Note that our NIFC fit is specific to cylinders with $\Ny =6$. { Here the uncertainty in the specific heat of the PPE models are estimated from an ensemble of fit curves created by bootstrapping the regression data \cite{efron1979computers}, as described in Appendix \ref{ErrorApp}.}
    }
\end{figure*}

\subsubsection{Torus lattices}
\label{subsubsec: flux_model}
We calculate  thermodynamic quantities on the torus lattices by a saddle point approximation of the effective free energy of the flux degrees of freedom,
\begin{align}
    F_{\phi}  &=  N_p{\E}(\nphi) - TS_{\phi}, 
     \label{eq: free_energy}
\end{align}
where ${\E}(\nphi)$ from now on denotes the average  flux energy density 
and  $ S_{\phi}$ is the flux entropy.
On a torus with $N_p$ plaquettes, there are ${N_p \choose N_\phi}=\frac{N_p!}{N_\phi !(N_p-N_\phi)!}$ possible configurations with a total of $N_{\phi}$ fluxes.
Hence the configurational entropy is equal to $S_{\phi} = \ln{N_p \choose N_\phi}$.  Using the Stirling's formula, for large $N_p$ and $N_{\phi}$ we find:
\begin{align} \label{Eq:Stirling}
    S_{\phi} & = N_p\left[-\nphi\ln{\nphi} - (1 - \nphi) \ln{(1-\nphi)}\right].
\end{align}
In the thermodynamic limit, the value of 
 the flux density $\nphi$ can be obtained by the minimizing the free energy:
\begin{align}
  \frac{1}{N_p}  \frac{d F_{\phi} }{d \nphi }&= \E'(\nphi) - T\ln{\frac{1-\nphi}{\nphi}}=0. \label{eq: flux_density_equation}
\end{align}
where $\E'(\nphi) = \partial \E (\nphi) / \partial n_\phi$.  
This yields the flux density:
\begin{align} \label{Eq:neq}
 \nphi(T)= \frac{1}{e^{\E'(\nphi) /T} + 1}.
\end{align}
 Since the energy density depends on $T$ only through its dependence on $\nphi$, the specific heat per site  arising from solely flux degrees of freedom can be expressed as:
 \begin{align}\label{eq: linear_C} 
    C_{\phi} &=\frac{1}{2} \E'(\nphi)\frac{\ud \nphi}{\ud T} \ .
\end{align}  
Explicitly,
\begin{align}
& \frac{ \partial \nphi(T)}{\partial T} = \\&-  \frac{e^{ \E' (\nphi)/T}}{( e^{ \E' (\nphi)/T} + 1)^2 } \left(- \frac{\E'(\nphi)}{T^2} + \frac{1}{T} \frac{\partial \E'(\nphi)}{\partial T} \right)  \nonumber \\\nonumber
&=\nphi(1-\nphi)  \left(\frac{\E'(\nphi)}{T^2} - \frac{1}{T} \frac{\partial \E'(\nphi)}{\partial \nphi}  \frac{ \partial \nphi(T)}{\partial T}  \right) .
\end{align}
Solving the above for $\partial \nphi(T)/\partial T$ gives the specific heat per site:
\begin{equation} \label{eq: poly_C}
    C_{\phi} =  \frac{1}{2} \left(  \frac{ \E'(\nphi) }{T}\right)^2 \onefrac{\onefrac{\nphi(1-\nphi)} + \frac{\E''(\nphi)}{T}},
 \end{equation}
 where $\E''(\nphi) = \partial^2  {\E(\nphi)}/\partial \nphi^2$.  
The denominator of \refeq{eq: poly_C} shows two distinct contributions to the specific heat. The first, from
 the term $\onefrac{\nphi(1-\nphi)}$, is entropic in origin and  comes from the curvature of the configurational entropy, $ S_{\phi}''/N_p = - \onefrac{\nphi(1-\nphi)}$.  
 The second contribution is due to the curvature  of the average flux energy density
${\E}(\nphi)$. For an attractive interaction, such as the one we found for our flux PPE in Sec. \ref{subsubsec: torus_PPE}, $\E''(\nphi) < 0$ and this term increases the  value of the specific heat.  
  Intuitively, this 
 is because the attractive interaction between the fluxes lowers the energy required to excite additional fluxes as the flux density increases,  increasing the number of thermal flux excitations at a given temperature.

We evaluate the flux contribution to the specific heat by  using the best-fit polynomial (\ref{fbest})  to describe the average flux energy density ${\E}(\nphi)$ in \refeq{eq: poly_C}. This gives us the specific heat of  the  interacting flux  model  (IFT) on the torus. For comparison,  we also compute the specific heat  \refeq{eq: poly_C} for the non-interacting flux model on the torus  (NIFT), for which the energy density is simply $ {\E}(\nphi) = \Delta_\phi n_\phi$,  with $\Delta_\phi = \overline{\Delta}_{\phi} $, and $\E''(\nphi) =0$.  In  Fig.\ \ref{fig: thermolimit_C}(a) we plot the specific heat for both the IFT (dark blue) and NIFT (light blue) models.
 Clearly, flux interactions have an important effect on the shape of the peak in specific heat:  the IFT model has a higher and narrower peak than the NIFT model due to the attractive flux interactions. The difference is on the order of $30 \%$, indicating that interaction effects are quantitatively important.   { We also find that the precise shape and height of the specific heat curve is sensitive to small changes in the flux PPE, resulting in significant error bars to our predictions for $C$ in spite of the relatively small error of our PPE fits.  (See Appendix \ref{ErrorApp} for details.)}
 
 Fig.\ \ref{fig: thermolimit_C}(a) also compares these flux-only specific heat curves  $C_\phi(T)$ to those obtained by the MC simulations on different-size torus lattices, which describe both flux and fermionic contributions to the specific heat $C(T)$.   We find reasonable agreement between the specific heat of the IFT model  and that of the MC simulations over most of the lower peak in $C(T)$, particularly for the largest lattice sizes simulated.  The two models do exhibit significant differences at low temperatures, with the specific heat falling off more quickly in the MC simulations than in our IFT model.  We conjecture that this is primarily due to finite-size effects, which decrease the average distance (and thus increase the average interaction) between fluxes at low flux numbers in our MC simulations.
Thus, as expected, we find that the low-temperature peak in $C(T)$ can be well accounted for by a model that includes only flux excitations, and ignores the dispersing fermions.

{ Moreover, the IFT  model based on the PPE  polynomial fit suggests
 that the low-temperature peak in specific heat is of finite width even in the thermodynamic limit.}  This confirms the expectation \cite{Nasu2015} that the gapless phase of the 2D Kitaev model does not have a finite-temperature phase transition, but rather a crossover from a state of vanishingly small flux density at temperatures well below the flux gap, to one with flux density $\nphi \approx 0.5$ at temperatures above the flux gap.  
To see why, we expand the exact expression for the free energy at large $N_p$ and small $N_{\phi}$, to find 
$
    F_{\phi} \approx (\E'(0) - T\ln N_p) N_\phi.
   $
This is minimized by $N_{\phi}>0$ at the flux onset temperature
\begin{align} \label{Eq:Tonset}
T_{\text{onset}} & \approx  \bar{\Delta}_{\phi}/\ln N_p \ , 
\end{align}
where $2 \bar{\Delta}_{\phi}$ is the average energy cost of inserting a pair of fluxes.   Thus in the thermodynamic limit, for any $T>0$ there will be some (finite) number of fluxes in the system. Although for $T \ll \bar{\Delta}_{\phi}$ the flux {\it density} is effectively 0, these few fluxes are sufficient to destroy the topological order, such that there are no singularities in $C(T)$.  Our numerics suggest that including the low-energy fermionic excitations in this analysis will not substantively alter $T_{\text{onset}}$, or the nature of the crossover.
We emphasize that $T_{\text{onset}}$ does not correspond to the position of the low-temperature peak in $C(T)$, which always occurs at a temperature scale set by $\bar{\Delta}_{\phi}$, irrespective of the system size.
On the finite-sized systems studied here, however, where the flux density can never be lower than $2/N_p$, $T_{\text{onset}}$ is a good proxy for where this  lower peak begins.

 In Fig.\ \ref{fig: thermolimit_C} (a), the specific heat curves of   the NIFT  and IFT models merge together at low temperatures, where the flux density is small.  This is because at low flux densities the energy density of the IFT model is dominated by the linear term, $\E(n_\phi) \approx  \E'(0) n_{\phi}$.   Thus taking the flux gap of the NIFT model to be $\bar{\Delta}_\phi = \E'(0)$, the two models exhibit very similar behavior for $C(T)$ at temperatures well below $\bar{\Delta}_\phi$, where the flux density is small.

\subsubsection{Cylinder lattices}

\begin{figure}
 \includegraphics[width = 1\columnwidth]{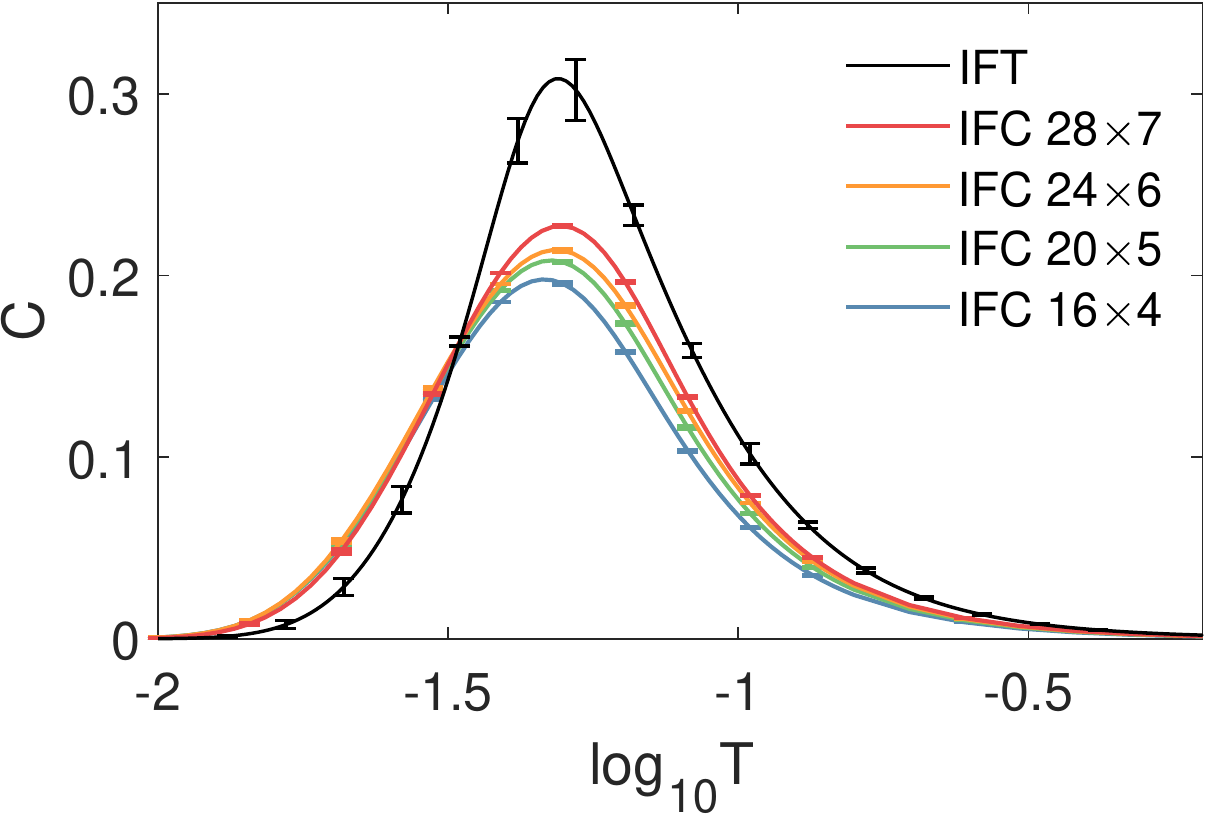}
\caption{\label{fig: C_cyltor} 
    Comparison  of  the specific heat  computed using an IFC model based on fits for various lattice sizes, with the specific heat computed using the IFT model in the thermodynamic limit.  
    {
       Error estimates are obtained using  the same method as in \reffg{fig: thermolimit_C}.
    }
}
\end{figure}

We now discuss the analogs of both the non-interacting (NIFC) and the interacting (IFC) flux models appropriate for cylinder lattices.
%
%
With the 2-variable model of the flux energy described in Sec. \ref{subsubsec: cyl_PPE},  the approach taken above to calculate $C(T)$ becomes cumbersome.  Instead, 
  we  start from the expression for the specific heat  in Eq.  (\ref{eq: C_formula}), and exploit the fact that the last term vanishes 
 in the absence of fermionic excitations, i.e. 
  \begin{align}
    C_{\phi} (T) &=\frac{\ud \langle  E^{(0)}_{\phi_p}
    \rangle}{\ud  T} =\onefrac{T^2}\left( \lefta ( E^{(0)}_{\phi_p})^2 \righta - \lefta  E^{(0)}_{\phi_p}\righta^2\right).
     \label{eq: band_C}
\end{align} 
 Here the expectation $\lefta \ldots \righta$ is taken over  different flux configurations characterized by the pairs of the flux numbers $(N^e_\phi, N^b_\phi)$, weighted by the probability
\begin{align}
    p(N^e_\phi, N^b_\phi) &= e^{-\beta E^{(0)}_{\phi_p}
     + \ln {N^e_{p} \choose N^e_\phi } + \ln {N^b_{p} \choose N^b_\phi} }. \label{eq: prob_flux_band_cyl}
\end{align}
On a finite-size cylinder, this allows us to compute the specific heat exactly by summing over the possible values of $N^e_\phi, N^b_\phi$:
\begin{align}
    C_{\phi}(T) & = \onefrac{N_s T^2}\sum_{\Nephi, \Nbphi} ( E^{(0)}_{\phi_p}  - \lefta  E^{(0)}_{\phi_p} \righta)^2  p(\Nephi, \Nbphi),
     \label{eq: exact_sum}
\end{align}
where the sum over $\Nephi$ ($\Nbphi$) runs from $1$ to $N^e_{p}$ ($1$ to $N^b_{p}$), where $N^e_p (N^b_p)$ is the number of plaquettes on the edge (in the bulk).
Using Eq. (\ref{eq: exact_sum}), we compute the specific heat  using  both the  best-fit PPE surface for the cylinders (IFC model) and
the  non-interacting flux model (NIFC), in which $ E^{(0)}_{\phi_p} = \Nephi \Delta_{\phi,e} +  \Nbphi \Delta_{\phi,b}$, with the gaps $\Delta_{\phi,e}$ and $\Delta_{\phi,b}$ given by Eqs.~(\ref{1}) and (\ref{2}).

\begin{figure*}
    \includegraphics[width = 1\textwidth]{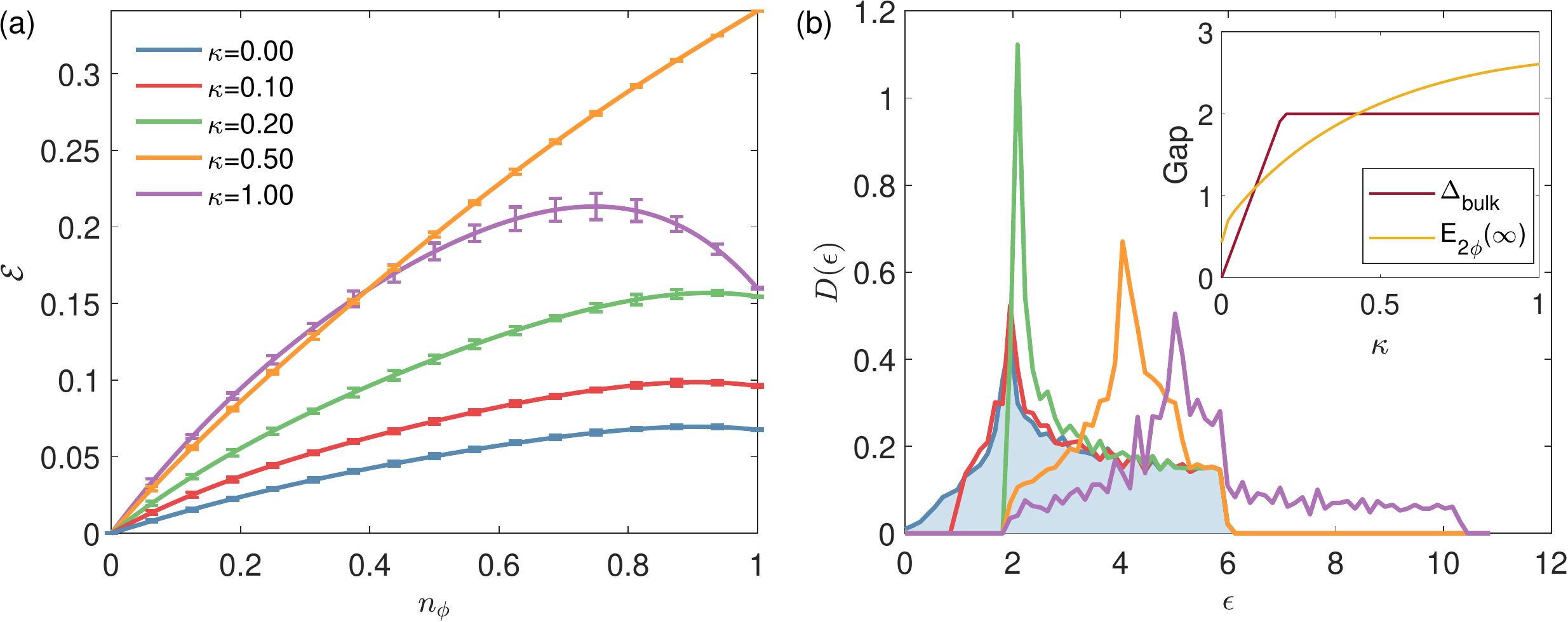}
\caption{\label{fig: flux_kap}
(a) The flux PPE for different $\kappa$ on the $\Nx\times\Ny = 16\times 16$ torus. The corresponding best-fit polynomials  are given in Appendix \ref{App:polynomials}. For $\kappa=0.5$, the nearly straight band and small standard deviation indicates that flux interactions are very weak. 
(b) The fermionic DOS, $D(\epsilon)$, in the flux-free sector obtained by numerically diagonalizing the Bloch Hamiltonian $H_\kv$ defined  in \refeq{HCH2} on the
$\Nx\times\Ny = 30\times 30$ torus.  The inset in (b) shows the two-flux gap   $E_{2\phi}(\infty)$  (dark yellow line) and the fermionic gap  $\Delta_{\rm bulk}$ (dark red line). Here  $E_{2\phi}(\infty)$ is defined as the flux energy when two fluxes are placed at the maximum separation. 
}    
\end{figure*}

Fig.\ \ref{fig: thermolimit_C}(b) shows the  specific heat of the NIFC and IFC models evaluated on a $ 24\times 6$ lattice along with the ones obtained by  the MC simulations on $ 16\times 4$, $ 20\times 5$  and $ 24\times 6$ lattices.  (Recall that the fit used in our IFC model is specific to cylinders with $\Ny = 6$).  Again, we see that flux interactions, which are attractive on average, play an important role: the specific heat in the NIFC model has a lower and broader peak than that  in the IFC model, with a difference in peak heights on the order of $20 \%$.    We also find good agreement between the specific heat in the IFC and our MC simulations in the vicinity of the low-temperature  peak, particularly for the two largest  system sizes.  Notice that there is still a small disagreement  between the IFC and the MC results on a  $24\times 6$ lattice: the peak temperature in the  IFC model is slightly higher than that obtained with  the MC simulation.  This is because the IFC model assigns the same energy to any flux not on the edge.  Though this leads to an under-estimate for the gap of fluxes deep in the bulk of the system, it also over-estimates the gap of fluxes  close to, but not on, the boundary (see Fig.\ \ref{fig: one_flux_cyl}).  Since these lower energy excitations contribute disproportionately to the specific heat at lower temperatures, our IFC model actually over-estimates the peak position.  Still, Fig.\ \ref{fig: thermolimit_C}(b) shows that  the IFC model  captures the specific heat due to thermal flux  excitations relatively well.

\subsubsection{Comparison  between the IFT and   IFC specific heats}

To illustrate the effect of the boundary fluxes on the thermodynamics, here we directly compare the specific heat on torus and cylinder lattices.  Fig.~\ref{fig: C_cyltor} presents the specific heat of the IFC model computed on a series of finite-size cylinder lattices and  that of the universal IFT model on the torus lattice.  It shows that the peak begins at a lower  temperature on the cylinder than on the torus, predominantly due to the smaller single-flux gap near the edge of the cylinder, which broadens the temperature range over which the flux entropy is released.    Because the total entropy released is the same in both cases, the peak height  for the cylinder lattice is correspondingly lower than that on the torus lattice.  

Additionally, as the lattice size increases, the specific heat peak of the cylinders  move slightly  towards right, with the right half of the peak appearing to tend towards the IFT curve.    This is because for an $a L \times L$ lattice ($a>1$), as the lattice size $L$ increases, the fraction of the plaquettes on the boundary falls off as $N^e_p/N_p = a/L$.  Thus any effects associated with the edge diminish in importance in the thermodynamic limit.

 \begin{figure*}
    \includegraphics[width = 1\textwidth]{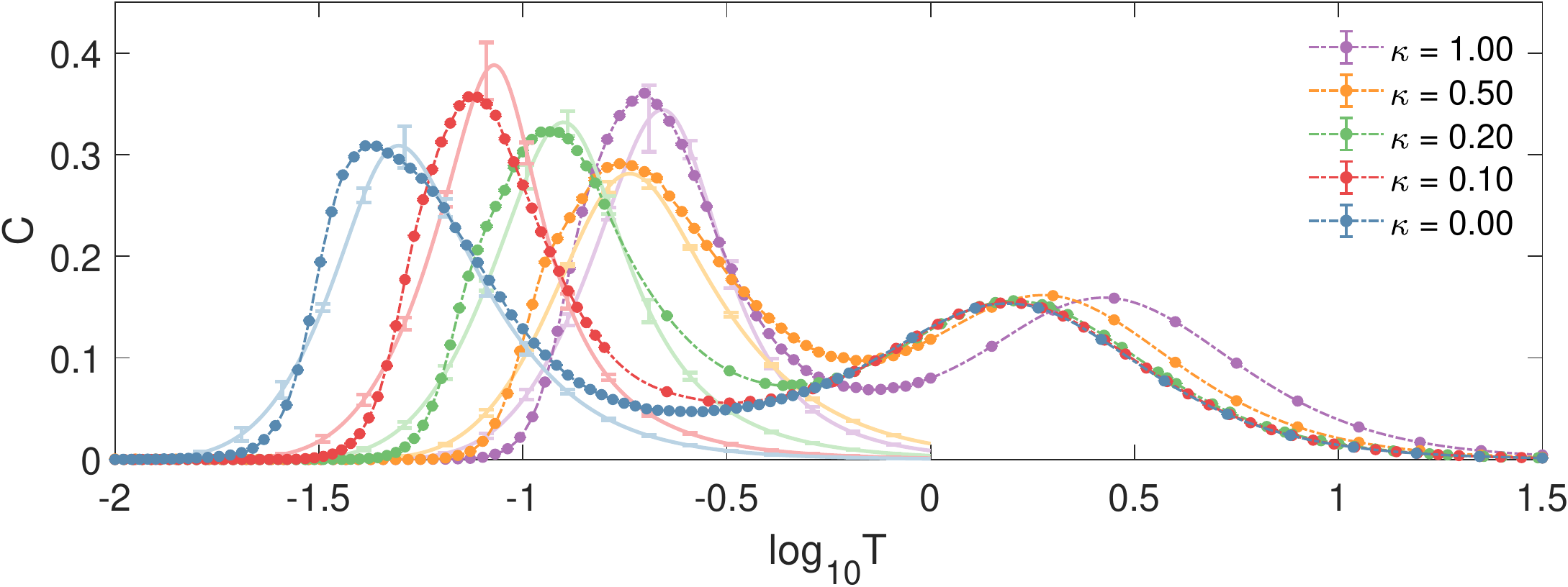}
    \caption{
        Comparison of the specific heat per site for various values of $\kappa$ on a $16\times 16$ torus lattice, computed by the MC simulations (the dots), and from the flux PPE model via \refeq{eq: poly_C} (the solid lines).  
        { 
            The uncertainty of the PPE model specific heat is estimated the same way as described in \reffg{fig: thermolimit_C}.
        }
    }
    \label{fig: C_kap}
\end{figure*}


\section{Flux energetics in the extended Kitaev model ($\kappa\neq 0$)}
\label{sec: mag_flux_energetics}
\subsection{Flux pseudopotential energy on a torus at $\kappa\neq 0$}\label{subsubsec: torus_PPE}

We now turn to the time-reversal symmetry broken Kitaev model and study the  energetics of the many-flux problem at finite $\kappa$.  
  As we discussed in Sec.~\ref{sec:kappanonzero}, the $\kappa$-term introduces  both an energy dispersion to the edge modes, and a bulk energy band gap.   Collectively, these lead to a slightly weaker  dependence  of  the single flux energy  on its proximity to the edge of the lattice; however, the main effects of the boundary on the flux energetics remain the same as in the time-reversal invariant case.  Here 
we therefore only consider  the flux energetics on torus lattices,
   for which we numerically obtain  best-fit PPE polynomials to the average flux energy as a function of the flux density $n_{\phi}$ and the strength of $\kappa$ (see Fig.\ref{fig: flux_kap} (a)).  Note that  we  treat the three-spin interaction as an independent time-reversal symmetry breaking term, rather than as perturbative effect of the magnetic field~\cite{Kitaev2006}, and  thus for curiosity  consider  values of $\kappa$ up to 1.

The value of $\kappa$ affects the flux energetics through its impact  on the fermionic spectrum.  In Fig.\ref{fig: flux_kap} (b)  we plot the fermionic DOS in the flux-free sector  for various values of $\kappa$.   Recall that in the extended Kitaev model,  Majorana fermions can hop to  nearest neighbor sites with an amplitude of $J$,  and to second neighbor sites with an amplitude of $\kappa$. 
    Two main effects of finite  $\kappa$ should be noticed:  first, as expected, $\kappa$ leads to a bulk  gap $\Delta_{\rm bulk}$ in
the fermionic spectrum, which grows  linearly with $\kappa$  up to $\kappa \simeq 0.2$.   The inset in (b) shows that both this fermionic energy gap (dark red line) and the flux gap $E_{2\phi}(\infty)$ (dark yellow line, defined as the flux energy with a pair of maximally separated fluxes on the torus), grow monotonically with $\kappa$.     
Second, for $\kappa > 1/ \sqrt{3}$, 
      the maximum fermionic energy is greater than its $\kappa=0$ value of $6J$, and increases linearly with $\kappa$, leading to the increased maximum fermionic energy that is apparent for $\kappa =1$.

Fig.\ref{fig: flux_kap} (a)  shows the  phenomenological models of flux energetics in the time-reversal symmetry broken case computed on the $\Nx\times\Ny = 16\times 16$ torus. The PPE polynomials are shown using solid lines, while the vertical bars show the mean and standard deviation of  60 random flux configurations at each flux density, with colors indicating the value of  $\kappa$ in each case.  As before, the best-fit polynomials (given explicitly in  Appendix \ref{App:polynomials}) are obtained by fitting  the mean energy as a function of $n_\phi$ for each value of $\kappa$.  We see that 
the slope of the PPE at low flux densities increases monotonically with $\kappa$, due to the increase in the
flux energy gap.  The explicit dependence of  $\bar{\Delta}_{\phi}$ on the strength of $\kappa$ is shown in  Fig. \ref{fig:gapkappa}, and is consistent with the trend for  $E_{2\phi}(\infty)$ shown in the inset of Fig.\ref{fig: flux_kap} (b).    
 For $\kappa \leq 0.5$ this difference in slope is clearly the most significant effect of increasing $\kappa$.   We also observe differences in the curvature of the PPE for different $\kappa$, indicating differences in the average importance of flux interactions, which we will attempt to understand with the minimal two-flux interaction model  in  Sec.\ref{Sec:interactions}.

  \subsection{Flux contribution to the specific heat  at $\kappa\neq 0$ }\label{subsubsec: flux_modelkappa}

The change of the flux energetics due to $\kappa$ is clearly  observable  in the behavior of the specific heat.   This is shown in Fig.\ \ref{fig: C_kap}, which compares the predictions of our flux-PPE model $C_\phi(T)$ with MC results  $C(T)$ on $16 \times 16$ tori.    The solid lines  show the specific heat  for various values of $\kappa$ as a function of temperature  using  the best-fit  flux PPE polynomials  explicitly given in Eq. (\ref{eq:polynomialskappa}) and shown in Fig. \ref{fig: flux_kap} (a).    Comparing these with the results for $C(T)$  obtained with MC simulations on the $16 \times 16$ torus shows that the specific heat peak positions are well approximated by the flux PPE model:   in  Fig.\ \ref{fig: C_kap}  we clearly see a rightward shift in the position of the low-temperature peak with increasing $\kappa$ in both the flux PPE models and the MC  simulation results.  This is due to the fact that  the flux gap increases with $\kappa$.

{  The peak's height and shape are more sensitive to the details of the PPE fit.  In particular, small changes in the values of the polynomial coefficients of the flux PPE can lead to large differences in the peak height and shape; this is apparent when comparing the fit at $\kappa =0$ in Fig. \ref{fig: C_kap} (obtained from a PPE derived from a single system size) to that in Fig. \ref{fig: thermolimit_C}, obtained from the universal best-fit PPE in Eq. (\ref{fbest}).   As a result, as discussed in detail in Appendix \ref{ErrorApp}, the small sampling error in the PPE fits translates to significant error bars near the peak of $C_\phi(T)$.  
Moreover,  the flux PPE model predicts $C_\phi(T)$  based on the average flux energy  for a given flux density, and does  not account for fluctuations in this energy over different configurations with the same flux number, which can also impact the specific shape of the peak.  
These factors, together with finite-size effects in the MC data, lead to differences between the PPE and MC predictions for the peak height and shape.  These are particularly apparent when comparing the PPE results for the specific heat  computed for $\kappa =0.1$ with corresponding MC result.
}

 For $\kappa \gtrapprox 0.2$, our MC simulations also show changes in the high-temperature (fermionic) peak in $C (T)$, which shifts to noticeably higher temperatures for $\kappa =1$.  This reflects the changes in the high-energy fermionic DOS for  larger values of $\kappa$ (see Fig. \ref{fig: flux_kap} (b)).

\section{Two-flux interactions} \label{Sec:interactions}

In the previous sections, we  presented phenomenological models of the average flux energy $\bar{\E}(n_{\phi})$ as a function of flux density $n_{\phi}$, that we argued were universal on sufficiently large tori, and universal for a fixed value of $\Ny$ on the cylinder.  We found that $\bar{\E}(n_{\phi})$ is a concave function of $n_{\phi}$, indicating  that on average,  interactions  between the fluxes are attractive.  For sufficiently large system sizes, we showed that the width  of the distribution of energies around  the best-fit PPE curve was narrow, indicating that at given flux density  the  flux  energy  depends only weakly on the specific flux configuration.
In this section, we examine the microscopics of the flux interactions, and argue that the essential features of our flux PPE models can be understood  by studying  the interaction between just two fluxes. We will also show how these interactions can be tuned by
the time-reversal symmetry breaking term $\kappa$. 

\begin{figure*}
    \includegraphics[width = 1\columnwidth]{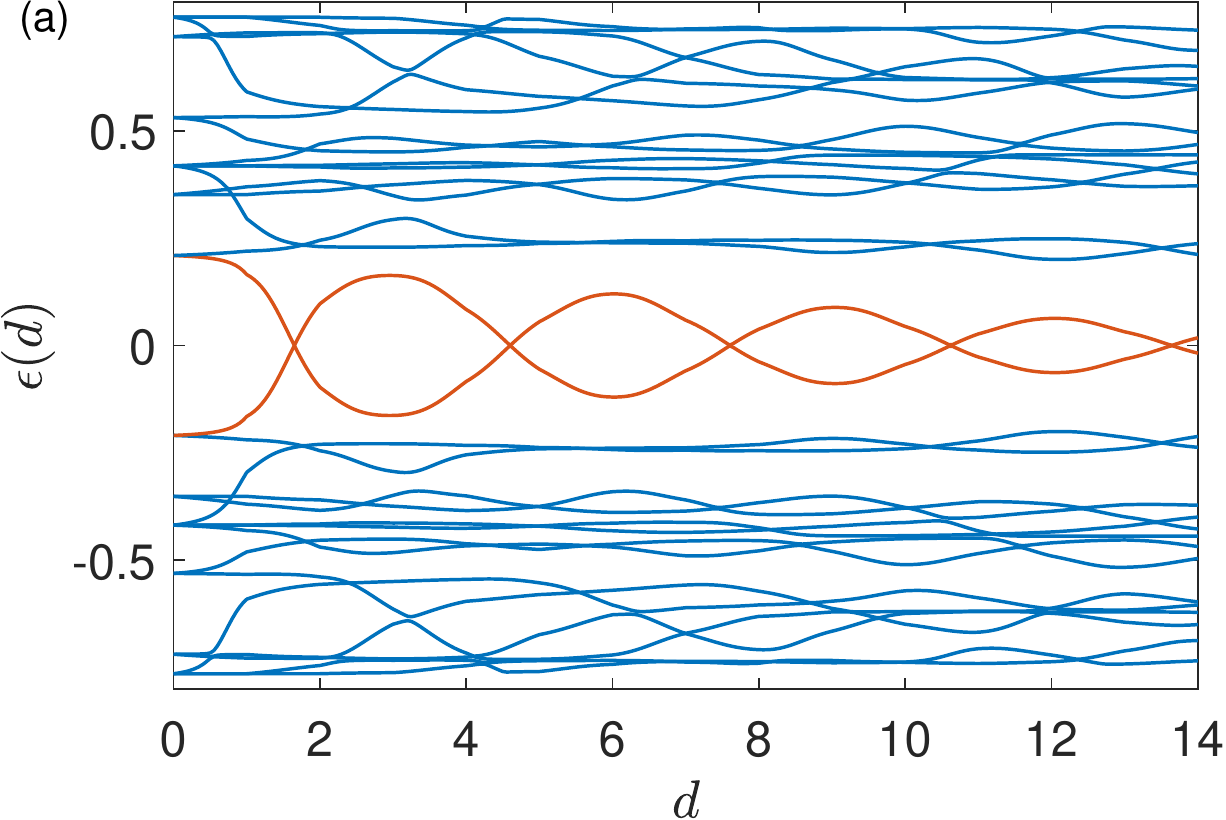}
    \includegraphics[width = 1\columnwidth]{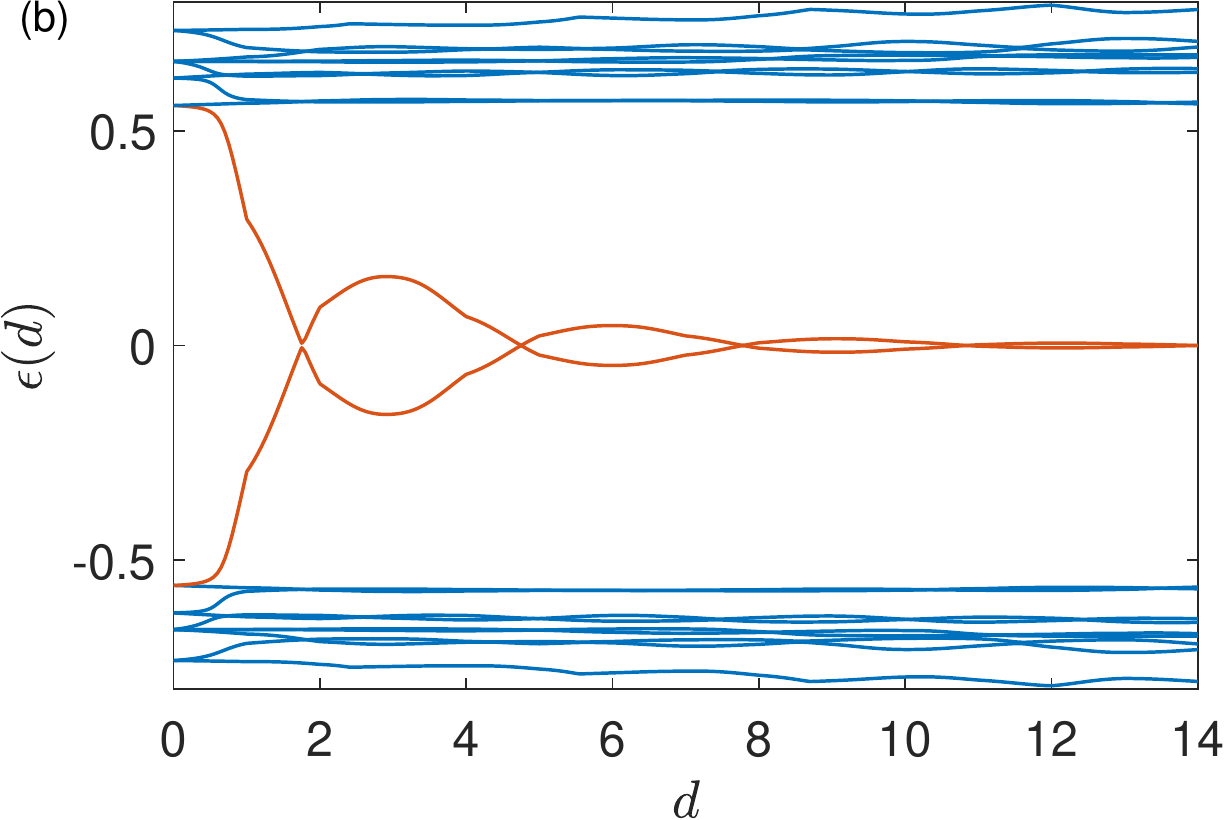}
   \caption{\label{fig: spectrum}  The fermion energy spectrum on $N_1 \times N_2 = 30\times20$ torus as a function of the two-flux separation $d$ along the $\nx$-direction computed  with (a) $\kappa=0$ and (b) $\kappa=0.05$. The plot shows the 40 lowest energy bands.  The lowest-energy fermionic band is associated with the Majorana zero modes bound to the flux pair, with the corresponding energy $\epsilon_0(d)$ shown by the red line.  In both cases $\epsilon_0(d)$  displays decaying oscillations as a function of $d$.  For $\kappa =0.05$ the rest of the fermionic modes are almost independent of $d$ for $d>1$; for $\kappa =0$ all modes exhibit $d$-dependence. }
   \label{fig: 1}
\end{figure*}

At finite $\kappa$, the interaction between fluxes  in the Kitaev model originates from the Majorana zero mode localized near each flux~\cite{Kitaev2006}, leading to flux interactions that are similar to those of vortices in chiral $p$-wave superconductors~\cite{Cheng2009}.  The resulting fermion energetics was studied numercially by Ref.~\cite{lahtinen2011interacting}, who obtained a phenomenological fit to the Majorana fermion energy as a function of the separation between the two fluxes.  Subsequently, similar effective Hamiltonians describing the interactions between Majorana zero modes have been used to study the Majorana  zero mode spectrum in the presence of various flux lattices, often called vison crystals~
 \cite{lahtinen2012topological,ludwig2011two,lahtinen2014perturbed,batista2019}.

 Specifically,  a pair of Majorana zero modes corresponds to two possible states, which can be characterized by their fermion parity.  Thus due to the bound Majorana zero modes, each pair of fluxes
 can have either even or odd fermion parity. As the separation $d$ between the two  fluxes approaches $\infty$,  the energy splitting $2 \epsilon_0(d)$ between these states approaches 0, and the two possibilities are energetically degenerate, whence the name ``zero mode".   At finite separation, however, the wave functions of the Majorana zero modes hybridize, leading to a non-vanishing value of $\epsilon_0(d)$, which can be quantitatively described by~\cite{lahtinen2011interacting}
\begin{align}
    \epsilon_0(d) &= \Delta_{\rm bulk} \cos\left(\frac{2\pi d}{\lambda}\right) e^{-\frac{d}{\xi}}, \label{eq: flux_interaction}
\end{align}
where $\lambda$ is a characteristic wavelength and  $\Delta_{\rm bulk}$ is the complex fermion  bulk energy gap. 
The coherence length $\xi$ is inversely proportional to the bulk fermion gap, i.e. $\xi \sim \Delta_{\rm bulk} ^{-1}$, indicating that  the length-scale over which the flux interactions are appreciable decreases with increasing $\kappa$.
In addition to the exponential decrease with separation, Eq.(\ref{eq: flux_interaction}) shows that $\epsilon_0(d)$ oscillates as a function of the separation $d$.

To show the relationship between Majorana zero modes and the two-flux interactions, we follow Ref. \cite{lahtinen2011interacting} and plot the  fermionic energy spectrum on the torus as a function of the two-flux separation $d$ along the $\nx$-direction, for both $\kappa=0$ (Fig.\ \ref{fig: spectrum} (a)) and $\kappa=0.05$
 (Fig.\ \ref{fig: spectrum} (b)).  
 In Fig.\ \ref{fig: spectrum}
we show the full particle-hole symmetric Majorana fermion spectrum; the   positive energy corresponding to a complex fermion at momentum $\kv$ is given by twice the energy of the upper band at $\kv$.  Note that throughout this paper, we use the symbol $\epsilon$ to denote this complex fermion energy.
We identify the two Majorana states with energies closest to 0 (red lines) as those associated with the Majorana zero mode pair, with energy $\pm \frac{1}{2} \epsilon_0(d)$.  
The two-flux interaction energy $  V_\tx{int}(d)$ is given by  the difference between the energy of the flux pair in the lower energy state at separation $d$ (see \refeq{eq: flux_energy}) and the energy  of the flux pair at infinite separation:
 \begin{align}
   V_\tx{int} (d) =  -  \onefrac{2} \epsilon_0(d) - \onefrac{2} \sum_{ i > 0}( \epsilon_i(d) - \epsilon_i(\infty) ) ,
 \end{align}
 where we have used the fact that $\epsilon_0(\infty) = 0$, and the second sum runs over all positive-energy fermionic states except the lowest-energy one.  
   It is clear that due to the oscillating behavior of  $\epsilon_0(d)$,  $V_\tx{int} (d)$ also  exhibits  short-range modulations.

 Fig.\ \ref{fig: spectrum} (b) shows  that for $\kappa=0.05$, all fermionic modes except the lowest one (shown in red) are almost independent of $d$. This is true for all non-zero $\kappa$, thus at finite $\kappa$ the two-flux interaction energy is due almost entirely to the hybridization of the Majorana zero modes, which determines $\epsilon_0(d) $.  
On the other hand at $\kappa=0$ (see Fig.\ \ref{fig: spectrum} (a)), the entire fermionic spectrum varies with the two-flux separation $d$, and we can no longer study $V_\tx{int} (d)$ merely by examining $\epsilon_0(d)$.    
Moreover at $\kappa =0$ the expression (\ref{eq: flux_interaction}) no longer holds in the thermodynamic limit, where the bulk fermion gap $ \Delta_{\rm bulk}=0$, and $\xi$ is infinite.  In this case we expect a power-law decay to replace the exponential envelope; for finite-size systems, we expect $\xi$ to be on the order of the system size.  
 This is apparent  if we compare Figs.\ \ref{fig: spectrum} (a) and (b), which  show a clear contrast in the length-scale over which $\epsilon_0(d)$ falls off. 
  Thus $V_\tx{int} (d)$ is effectively long-ranged for $\kappa =0$, but short-ranged (for sufficiently large system sizes) when $\kappa>0$.

\begin{figure*}
    \includegraphics[width = 1\textwidth]{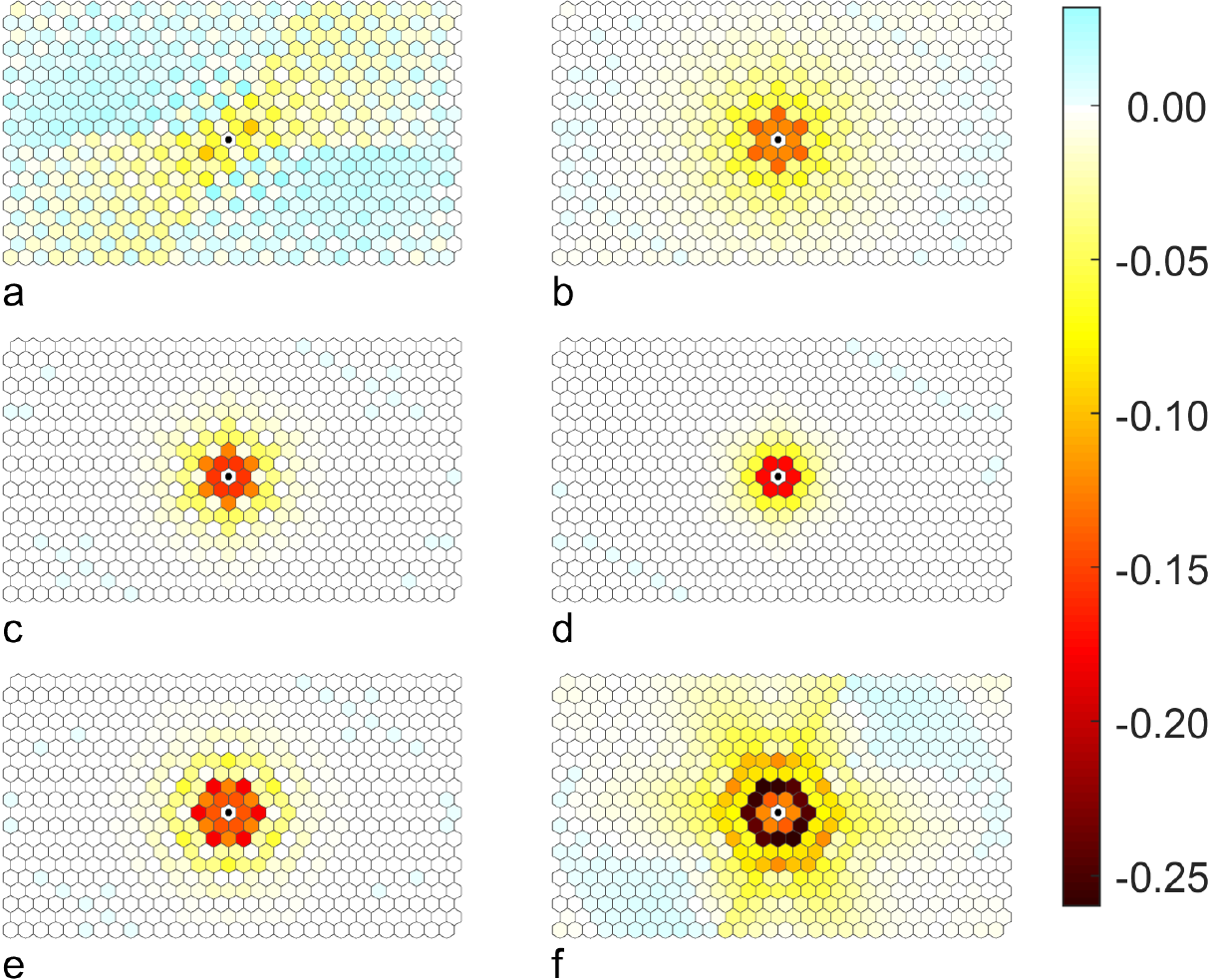}
  \caption{\label{fig: inter_map} The two-flux interaction map on a $30\times 20$ torus for various values of $\kappa$: (a) $\kappa = 0$, (b) $\kappa = 0.05$, (c) $\kappa = 0.1$, (d) $\kappa = 0.2$, (e) $\kappa = 0.5$, (f) $\kappa = 1$. $E_{\phi, 2}(\infty)$ is taken as the energy where separation of the two fluxes is 15 plaquettes.}    
\end{figure*}

\begin{figure*}
    \includegraphics[width = 1\textwidth]{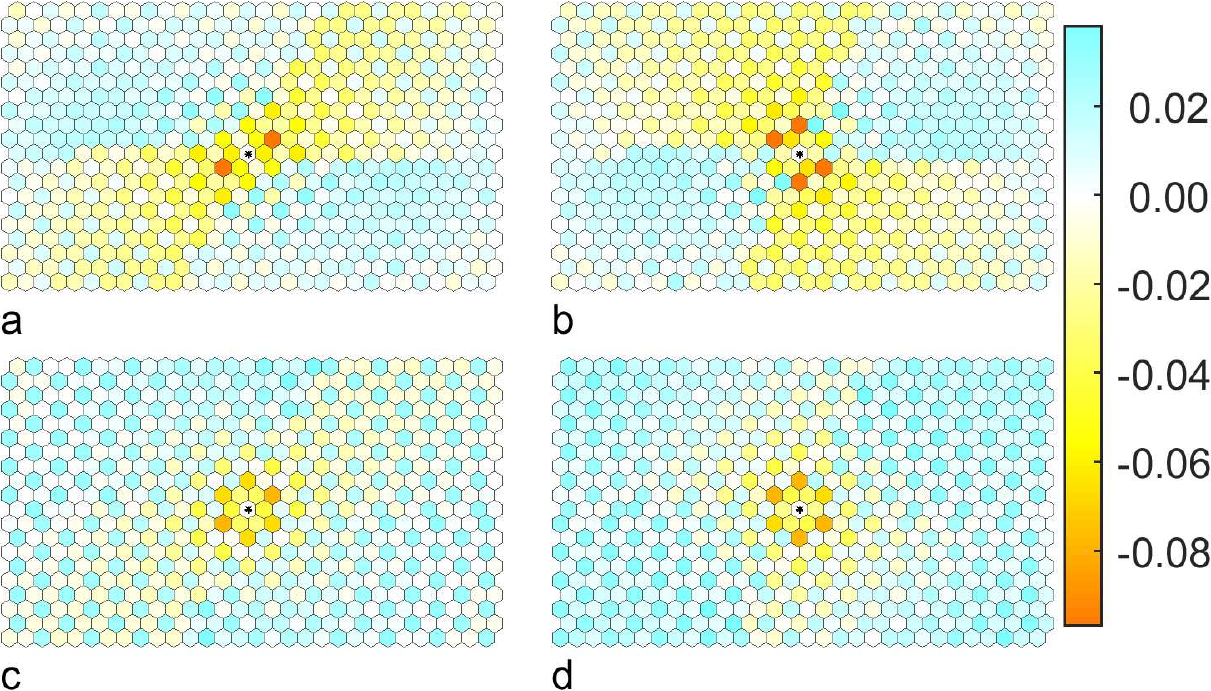}
   \caption{\label{fig: inter_map_4} The two-flux interaction map on a $30\times20$ lattice for the four topological sectors at $\kappa=0$ obtained by: {(a) is generated by taking bonds along the path described in Fig. 19 a to be $-1$, with all other bonds equal to $+1$. 
Bonds in (b) and (c) are obtained from bonds in (a) by flipping the $z$ bonds along a loop in the $\nx$ direction, and the $y$-bonds along a loop in the $\ny$ direction, respectively. In (d) we have flipped the appropriate bonds along both $\nx$ and $\ny$ directions.} $E_{\phi, 2}(\infty)$ is taken as the energy where separation of the two fluxes is 15 plaquettes. 
   }
\end{figure*}

To demonstrate the evolution of the two-flux  interaction  with increasing $\kappa$, Fig.\ \ref{fig: inter_map}  shows  spatial maps of $V_\tx{int} (d)$ on a $30\times20$ torus  for  $\kappa$ ranging from $  0$ (panel (a)) to $1$ (panel (f)). For $\kappa =0$, where the coherence length $\xi$ is comparable to the system size, we see appreciable interactions on virtually all plaquettes in the system. Moreover, they can be either attractive (negative)  or repulsive (positive), depending on the relative positions of the two plaquettes.  
In addition, { at the system sizes studied here}, the spatial pattern of flux interactions depends strongly on the topological sector. This is clearly seen in \reffg{fig: inter_map_4}, which shows the two-flux interaction maps for the four topological sectors.  Here fixing the topological sector effectively determines which of the possible configurations of bonds with $\hat{u}_{ij} = -1$ connect our flux pair (see Appendix \ref{MC} for our conventions);  these differences lead to a significant  anisotropy in the pattern of spatial interactions, as seen in   Fig.\ \ref{fig: inter_map_4}.   
 As a consequence, on finite-size tori a fixed flux configuration corresponds to multiple distinct energies, depending on the underlying topological sector; this is responsible for a significant fraction of the variance in the energy at a given flux sector observed in Fig. \ref{fig: fluxenergydensity} (a).  {
  We emphasize that the anisotropy in the pattern of spatial interactions seen in   Fig.\ \ref{fig: inter_map_4} is due to finite-size effects: choosing a different topological sector in the $\nx$ or $\ny$ direction corresponds to interchanging periodic and anti-periodic boundary conditions in $\nx$ or $\ny$, respectively, for the itinerant fermions. On small tori, where the fermion momenta are discrete, such changes in boundary conditions can lead to significant changes in the energy.   In the thermodynamic limit, however, these differences vanish with the inverse system size, and the four topological sectors are energetically degenerate. This highlights the fact} that finite-size effects due to these topological sectors are present in all of the system sizes shown in Fig. \ref{fig: fluxenergydensity} (b).

 When $\kappa$ increases, the coherence length $\xi$  decreases,
and already for $\kappa = 0.05$, we see that $\xi$ is less than the maximal separation between fluxes on the $30 \times 20$ lattice shown in  \reffg{fig: inter_map}.  Consequently,  the anisotropy corresponding to different topological sectors for $\kappa =0$ is no longer present, indicating that the impact of finite-size effects on flux energetics is already small for this system size. 

Fig.\ \ref{fig: inter_map}  also shows how the two-flux interaction depends on the strength of $\kappa$. 
For $\kappa$ up to $0.2$, we see that $\xi$ decreases with increasing $\kappa$, as anticipated above.  
    For  larger values of $\kappa$ (see
\reffg{fig: inter_map} (d)-(f)), however, we can see  an interesting transition.  As was discussed in Ref.\ \onlinecite{lahtinen2011interacting},
 for $\kappa \gtrapprox 0.2$, the coherence length $\xi$ is on the order of a single plaquette, and cannot decrease further with increasing $\kappa$; this corresponds to the point at which the perturbative treatment used to obtain $\xi \sim \Delta_{\rm bulk}^{-1}$ is no longer valid.  Instead, by  $\kappa \sim 0.5$, we find that the coherence length has begun to {\it increase} with $\kappa$, with a flux interaction pattern that is qualitatively different from the one at small $\kappa$.   

We now connect our findings about the pairwise flux  interactions to the character of the flux energetics when the flux density is finite.  
 For $\kappa =0$ the interactions are longer ranged and hence stronger in magnitude on average than for finite $\kappa$; however from Fig. \ref{fig: inter_map_4} we see that they can be either attractive or repulsive, depending on the relative positions of the two plaquettes.  Thus, we do not see a  higher curvature for our flux PPE (which averages over all positions) for $\kappa =0$ than for $\kappa =0.1,0.2$ shown in Fig.\ \ref{fig: flux_kap}. As noted above, somewhere between $\kappa =0.2$ and $\kappa =0.5$ the coherence length begins to increase with $\kappa$.  Since interactions in this regime are uniformly attractive, this  leads to an increase in curvature of the flux PPE, at least at low flux densities.   However this does not carry over to high flux density: near $n_{\phi} =1$ the flux PPE for $\kappa= 0.5$ shows the lowest curvature of any of our PPE curves, and is relatively close to the straight line expected for non-interacting fluxes.  This indicates that the two-flux interaction does not fully capture the flux energetics at high densities.   In contrast, for $\kappa=1$ we observe substantial curvature, both at low flux densities (in agreement with the longer coherence length observed in Fig. \ref{fig: inter_map} (f)) and at high flux densities.  This is consistent with the 2-flux interaction map, which indicates significantly longer ranged, predominantly attractive interactions at this value of $\kappa$.


\section{Summary}\label{sec: summary}

In this work, we address the question of whether the physics characteristic of the boundary of the Kitaev honeycomb spin liquid has observable thermodynamic signatures.  We consider two aspects of this question: first, in the ultra-low temperature regime, we consider possible thermodynamic signatures of the gapless boundary fermion modes associated with the topological Majorana band structure.  Second, at temperatures on the order of the bulk flux gap, we study the impact of boundary conditions, as well as time-reversal symmetry breaking, on flux energetics of the Kitaev model, and discuss the resulting quantitative impact on the specific heat.


We find that in realistic experiments on currently available sample sizes, the topological boundary flat-band cannot be seen in the specific heat; rather, it can only be detected indirectly through a contribution to the residual entropy.  Because the boundary flat band is topological in nature, this residual entropy is a robust feature of the gapless spin liquid phase -- though at the Kitaev point, there is an additional contribution due to boundary operators that commute with the Kitaev Hamiltonian.  The chiral edge modes that arise in the presence of time-reversal symmetry breaking, on the other hand, are in principle observable in the specific heat, though the temperature scale for this is set by the parameter $\kappa$.  In most experiments this term would be generated by a magnetic field, which must be small compared to the flux gap in order for our analysis to be valid.  We note that though in the pure Kitaev model the temperature scale  of the flux gap is  $\Delta_{\phi} \sim15 K$ for $J \sim 100 K$, in real materials we expect this value to be  smaller.  

We also find that boundary conditions have a significant impact on flux excitations in the Kitaev model.  We quantify this impact on both  cylinder and torus  geometries by using a polynomial fit (the PPE) for the average flux energy as a function of flux density.  We show that a single universal curve provides a good fit for tori of various sizes and aspect ratios  for both the time-reversal symmetric and time-reversal broken cases.  On the cylinder the best-fit depends on  both the density of fluxes in the bulk and on the edges and can be described by a single  universal surface. Finally, we use these universal  best-fits to compute the flux contribution to the low-temperature specific heat in the thermodynamic limit, and show that these agree well with the specific heat obtained in Monte Carlo simulations at  finite system sizes.  
We show that the flux interactions captured by  { the PPE models} have important quantitative { fingerprints on  the shapes  and positions of the low-temperature peaks in the specific heat computed   in different geometries. In particular, in the cylinder, the lower average energy of fluxes near the boundary leads to a broadening of the specific heat peak.  Finally, we analyze how a non-zero value of  the  time-reversal symmetry breaking  strength $\kappa$  tunes the flux energetics, and thus affects the specific heat, finding that the flux energetics are qualitatively different between large and small values of $\kappa$. }

 \vspace*{0.3cm} 
\noindent{\it  Acknowledgments:} 
We thank  Brent Perreault and Wen-Han Kao for helpful discussions. The work of  N.B.P.\ was supported by the  National Science Foundation  under Award No.\ DMR 1929311.   F. J. B. are grateful for the support of the  National Science Foundation under Award No.\ DMR  1352271.
K. F.  was partly supported by  the  National Science Foundation under Awards No.\ DMR  1352271 and partly by  DMR 1929311. 

\appendix
\section{ Majorana fermion spectrum in a given flux sector}\label{App:MFspectrum}
Within a given flux sector,  the $\hat{u}_{ij}$ operators in the Hamiltonian (\ref{eq: Hamiltonian}) are replaced by  the corresponding eigenvalues ${u}_{ij}$, so the Hamiltonian (\ref{eq: Hamiltonian}) becomes  quadratic in the Majorana fermion operators. 
Exploiting the bipartite nature of the honeycomb lattice, and noting
that each unit cell $l$ has two sites $\mathbf{r}_{A,l}$ and
$\mathbf{r}_{B,l}$ in the two sublattices $A$ and $B$, the Hamiltonian (\ref{eq: Hamiltonian}) can be written as 
\begin{align}\label{HamMmatrix}
H =&
\sum_{\langle l,l' \rangle } i M_{ll'} c_{A,l} c_{B,l'} +\\\nonumber& \sum_{\langle\langle l,l' \rangle\rangle} i {\tilde M}_{ll'} (c_{A,l} c_{A,l'} +c_{B,l} c_{B,l'}),
\end{align}
 where  the first term describes the nearest neighbor hopping of the Majorana fermions with $M_{ll'} = -J^\alpha
{u}_{AB,ll'}$ if $\mathbf{r}_{A,l}$
and $\mathbf{r}_{B,l'}$ are connected by $\alpha$-bond and $M_{ll'} = 0$ otherwise. The second term describes  the second neighbor hopping
between two sites of the A  (or B) subblatice with  ${\tilde M}_{ll'} = -\kappa
{u}_{AB,ll^{"}} {u}_{AB,l'l^{"}}$.     Using the singular-value decomposition $M
= U \cdot \Lambda \cdot V^T$, the resulting  free-fermion Hamiltonian can be written in the canonical form as
\begin{align}\label{eq:free-fermionHamiltonian}
    H  = \sum_n\varepsilon_n (\psi_n^{\dag} \psi_n^{\phantom{\dag}} - 1/2),
\end{align}
 where the fermions $\psi_n = (\gamma_{A,n} + i \gamma_{B,n}) / 2$ are expressed in
terms of Majorana fermions $\gamma_{A,n} = \sum_l U_{ln} c_{A,l}$ and $\gamma_{B,n} =
\sum_l V_{ln} c_{B,l}$ on sublattices $A$ and $B$, respectively, and  $\varepsilon_n = 2
\Lambda_{nn}$ are their energies.
Since $M$ is a real matrix,
$U$ and $V$ are real orthogonal matrices, while $\Lambda$ is a
diagonal matrix with non-negative (real) entries.

\section{Statistical Mechanics of the Kitaev model} \label{App:Kitaevthermodynamics}

Here we
briefly  outline the distinctive aspects of the thermodynamics 
of the Kitaev model \cite{Nasu2015,Nasu2016,Motome2019}. Given the exact solution of the model~\cite{Kitaev2006},
both $\mathbb{Z}_2$ fluxes and fermionic fractional excitations contribute to the thermodynamic behavior of the system. In a given flux configuration, $\phi_p$, with a given fermionic occupation number configuration, $\{n_i\}$, the corresponding energy of the system is $\sum_i \epsilon_{i,\phi_p} (n_i- \frac{1}{2})$, where the fermionic energy levels $\epsilon_{i,\phi_p}$ are obtained by diagonalizing the Majorana fermion Hamiltonian in a given flux configuration $\phi_p$ (see details in Appendix \ref{App:MFspectrum}).  The energy of the  lowest-energy state in a given flux configuration,
\begin{equation}
E^{(0)}_{\phi_p}\equiv -\frac{1}{2} \sum_{i}\epsilon_{i,\phi_p},  \label{eq: flux_energy}
 \end{equation}
  which corresponds to all unoccupied fermionic states, is now  associated with the energy of  a corresponding flux sector.
   
   The partition function of the system is given by:
 \begin{align}
   Z= \sum_{\phi_p, n_i}e^{-\beta\sum_{i}\epsilon_{i,\phi_p} (n_i- \frac{1}{2})}= \sum_{\phi_p} Z_{\phi_p} \label{eq:Z},
   \end{align}
   with
    \begin{align}
   Z_{\phi_p} &= e^{-\beta  E^{(0)}_{\phi_p}}\prod_i(1+e^{-\beta\epsilon_{i, \phi_p}}), \label{eq:Zp}
\end{align}
where  we denote $\beta = 1/T$. The expectation value of the total energy  at a given temperature  is then given by 
\begin{align}
  \langle E\rangle& = \onefrac{Z}\sum_{\phi_p, n_i}\sum_i \epsilon_{i,\phi_p} (n_i- \frac{1}{2}) e^{-\beta \sum_i \epsilon_{i,\phi_p} (n_i- \frac{1}{2})}  \nonumber \\
 & =
 \onefrac{Z}
     \sum_{\phi_p}
E_{\phi_p}  Z_{\phi_p} 
\end{align}
where 
$
E_{\phi_p} = \sum_i \epsilon_{i, \phi_p} n_F( \beta\epsilon_{i, \phi_p})  +  E^{(0)}_{\phi_p}
 $
and $n_F ( \beta\epsilon_{i, \phi_p}) = 1/(e^{\beta\epsilon_{i, \phi_p}}+1)$ is the Fermi-Dirac distribution function.
The specific heat of the Kitaev model is thus given by:
\begin{flalign}\label{eq: C_formula}
    C &\!=\! \frac{\ud   \langle E\rangle  }{\ud T}\!\\&\! =\! -\onefrac{T^{2}} \!\sum_{\phi_p} \! \left( E_{\phi_p} \frac{\partial (Z_{\phi_p} / Z)}{\partial \beta}\!+\! \frac{Z_{\phi}}{Z} \frac{\partial E_{\phi_p}}{\partial \beta} \!\right)\! \nonumber\\\nonumber
 &\!=\!    \onefrac{T^{2}}\left(\langle 
    E_{\phi_p}^{2}\rangle-\langle E _{\phi_p}\rangle^{2} 
- \lefta\frac{\partial E_{\phi_p}}{\partial \beta}\righta \right)\, ,
\end{flalign}
where we have simplified
the first term in the second line  via:
\begin{flalign} 
    &  \onefrac{Z} \sum_{\phi_p} E_{\phi_p} \frac{\partial Z_{\phi_p}}{\partial \beta} - \onefrac{Z^{2}} \frac{\partial Z}{\partial \beta} \sum_{\phi_p} E_{\phi_p} Z_{\phi_p}  \label{eq: 5}  \\ 
      &=\!-\onefrac{Z} \sum_{\phi_p}  E_{\phi_p}^{2} Z_{\phi_p}\!+\!   \langle E_{\phi_p}\rangle  \onefrac{Z} \sum_{\phi_p} E_{\phi_p} Z_{\phi_p} \label{eq: 6} \\
     &= -\langle E_{\phi_p}^{2}\rangle + \langle E_{\phi_p}\rangle^{2}, \label{eq: 7}
\end{flalign}
i.e., it is simply the variance of the fermionic energy.

\section{Monte Carlo  method for the flux-fermion model}
\label{MC}

%
\begin{figure}[!t]
    \includegraphics[width = 1\columnwidth]{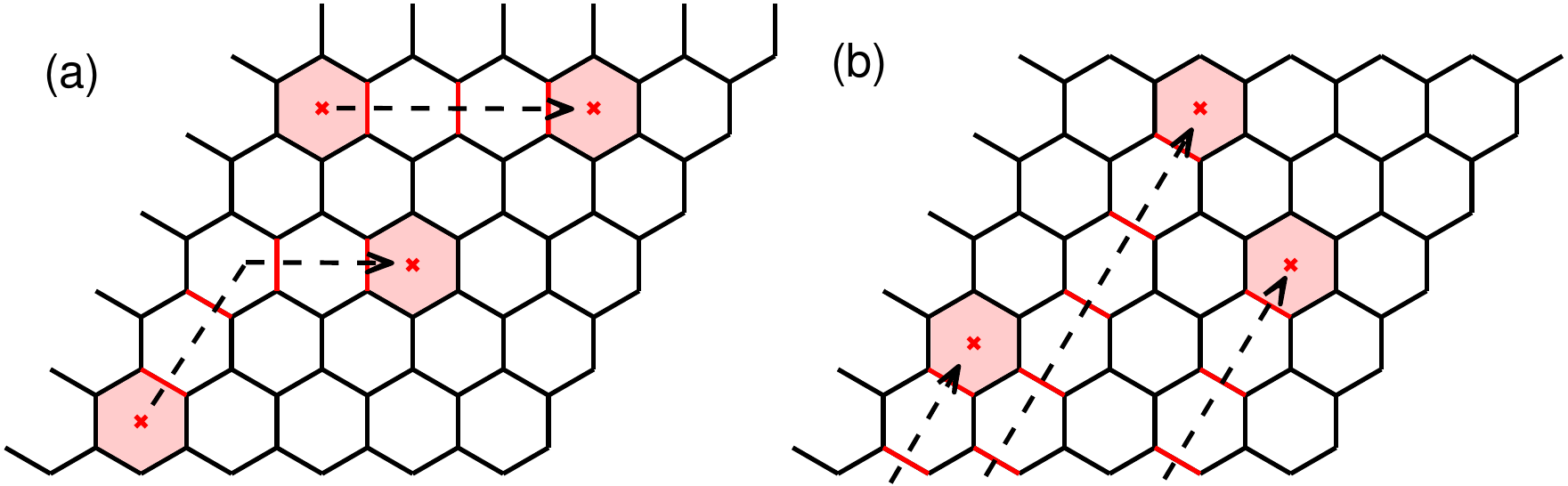}
\caption{Gauge choice used to generate a given flux configuration on (a): torus (b) cylinder in our MC algorithm. The red bonds correspond to $u_{ij} = -1$. The fluxes are located on the red plaquettes.}
\label{fig: flux_gen}
\end{figure}

 In this appendix, we discuss  details  of the implementation of the MC   algorithm~\cite{nasu2014vaporization}, which has been used  for  the computation of the specific heat of the flux-fermion model  (\ref{eq: C_formula}). The basic idea of this MC algorithm is that we can perform sampling over flux configurations $\{\phi_p\}$ classically by exploiting the fact that the energy  of  each  flux configuration, $  E^{(0)}_{\phi_p}$,  can be  computed exactly by  diagonalizing  the  quadratic  Majorana Hamiltonian (\ref{eq:free-fermionHamiltonian}).  From Eqs.(\ref{eq:Z}) and (\ref{eq:Zp}), the probability distribution function for fluxes  is defined as  
\begin{align}
    p(\phi_p) =\frac{ Z_{\phi_p}}{Z}=
 \onefrac{Z} e^{-\beta  E^{(0)}_{\phi_p}} \prod_i(1 + e^{-\beta\epsilon_i}). \label{eq: distribution}
\end{align}

 In practice, to implement a given flux configuration  $\phi_p$, we must choose one of the many possible bond configurations $\{ u_{ij} \}$ that lead to fluxes on the desired plaquettes. Fig.\ \ref{fig: flux_gen} shows the convention that we use for our MC simulations. On a torus lattice, a pair of fluxes is generated as follows. (i) Choose a pair of plaquettes, which are separated by $a \nx + b \ny$.  (ii) Choose a path on the dual lattice that connects these two plaquettes by first crossing $a$ bonds in the $+\nx$ direction, and then $b$ bonds in the $+\ny$ direction, and flip the sign of $u_{ij}$ on all bonds that this path crosses.  The same convention can be used to move a flux between two plaquettes.
To create a more general flux configuration, steps (i)-(ii)  can be repeated.
  On a cylinder lattice, we can create a single flux on the plaquette $p$ by choosing a path along the $+\ny$ direction connecting $p$ to the lower boundary of the cylinder,  and flipping the sign of $u_{ij}$ on all of the bonds crossed by this path. Flux annihilation is carried out using the same conventions.  On the cylinder, we move a flux from plaquette $p$ to plaquette $p'$ by annihilating the flux on plaquette $p$, and creating a new flux on $p'$.
  
 All the simulations for the  specific heat reported in the main text were done in the following way.  The simulations were performed  on finite-size  lattices with  spanning vectors   $(\mathbf{R}_1,\mathbf{R}_2)=(N_1 \nx, N_2 \ny)$
 with  either periodic boundary conditions, which we refer to as torus lattices, or with open boundary conditions in the $ \ny$ direction and periodic boundary conditions in the $ \nx$ direction, which we refer to as cylinder lattices.  
 We initially perform  1,000 MC steps for thermalization and then 100,000 steps for measurement on a lattice of up to $2\times 24 \times 6= 288$ sites.
 
 \begin{figure}[!b]
    \includegraphics[width=1\columnwidth]{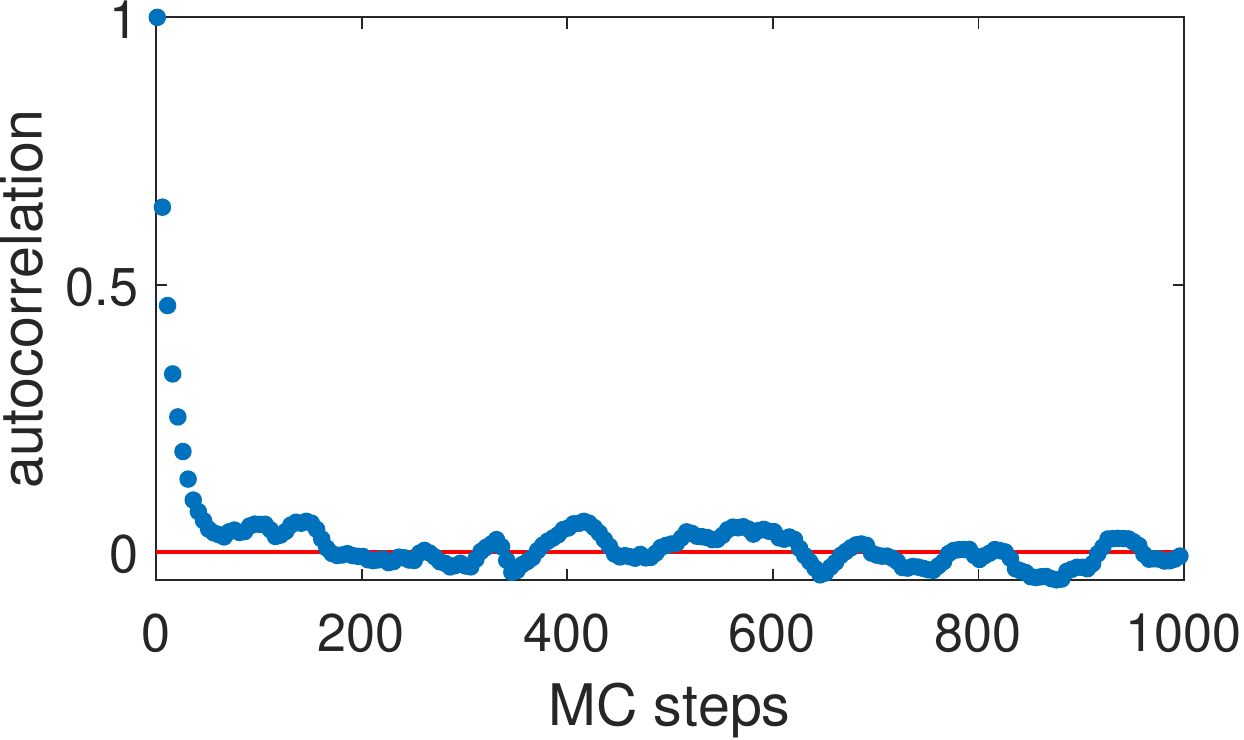}
    \caption{The autocorrelation as a function of the number MC steps on an $N_1\times N_2 = 16\times 16$ torus lattice, at the temperature of $10^{-1.3}$.  This is approximately the temperature of the low-temperature peak in the specific heat shown in \reffg{fig:CS_cyl_tor}}
    \label{fig:autcorr}
\end{figure}

Three kinds of MC moves are implemented: (1) Randomly shuffle all the fluxes while keeping the total flux number fixed. (2) Count the current total flux number, change the total flux number by two, and then randomly place the updated number of fluxes on the plaquettes. (3) Flip all the bonds crossed by a path encircling the system in (one of) the periodic directions, which will change the topological sector while keeping the flux configuration fixed.  The transition (or acceptance) probability, satisfying the detailed balance  condition, is then defined as
 \begin{flalign}
    T(x \to y) &= \min \left(1, \frac{p(y)n(y)}{p(x)n(x)}\right), \label{eq: transition} 
\end{flalign}
 where $x$ represents a certain flux configuration $\{\phi_p\}$ and  $n(x)$ is the number of the candidate flux configurations in a proposal (see Ref. \onlinecite{geyer1992practical} for details). Since these MC moves are all global, our algorithm does not struggle with local minima, and we find good convergence without requiring parallel tempering. 
In order to demonstrate the quality of convergence, in \reffg{fig:autcorr} we plot the autocorrelation function of our MC algorithm for $N_1\times N_2=16\times16$ torus lattice at temperature equal to $10^{-1.3} \approx 0.05 $, a typical temperature near the low-temperature peak of specific heat.  (We have found that this is the temperature regime where MC convergence is most difficult). The autocorrelation function  reaches zero in around 200 MC update steps, indicating that our MC moves lead to thermalization on a relatively rapid time-scale.

 A note of caution is in order here. For  lattices with periodic boundaries, there is a parity constraint \cite{pedrocchi2011physical} on the total fermion parity $\prod_i (-1)^{n_i}$. Specifically, for a given  lattice shape with fixed boundary conditions and gauge field (bond operator) configurations, we find that the physical states have fixed fermion parity; 
 states in the flux-fermion model with opposite fermion parity are unphysical, in the sense that they do not correspond to states in the spin Hilbert space \cite{zschocke2015physical}. 
In the thermodynamic limit, this constraint has little relevance for the model's spectrum, and hence its thermodynamics; however it can lead to substantial finite-size corrections to the energy spectrum, which in turn can be relevant for thermodynamic quantities such as the specific heat.

One way to circumvent the parity constraint is to remove a single bond from the lattice \cite{nasu2014vaporization}. Exactly as for the open cylinder boundary discussed in the main text, this creates a pair of dangling Majorana fermions, and a corresponding zero-energy state.  This ensures that for every state with odd fermion parity, there is a state of identical energy with even fermion parity; in this case we do not need to know which of the two states is physical to compute thermodynamic quantities exactly.  (We do, however, need to account for the impact of this zero-mode on the entropy $S_{\infty}$).  
Though the numerical results shown here are evaluated with all bonds present in the lattice, we have checked that the differences caused by removing the single bond is within the statistical error of our Monte Carlo simulations.

\section{Extrapolating the flux gap energy to thermodynamic limit}\label{App:extrapolation}
\begin{figure}
    \includegraphics[width=0.49\columnwidth]{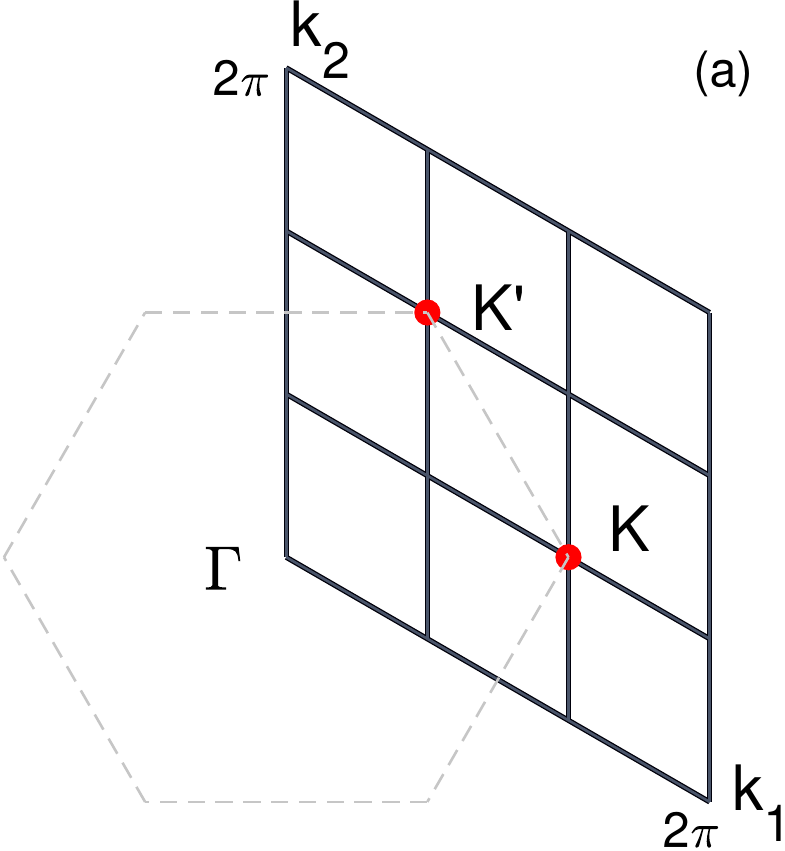}
    \includegraphics[width=0.49\columnwidth]{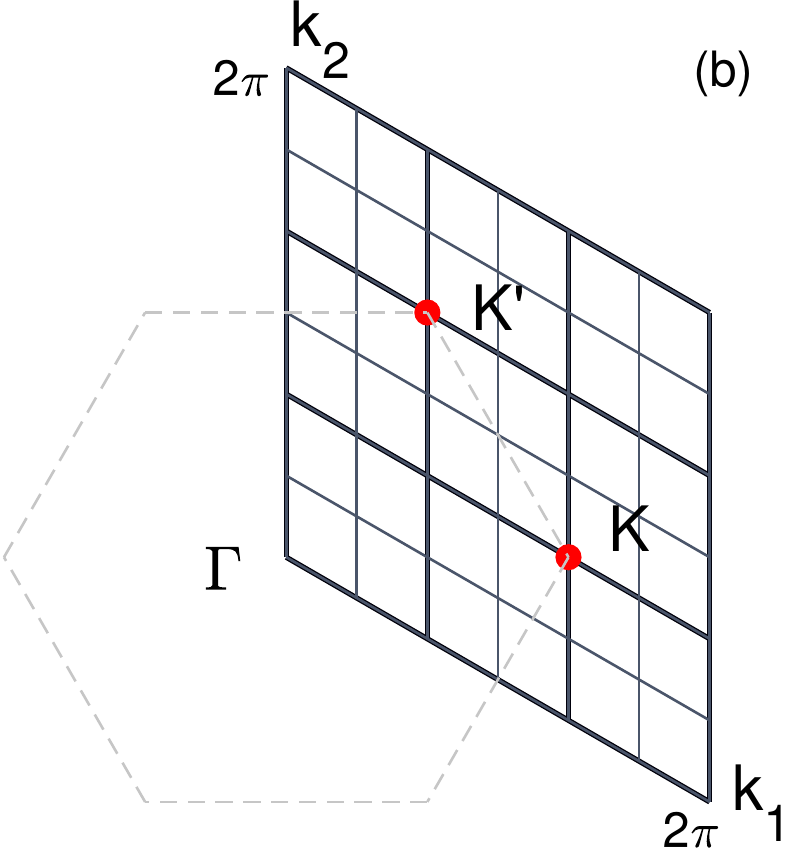}
    \caption{The reciprocal lattices that correspond to two real-space lattice sizes: (a) $L=3$, (b) $L=6$, where $N_1 = N_2 = L$. The dashed honeycomb is the Brillouin zone. The two red points $K$, $K'$ are the two nonequivalent Dirac points.  This shows that only when $L = 3k$ do the Dirac points sit on the sites of reciprocal lattice.}
    \label{fig: discrt_Bzone}
\end{figure}

As we discussed in Sec.\ref{subsec: one_flux gap} of the main text, when extrapolating the flux gap energy to infinitely large lattice size, generally the energies display period-3 oscillatory finite-size behavior \cite{Kitaev2006}.   In order to understand this behavior, we consider the flux-free sector, where the 2D reciprocal space is well defined. Fig.\ \ref{fig: discrt_Bzone} shows the reciprocal lattices corresponding to two real-space lattice sizes: $L = 3$ and $L=6$.  We see that for $L = 3k$ can the two Dirac points $K$, $K'$ sit on reciprocal lattice sites, while for $L = 3k + 1$ and $L = 3k+2$, the Dirac point is off the reciprocal lattice by a distance of order $1/L$.  This difference in the fermionic spectrum $\{\epsilon_i\}$ leads to a different series of flux gaps $ E^{(0)}_{\phi_p}$ for $L = 3k, 3k+1$, and $3k+2$ respectively. 

\begin{figure}[!h]
    \includegraphics[width=0.9\columnwidth]{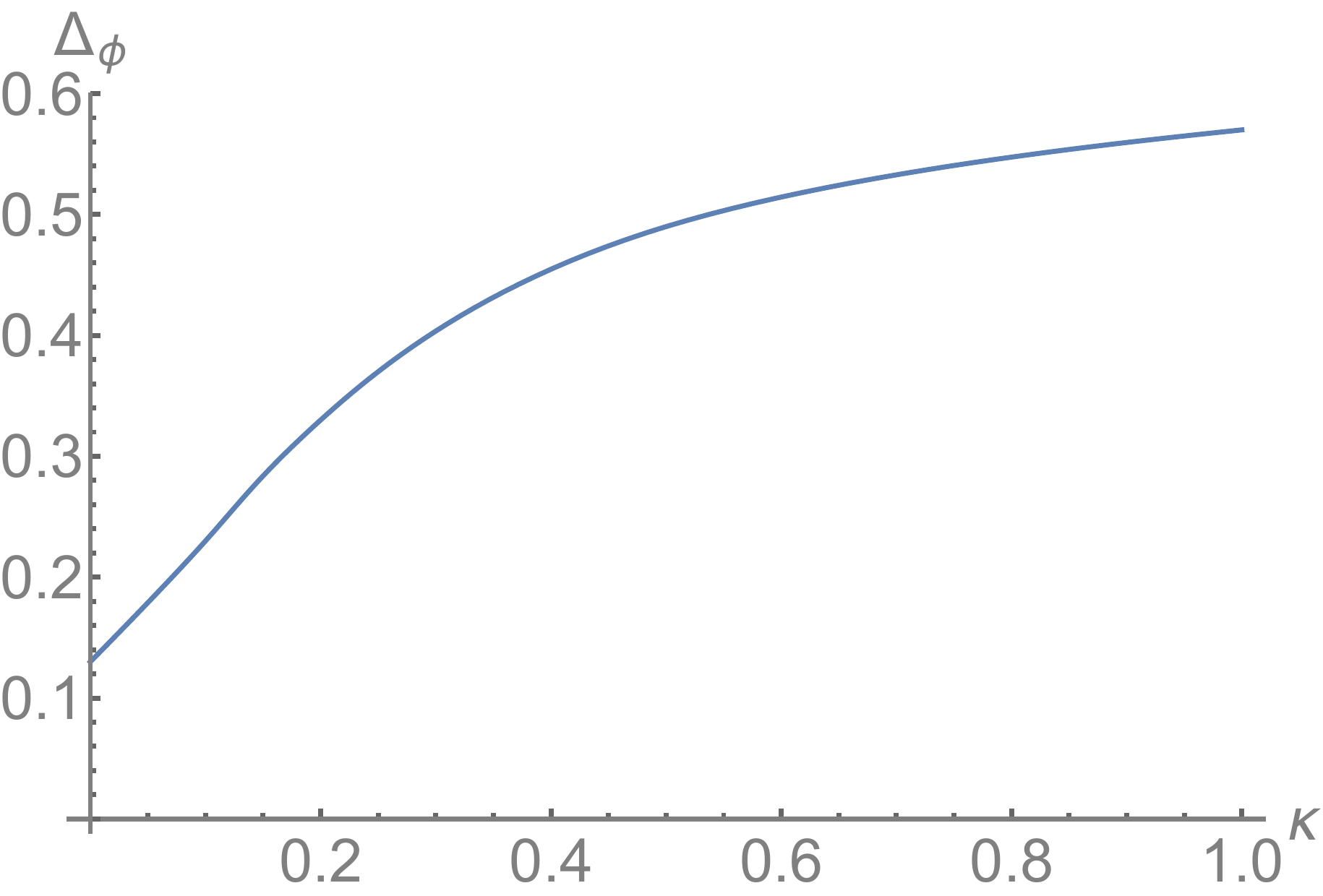}
    \caption{The dependence of the flux gap $\Delta_{\phi}$ on $\kappa$.}
    \label{fig:gapkappa}
\end{figure}

\section{
The best-fit flux  PPE  polynomials for different $\kappa$ on the $\Nx\times\Ny = 16\times 16$ torus}\label{App:polynomials}

{ The equation of the best-fit polynomials   $\mathcal{E}(n_\phi)$ for various $\kappa$  shown in Fig.\ref{fig: flux_kap} are:
\begin{widetext}
\begin{align}\label{eq:polynomialskappa}
    \kappa = 0.0, &\quad
    \mathcal{E}(n_\phi) = -0.309 n_\phi^6 +0.814 n_\phi^5 -0.859 n_\phi^4 +0.456 n_\phi^3 -0.176 n_\phi^2 +0.141 n_\phi \nonumber\\
    \kappa = 0.1, &\quad
    \mathcal{E}(n_\phi) = -0.024n_\phi^5 -0.115 n_\phi^4 +0.286 n_\phi^3 -0.282 n_\phi^2 +0.232n_\phi \nonumber\\
    \kappa = 0.2, &\quad
    \mathcal{E}(n_\phi) = -0.055 n_\phi^5 -0.075 n_\phi^4 + 0.270 n_\phi^3 -0.316 n_\phi^2 + 0.330 n_\phi \nonumber\\
    \kappa = 0.5, & \quad
    \mathcal{E}(n_\phi) = 0.239 n_\phi^5 -0.652 n_\phi^4 + 0.679 n_\phi^3 -0.403 n_\phi^2 + 0.489 n_\phi \nonumber\\
    \kappa = 1.0, & \quad
    \mathcal{E}(n_\phi) = -0.157 n_\phi^5 -0.062 n_\phi^4 + 0.348 n_\phi^3 -0.536 n_\phi^2 + 0.566 n_\phi 
\end{align}
\end{widetext}
We find that while sixth-order polynomials give the best fit to C for $\kappa=0$, for non-zero kappa a better fit is achieved with fifth-order polynomials.  We note, however, that it is difficult to distinguish the fifth- and sixth-order fits based on residual errors, which are comparable in both cases.  This reflects the extreme sensitivity of C to the details of the PPE curve used.}

The  dependence of the flux gap ${\Delta}_{\phi}$ computed by the first derivatives of the PPE  polynomials on the strength of $\kappa$ is shown in  Fig. \ref{fig:gapkappa}.

{
\section{ The computation of the error bars for PPE curves} \label{ErrorApp}

To obtain the error estimates in Figs. \ref{fig: thermolimit_C}, \ref{fig: C_cyltor}, and  \ref{fig: C_kap}, we generate 60 PPE-like curves by randomly sampling from the data used to generate the initial PPE curve, choosing one value of the energy density for each value of the flux density.  For each set of data, we generate a PPE curve using the best-fit polynomial obtained from the mean of the data.  Then, from each such PPE curve, we generate a specific heat curve using Eq. (\ref{eq: poly_C}) (or, for cylinder lattices, Eq. (\ref{eq: exact_sum}).  The error bars at each point show one standard deviation of the resulting curves.  
Some results are shown in Fig. \ref{fig:errorkappa}.   

\begin{figure*}
    \includegraphics[width=0.95\textwidth]{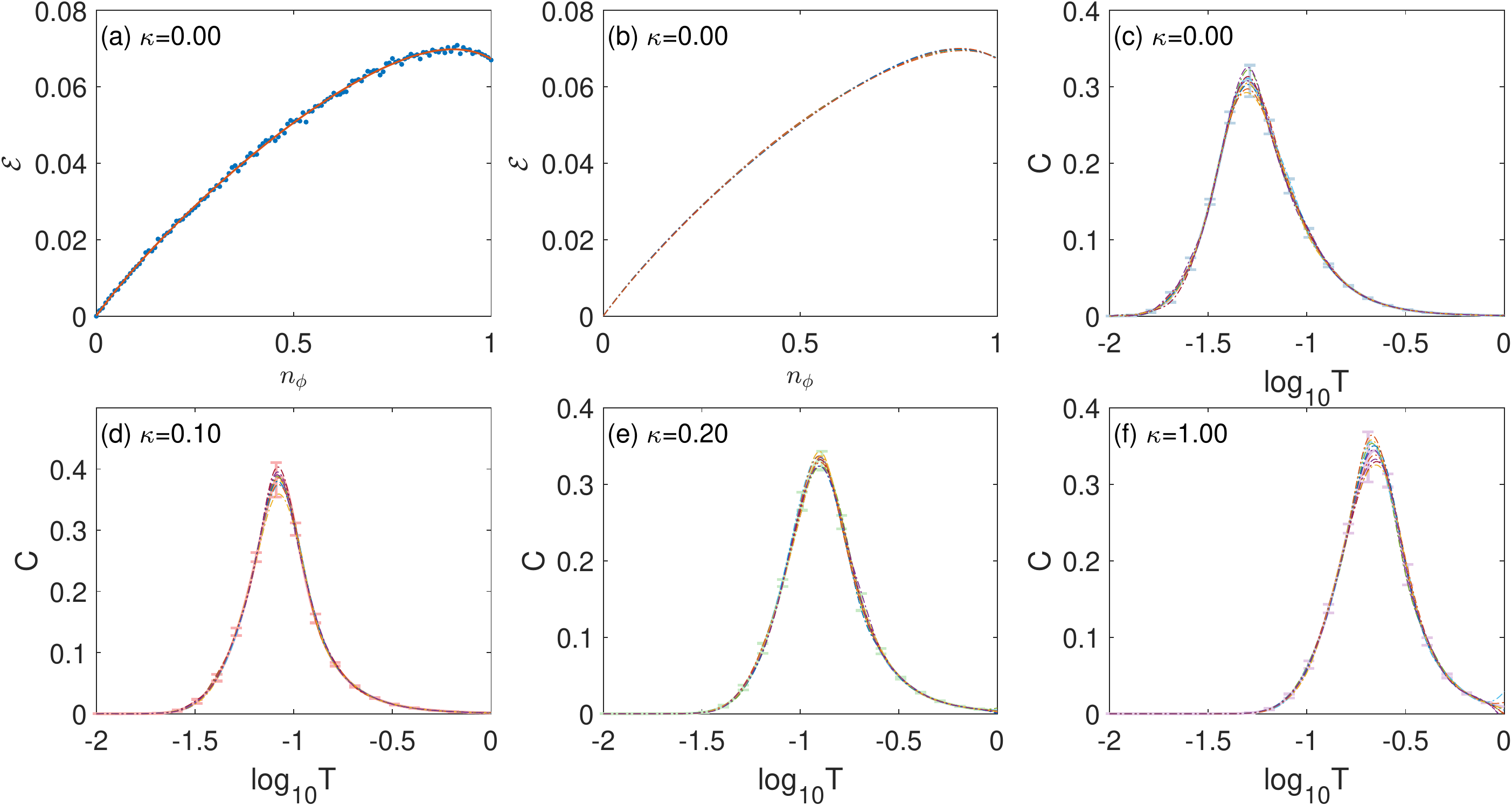}
    \caption{ {   Error estimates for of specific heat curves from flux PPE fits.  (a) One random sample drawn from the data, shown together with the corresponding flux PPE curve.  (b) An ensemble of a 10 selected flux PPE curves generated from these random samples.  (c)-(f)  Plots of the corresponding specific heat, calculated using Eq. (\ref{eq: poly_C}).   We see that very small variations in the flux PPE correspond to relatively large variations in the precise shape of the specific heat peak. }}
    \label{fig:errorkappa}
\end{figure*}
}

\bibliography{thesis}
\end{document}